\documentclass[11pt,a4paper]{article}
\usepackage{psfrag}
\usepackage{graphics}
\usepackage{graphicx}
\usepackage{float}
\usepackage{amssymb,epsfig,amsmath,euscript,array}
\usepackage{cite}
\usepackage{multicol}
\usepackage{subfig}
\usepackage{fancyhdr}
\usepackage{color}
\usepackage{scalerel,stackengine}
\usepackage{pdflscape}
\usepackage{hyperref}

\makeatletter
\@addtoreset{equation}{section}
\makeatother

\usepackage[us,24hr]{datetime}

\newcounter{multieqs}

\newcommand{\be}{\begin{equation}}
\newcommand{\ee}{\end{equation}}

\newcommand{\bm}[1]{\mbox{\boldmath $#1$}}

\def\bd{\begin{document}}
\def\ed{\end{document}}
\def\nn{\nonumber}
\def\bea{\begin{eqnarray}}
\def\eea{\end{eqnarray}}
\let\bm=\bibitem
\let\la=\label

\def\npb#1#2#3{Nucl. Phys. {\bf{B#1}} #3 (#2)}
\def\plb#1#2#3{Phys. Lett. {\bf{#1B}} #3 (#2)}
\def\prl#1#2#3{Phys. Rev. Lett. {\bf{#1}} #3 (#2)}
\def\prd#1#2#3{Phys. Rev. {D \bf{#1}} #3 (#2)}
\def\cmp#1#2#3{Comm. Math. Phys. {\bf{#1}} #3 (#2)}
\def\cqg#1#2#3{Class. Quantum Grav. {\bf{#1}} #3 (#2)}
\def\nppsa#1#2#3{Nucl. Phys. B (Proc. Suppl.) {\bf{#1A}}#3 (#2)}
\def\ap#1#2#3{Ann. of Phys. {\bf{#1}} #3 (#2)}
\def\ijmp#1#2#3{Int. J. Mod. Phys. {\bf{A#1}} #3 (#2)}
\def\rmp#1#2#3{Rev. Mod. Phys. {\bf{#1}} #3 (#2)}
\def\mpla#1#2#3{Mod. Phys. Lett. {\bf A#1} #3 (#2)}
\def\jhep#1#2#3{J. High Energy Phys. {\bf #1} #3 (#2)}
\def\atmp#1#2#3{Adv. Theor. Math. Phys. {\bf #1} #3 (#2)}

\newcommand{\EQ}[1]{\begin{equation} #1 \end{equation}}
\newcommand{\AL}[1]{\begin{subequations}\begin{align} #1 \end{align}\end{subequations}}
\newcommand{\SP}[1]{\begin{equation}\begin{split} #1 \end{split}\end{equation}}
\newcommand{\ALAT}[2]{\begin{subequations}\begin{alignat}{#1} #2 \end{alignat}\end{subequations}}
\def\beq{\begin{equation}}
\def\eeq{\end{equation}}

\def\N{{\cal N}}
\def\sst{\scriptscriptstyle}
\def\thetabar{\bar\theta}
\def\Tr{{\rm Tr}}
\def\one{\mbox{1 \kern-.59em {\rm l}}}
 \def\Nh{\hat{N}}

\def\a{\alpha}      \def\da{{\dot\alpha}}
\def\b{\beta}       \def\db{{\dot\beta}}
\def\c{\gamma}  \def\G{\Gamma}  \def\cdt{\dot\gamma}
\def\d{\delta}  \def\D{\Delta}  \def\ddt{\dot\delta}
\def\e{\epsilon}        \def\vare{\varepsilon}
\def\f{\phi}    \def\F{\Phi}    \def\vvf{\f}
\def\h{\eta}
\def\k{\kappa}
\def\l{\lambda} \def\L{\Lambda}
\def\m{\mu} \def\n{\nu}
\def\o{\omega}
\def\p{\pi} \def\P{\Pi}
\def\r{\rho}
\def\s{\sigma}  \def\S{\Sigma}
\def\t{\tau}
\def\th{\theta} \def\Th{\Theta} \def\vth{\vartheta}
\def\X{\Xeta}
\def\z{\zeta}

\def\cA{{\cal A}} \def\cB{{\cal B}} \def\cC{{\cal C}}
\def\cD{{\cal D}} \def\cE{{\cal E}} \def\cF{{\cal F}}
\def\cG{{\cal G}} \def\cH{{\cal H}} \def\cI{{\cal I}}
\def\cJ{{\cal J}} \def\cK{{\cal K}} \def\cL{{\cal L}}
\def\cM{{\cal M}} \def\cN{{\cal N}} \def\cO{{\cal O}}
\def\cP{{\cal P}} \def\cQ{{\cal Q}} \def\cR{{\cal R}}
\def\cS{{\cal S}} \def\cT{{\cal T}} \def\cU{{\cal U}}
\def\cV{{\cal V}} \def\cW{{\cal W}} \def\cX{{\cal X}}
\def\cY{{\cal Y}} \def\cZ{{\cal Z}}

\def\ua{\underline{\alpha}}
\def\ub{\underline{\phantom{\alpha}}\!\!\!\beta}
\def\uc{\underline{\phantom{\alpha}}\!\!\!\gamma}
\def\um{\underline{\mu}}
\def\ud{\underline\delta}
\def\ue{\underline\epsilon}
\def\una{\underline a}\def\unA{\underline A}
\def\unb{\underline b}\def\unB{\underline B}
\def\unc{\underline c}\def\unC{\underline C}
\def\und{\underline d}\def\unD{\underline D}
\def\une{\underline e}\def\unE{\underline E}
\def\unf{\underline{\phantom{e}}\!\!\!\! f}\def\unF{\underline F}
\def\unm{\underline m}\def\unM{\underline M}
\def\unn{\underline n}\def\unN{\underline N}
\def\unp{\underline{\phantom{a}}\!\!\! p}\def\unP{\underline P}
\def\unq{\underline{\phantom{a}}\!\!\! q}
\def\unQ{\underline{\phantom{A}}\!\!\!\! Q}
\def\unH{\underline{H}}

\def\As {{A \hspace{-6.4pt} \slash}\;}
\def\bs {{b \hspace{-6.4pt} \slash}\;}
\def\Ds {{D \hspace{-6.4pt} \slash}\;}
\def\ds {{\del \hspace{-6.4pt} \slash}\;}
\def\ss {{\s \hspace{-6.4pt} \slash}\;}
\def\ks {{ k \hspace{-6.4pt} \slash}\;}
\def\ps {{p \hspace{-6.4pt} \slash}\;}
\def\pas {{{p_1} \hspace{-6.4pt} \slash}\;}
\def\pbs {{{p_2} \hspace{-6.4pt} \slash}\;}

\def\Fh{\hat{F}}
\def\Vh{\hat{V}}
\def\Xh{\hat{X}}
\def\ah{\hat{a}}
\def\xh{\hat{x}}
\def\yh{\hat{y}}
\def\ph{\hat{p}}
\def\xih{\hat{\xi}}

\def\psit{\tilde{\psi}}
\def\Psit{\tilde{\Psi}}
\def\tht{\tilde{\th}}
\def\lt{\tilde{\lambda}}
\def\At{\tilde{A}}
\def\Qt{\tilde{Q}}
\def\Rt{\tilde{R}}
\def\Nt{\tilde{N}}

\def\at{\tilde{a}}
\def\st{\tilde{s}}
\def\ft{\tilde{f}}
\def\pt{\tilde{p}}
\def\qt{\tilde{q}}
\def\vt{\tilde{v}}
\def\nt{\tilde{n}}

\def\delb{\bar{\partial}}
\def\bz{\bar{z}}
\def\bD{\bar{D}}
\def\bB{\bar{B}}

\def\bk{{\bf k}}
\def\bl{{\bf l}}
\def\bp{{\bf p}}
\def\bq{{\bf q}}
\def\br{{\bf r}}
\def\bx{{\bf x}}
\def\by{{\bf y}}
\def\bR{{\bf R}}
\def\bV{{\bf V}}

\def\d{\delta}\def\D{\Delta}\def\ddt{\dot\delta}

\def\pa{\partial} \def\del{\partial}
\def\xx{\times}
\def\uno{\mbox{1 \kern-.59em {\rm l}}}

\def\trp{^{\top}}
\def\inv{^{-1}}
\def\dag{{^{\dagger}}}
\def\pr{^{\prime}}
\def\lan{\langle}
\def\ran{\rangle}

\def\rar{\rightarrow}
\def\lar{\leftarrow}
\def\lrar{\leftrightarrow}

\newcommand{\0}{\,\!}      
\def\one{1\!\!1\,\,}
\def\im{\imath}
\def\jm{\jmath}

\newcommand{\tr}{\mbox{tr}}
\newcommand{\slsh}[1]{/ \!\!\!\! #1}

\def\vac{|0\rangle}
\def\lvac{\langle 0|}

\def\hlf{\frac{1}{2}}
\def\ove#1{\frac{1}{#1}}

\def\Box{\square}
\def\ZZ{\mathbb{Z}}
\def\CC#1{({\bf #1})}
\def\bcomment#1{}
\def\bfhat#1{{\bf \hat{#1}}}
\def\VEV#1{\left\langle #1\right\rangle}

\newcommand{\ex}[1]{{\rm e}^{#1}} \def\ii{{\rm i}}

\def\rr{{\rm r}} \def\rs{{\rm s}}\def\rv{{\rm v}}
\def\ri{{\rm i}}\def\rj{{\rm j}}

\newcommand{\lrbrk}[1]{\left(#1\right)}
\newcommand{\sfrac}[2]{{\textstyle\frac{#1}{#2}}}

\newcommand\equalhat{\mathrel{\stackon[1.5pt]{=}{\stretchto{%
    \scalerel*[\widthof{=}]{\wedge}{\rule{1ex}{3ex}}}{0.5ex}}}}

\font\mybb=msbm10 at 12pt
\def\bb#1{\hbox{\mybb#1}}

\font\myBB=msbm10 at 18pt
\def\BB#1{\hbox{\myBB#1}}

\newcommand{\tclr}{\textcolor}
\newcommand{\bpmat}{\begin{pmatrix}}
\newcommand{\epmat}{\end{pmatrix}}
\newcommand{\mrm}[1]{\mathrm{#1}}
\newcommand{\mrs}[1]{\scriptscriptstyle{\mathrm{#1}}}
\newcommand{\vct}[1]{\boldsymbol{#1}}
\newcommand{\hf}{\frac{1}{2}}
\newcommand{\x}{\times}
\newcommand{\pd}{\partial}
\newcommand{\dslash}{\displaystyle{\not}}
\newcommand{\ol}[1]{\overline{#1}}
\newcommand{\abs}[1]{\vert{#1}\vert}
\newcommand{\chiSqM}{\chi^2_{\mrm{min}}}
\newcommand{\chiSqMDof}{\chi^2_{\mrm{min}}/\mrm{d.o.f.}}
\newcommand{\om}{\omega}
\newcommand{\Lag}{\mathcal{L}}
\newcommand{\ord}{\mathcal{O}}
\newcommand{\eps}{\epsilon}
\newcommand{\beFrac}{\frac{1-\be}{1+\be}}
\newcommand{\beFracI}{\frac{1+\be}{1-\be}}
\newcommand{\amu}{a_{\mu}}
\newcommand{\damu}{\delta\amu}
\newcommand{\Damu}{\Delta\amu}
\newcommand{\amuUnit}{10^{-10}}
\newcommand{\mmu}{m_{\mu}}
\newcommand{\amuQED}{\amu^{\mrm{QED}}}
\newcommand{\amuEW}{\amu^{\mrm{EW}}}
\newcommand{\amuEWl}{\amu^{\mrm{EW,}\,1l}}
\newcommand{\amuEWll}{\amu^{\mrm{EW,}\,2l}}
\newcommand{\amuh}{\amu^{\mrm{had}}}
\newcommand{\amuhLO}{\amu^{\text{had, LOVP}}}
\newcommand{\amuhHO}{\amu^{\text{had, HOVP}}}
\newcommand{\amuhHOa}{\amu^{\text{had, HOVP(a)}}}
\newcommand{\amuhHOb}{\amu^{\text{had, HOVP(b)}}}
\newcommand{\amuhHOc}{\amu^{\text{had, HOVP(c)}}}
\newcommand{\amuhLbL}{\amu^{\text{had, LbL}}}
\newcommand{\ff}[3]{\mathcal{F}_{\pi^{0{#1}}\gamma^{#2}\gamma^{#3}}}
\newcommand{\alps}{\alpha_s}
\newcommand{\asmz}{\alpha_s(M_Z^2)}
\newcommand{\amz}{\alpha(M_Z^2)}
\newcommand{\aqmz}{\alpha_{\mrm{QED}}(M_Z^2)}
\newcommand{\delAlp}{\Delta\alpha}
\newcommand{\dAlpL}{\delAlp_{\mrm{lep}}}
\newcommand{\dAlpT}{\delAlp_{\mrm{top}}}
\newcommand{\dAlpH}{\delAlp_{\mrm{had}}}
\newcommand{\dAlpHF}{\dAlpH^{(5)}}
\newcommand{\dAlpHFmz}{\dAlpHF(M_Z^2)}
\newcommand{\tmin}{t_{\mrm{min}}}
\newcommand{\sTh}{s_{\mrm{th}}}
\newcommand{\eTh}{\sqrt{\sTh}}
\newcommand{\Ekmi}{E^{\,(k,m)}_i}
\newcommand{\Nkm}{N^{(k,m)}}
\newcommand{\Nkn}{N^{(k,n)}}
\newcommand{\Nexp}{N_{\mrm{exp}}}
\newcommand{\Nclu}{N_{\mrm{clu}}}
\newcommand{\Ntot}{N_{\mrm{tot}}}
\newcommand{\Rkmi}{R^{\,(k,m)}_i}
\newcommand{\Rknj}{R^{\,(k,n)}_j}
\newcommand{\dRkmi}{\mrm{d}\Rkmi}
\newcommand{\dRtkmi}{\mrm{d}\tilde{R}^{\,(k,m)}_i}
\newcommand{\BR}[2]{\mathcal{B}(#1\to #2)}
\newcommand{\decay}[2]{#1\to #2}
\newcommand{\UpsIVs}{\Upsilon(4S)}
\newcommand{\Gee}{\Gamma_{ee}}
\newcommand{\Gtot}{\Gamma_{\mrm{tot}}}
\newcommand{\ppC}{\pi^+\pi^-}
\newcommand{\ppN}{\pi^0\pi^0}
\newcommand{\pppC}{\pi^+\pi^-\pi^0}
\newcommand{\kkC}{K^+K^-}
\newcommand{\kskl}{K^0_S K^0_L}
\newcommand{\ksks}{K^0_S K^0_S}
\newcommand{\klkl}{K^0_L K^0_L}
\newcommand{\kskp}{K^0_S K^{\pm}\pi^{\mp}}
\newcommand{\eeMuMu}{e^+e^-\to\mu^+\mu^-}
\newcommand{\eeHadr}{e^+e^-\to\mrm{hadrons}}
\newcommand{\eeGhadr}{e^+e^-\to\gamma^*\to\mrm{hadrons}}
\newcommand{\tauNuHadr}{\tau\to\nu_{\tau}+\mrm{hadrons}}
\newcommand{\eeGPiPi}{e^+e^-\to\gamma^*\to\pi^+\pi^-}
\newcommand{\tauNuWNuPiPi}{\tau\to\nu_{\tau}W\to\nu_{\tau}\pi\pi^0}
\newcommand{\eeGIncl}{e^+e^-\to\gamma^*\to\mrm{all\,hadrons}}
\newcommand{\eeIncl}{e^+e^-\to\mrm{all\,hadrons}}
\newcommand{\eePiG}{e^+e^-\to\pi^0\gamma}
\newcommand{\eePiPi}{e^+e^-\to\pi^+\pi^-}
\newcommand{\eePiPiPi}{e^+e^-\to\pi^+\pi^-\pi^0}
\newcommand{\eeKK}{e^+e^-\to K^+K^-}
\newcommand{\ch}{\mrm{ch}}
\newcommand{\iso}{\mrm{iso}}
\newcommand{\noeta}{\text{no }\eta}
\newcommand{\kkr}{K\bar{K}\rho}
\newcommand{\kkp}{K\bar{K}\pi}
\newcommand{\kkpp}{K\bar{K}2\pi}
\newcommand{\kkppp}{K\bar{K}3\pi}
\newcommand{\isoAA}{(2\pi^+2\pi^-\pi^0)_{\mrm{no}\,\eta}}
\newcommand{\isoAB}{(\pi^+\pi^-3\pi^0)_{\mrm{no}\,\eta}}
\newcommand{\isoAC}{\omega(\to\mrm{npp})2\pi}
\newcommand{\isoACf}{\omega(\to\text{non-pure pionic states})2\pi}
\newcommand{\isoAD}{\eta\pi^+\pi^-}
\newcommand{\isoBA}{(2\pi^+2\pi^-2\pi^0)_{\mrm{no}\,\eta}}
\newcommand{\isoBB}{(\pi^+\pi^-4\pi^0)_{\mrm{no}\,\eta}}
\newcommand{\isoBC}{3\pi^+3\pi^-}
\newcommand{\isoBD}{\omega(\to\mrm{npp})3\pi}
\newcommand{\isoBDf}{\omega(\to\text{non-pure pionic state})3\pi}
\newcommand{\isoBE}{\eta\omega}
\newcommand{\isoEA}{\kkppp}
\newcommand{\isoEAa}{(K^+K^-\pi^+\pi^-\pi^0)_{\mrm{no}\,\eta}}
\newcommand{\isoEAb}{(K^0\bar{K}^0\pi^+\pi^-\pi^0)_{\mrm{no}\,\eta}}
\newcommand{\isoEB}{\omega(\to\mrm{npp})K\bar{K}}
\newcommand{\isoEBf}{\omega(\to\text{non-pure pionic states})K\bar{K}}
\newcommand{\isoEC}{\eta\phi}
\newcommand{\isoFA}{\eta2\pi^+2\pi^-}
\newcommand{\isoFB}{\eta\pi^+\pi^-2\pi^0}
\newcommand{\sigEEhadr}{\sigma(\eeHadr)}
\newcommand{\sigHad}{\sigma_{\mrm{had}}}
\newcommand{\sigHadB}{\sigHad^0}
\newcommand{\sigPt}{\sigma_{\mrm{pt}}}
\newcommand{\Rhad}{R_{\mrm{had}}}

\begin{document}

\setcounter{page}{1}
\thispagestyle{empty}
\begin{flushright}
LTH 1153 \\
KEK-TH-2035 \\
YITP-18-09 \\
 \
\\
6th April 2018
\end{flushright}
\begin{center}

\hspace{150cm}
\
\\
\
\\
\
\\

{\LARGE{\bf The muon $g-2$ and $\alpha(M_Z^2)$: a new data-based analysis} \\}
\
\\
\
\
\\
\
\\
{\large Alexander Keshavarzi$^1$, Daisuke Nomura$^{2,3}$ and Thomas Teubner$^{4}$}
\\
\
\\
{\small \em $^1$Department of Mathematical Sciences, University of Liverpool, Liverpool L69 3BX, United Kingdom} \\
{\small \em Email: a.i.keshavarzi@liverpool.ac.uk}
{\small \em \center $^2$KEK Theory Center, Tsukuba, Ibaraki 305-0801, Japan} \\
{\small \em $^3$Yukawa Institute for Theoretical Physics, Kyoto University, Kyoto 606-8502, Japan} \\
{\small \em Email: dnomura@post.kek.jp}
{\small \em \center $^4$Department of Mathematical Sciences, University of Liverpool, Liverpool L69 3BX, United Kingdom} \\
{\small \em Email: thomas.teubner@liverpool.ac.uk}
\
\\
\
\\
\
\\
\
\\


{\normalsize \bf Abstract}
\end{center}
This work presents a complete re-evaluation of the hadronic vacuum polarisation contributions to the anomalous magnetic moment of the muon, $a_{\mu}^{\rm had, \, VP}$ and the hadronic contributions to the effective QED coupling at the mass of the $Z$ boson, $\Delta\alpha_{\rm had}(M_Z^2)$, from the combination of $e^+e^-\rightarrow {\rm hadrons}$ cross section data. Focus has been placed on the development of a new data combination method, which fully incorporates all correlated statistical and systematic uncertainties in a bias free approach. All available $e^+e^-\rightarrow {\rm hadrons}$ cross section data have been analysed and included, where the new data compilation has yielded the full hadronic $R$-ratio and its covariance matrix in the energy range $m_{\pi}\leq\sqrt{s}\leq 11.2$ GeV. Using these combined data and perturbative QCD above that range results in estimates of the hadronic vacuum polarisation contributions to $g-2$ of the muon of $a_{\mu}^{\rm had, \, LO \, VP} = (693.26 \pm 2.46)\times 10^{-10}$ and $a_{\mu}^{\rm had, \, NLO \, VP} = (-9.82 \pm 0.04)\times 10^{-10}$. The new estimate for the Standard Model prediction is found to be $a_{\mu}^{\rm SM}  =  (11\ 659 \ 182.04 \pm 3.56) \times 10^{-10}$, which is $3.7\sigma$ below the current experimental measurement. The prediction for the five-flavour hadronic contribution to the QED coupling at the $Z$ boson mass is  $\Delta\alpha_{\rm had}^{(5)}(M_Z^2)= (276.11 \pm 1.11)\times 10^{-4}$, resulting in $\alpha^{-1}(M_Z^2) = 128.946 \pm 0.015$. Detailed comparisons with results from similar related works are given.


\newpage

\tableofcontents

\section{Introduction}

The anomalous magnetic moment of the muon, $a_{\mu} = (g-2)_{\mu}/2$,
stands as an enduring test of the Standard Model (SM), where the
$\sim3.5\sigma$  (or higher) discrepancy between the experimental
measurement $a_{\mu}^{\rm exp}$ and the SM prediction 
$a_{\mu}^{\rm SM}$ could be an indication of the existence of new
physics beyond the SM. For $a_{\mu}^{\rm exp}$, the
value is dominated by the measurements made at the Brookhaven
National Laboratory
(BNL)~\cite{Bennett:2002jb,Bennett:2004pv,Bennett:2006fi}, resulting
in a world average of~\cite{PDG2016} 
\beq \label{amuExp}
a_{\mu}^{\rm exp} = 11\ 659 \ 209.1 \ (5.4) \ (3.3)  \times 10^{-10}\,.
\eeq
Efforts to improve the experimental estimate at Fermilab
(FNAL)~\cite{Grange:2015fou} and at J-PARC~\cite{Mibe:2010zz} aim to
reduce the experimental uncertainty by a factor of four compared to
the BNL measurement. It is therefore imperative that the SM prediction
is also improved to determine whether the $g-2$ discrepancy is well
established. 

The uncertainty of $a_{\mu}^{\rm SM}$ is completely dominated by the
hadronic contributions, $a_{\mu}^{\rm had}$, attributed to the
contributions from the non-perturbative, low energy region of hadronic
resonances. The hadronic contributions are divided into the hadronic
vacuum polarisation (VP) and hadronic light-by-light (LbL) contributions, which are
summed to give 
\beq \label{eq:amu_had}
a_{\mu}^{\rm had} = a_{\mu}^{\rm had,\,VP} + a_{\mu}^{\rm had,\,LbL}\,. 
\eeq
This analysis, KNT18, is a complete re-evaluation, in line with
previous works~\cite{Hagiwara:2003da,Hagiwara:2006jt,Hagiwara:2011af}, of the
hadronic vacuum polarisation contributions, $a_{\mu}^{\rm had, \,
  VP}$. The hadronic vacuum polarisation contributions can be
separated into the leading-order (LO) and higher-order contributions,
where the LO and next-to-leading order (NLO) contributions are
calculated in this work.\footnote{The next-to-next-to-leading order
  (NNLO) contributions have recently been determined
  in~\cite{Kurz:2014wya} and are included as part of $a_{\mu}^{\rm
    SM}$ below.} These are calculated utilising dispersion integrals
and the experimentally measured cross section, 
\beq
\sigma^0_{{\rm had},\gamma} (s) \equiv \sigma^0(e^+e^-\rightarrow
\gamma^* \rightarrow \text{hadrons} + \gamma) 
\eeq
where the superscript 0 denotes the bare cross section (undressed of
all vacuum polarisation effects) and the subscript $\gamma$ indicates
the inclusion of effects from final state photon radiation. At
LO, the dispersion relation reads 
\beq \label{eq:amu}
a_{\mu}^{\rm had,\,LO\,VP} =
\frac{\alpha^2}{3\pi^2}\int^{\infty}_{s_{th}} \frac{{\rm d}s}{s}
R(s)K(s) 
\eeq
where $\alpha = \alpha(0)$, $s_{th} = m_{\pi}^2$, $R(s)$ is the
hadronic $R$-ratio given by 
\beq \label{eq:R(s)}
R(s) = \frac{\sigma^0_{{\rm had},\gamma} (s)}{\sigma_{\rm pt}(s)}
\equiv \frac{\sigma^0_{{\rm had},\gamma} (s)}{4\pi\alpha^2/(3s)} 
\eeq
and $K(s)$ is a well known kernel function~\cite{Brodsky:1967sr,Lautrup:1969fr}, given
also in equation (45) of~\cite{Hagiwara:2003da}, but differing by a
normalisation factor of $m_{\mu}^2/(3s)$. This kernel function (which behaves as 
$K(s) \sim m_\mu^2/(3s)$ at low energies), coupled with the factor of $1/s$
 in the integrand of equation~\eqref{eq:amu}, causes the hadronic vacuum polarisation
contributions to be dominated by the low energy domain. At NLO, the
data input is identical, with corresponding dispersion integrals and
kernel functions~\cite{Krause:1996rf} (see also~\cite{Hagiwara:2003da}). In the previous
analysis~\cite{Hagiwara:2011af} (denoted as HLMNT11), the LO hadronic vacuum
polarisation contributions were found to be 
\beq
a_{\mu}^{\rm had,\,LO\,VP}\big({\rm HLMNT11}\big) = (694.91 \pm 4.27)
\times 10^{-10}\,, 
\eeq
which resulted in the total SM prediction of
\beq \label{amuSM_HLMNT11}
a_{\mu}^{\rm SM}\big({\rm HLMNT11}\big) = (11 \ 659 \ 182.8 \pm 4.9)
\times 10^{-10}\,. 
\eeq
This, compared with the experimental measurement, gave a $g-2$
discrepancy of $3.3\sigma$. 

In addition to calculating $a_{\mu}^{\rm had,\,VP}$, the combination of
hadronic cross section data is also used to calculate the running
(momentum dependent) QED
coupling, $\alpha(q^2)$. This, in particular, is then used to determine the effective QED coupling
at the $Z$ boson mass, $\alpha(M_{Z}^2)$, which is the least precisely
known of the three fundamental electro-weak (EW) parameters of the SM
(the Fermi constant $G_F$, $M_Z$ and $\alpha(M_{Z}^2)$) and
hinders the accuracy of EW precision fits. Using an identical data
input as that used for $a_{\mu}^{\rm had,\,VP}$, the hadronic
contributions to the effective QED coupling are determined from the
dispersion relation 
\beq \label{eq:delAlpha}
\Delta\alpha_{\rm had}^{(5)}(q^2) = -\frac{\alpha q^2}{3\pi}{\rm P}
\int^{\infty}_{s_{th}} {\rm d}s\frac{R(s)}{s(s-q^2)}\,,
 \eeq
where the superscript (5) indicates the contributions from all quark flavours except the top quark, which is added separately. Together with the leptonic contributions, this is used to determine
the running coupling $\alpha(q^2)=\alpha/\big(1-\Delta\alpha_{\rm
  had}(q^2)-\Delta\alpha_{\rm lep}(q^2)\big) $.  

The structure of this paper is as
follows. Section~\ref{Changessincelastanalysis} describes all changes
to the treatment and combination of hadronic cross section data
since~\cite{Hagiwara:2011af}. Section~\ref{amuanddeltaalpha} gives
details concerning the contributions to $a_{\mu}^{\rm had,\,LO\,VP}$
and $\Delta\alpha_{\rm had}(M_{Z}^2)$ from the different hadronic
final states and energy regions, culminating in updated estimates of
both $a_{\mu}^{\rm had,\,LO\,VP}$, $a_{\mu}^{\rm had,\,NLO\,VP}$ and $\Delta\alpha_{\rm
  had}(M_{Z}^2)$. These results are then compared with other works and the new prediction of $a_{\mu}^{\rm had,\,VP}$ is
combined with all other SM contributions to determine the SM 
prediction $a_{\mu}^{\rm SM}$, with details given concerning the consequences of this
on the existing $g-2$ discrepancy and the new result for $\alpha(M_{Z}^2)$. Finally, the conclusions of this work, along with discussions of the prospects
for $g-2$ in the future and potential improvements for
$a_{\mu}^{\rm SM}$, are presented in
Section~\ref{Conclusions}.

\section{Changes since the last analysis (HLMNT11)} \label{Changessincelastanalysis}

\subsection{Radiative corrections}\label{radcorr}

\subsubsection{Vacuum polarisation corrections}\label{VPcorrections}

Equations \eqref{eq:amu} and \eqref{eq:delAlpha} require the
experimental cross section to be undressed of all vacuum polarisation
(VP) effects as VP (running coupling) corrections to the hadronic
cross section are counted as part of the higher order
contributions to $a_\mu^{\rm had, VP}$. Any new and old data that have not been corrected for VP effects
require undressing. However, recent data are more commonly undressed in the experimental analyses already, removing the need
to apply a correction to these data sets. This benefits the data
combination as new, more precise data undressed of VP effects are
dominating the fit for many channels which, in turn, reduces the impact of
the extra radiative correction uncertainty which is applied to each channel. 

Following the previous analyses, the `bare' cross section
$\sigma^0_{\rm had}$ in equation \eqref{eq:R(s)} is determined to be 
\beq
\sigma^0_{\rm had} = C_{\rm vp}\sigma_{\rm had} \ ,
\eeq
where $C_{\rm vp}$ is a multiplicative VP correction. For this
purpose, the self-consistent KNT18 vacuum polarisation routine,
\texttt{vp\_knt\_v3\_0}, has been updated via an iteration of the new
data input.\footnote{This routine is available for use by contacting
  the authors directly.} An in-depth discussion of the determination
of the vacuum polarisation and the corresponding routine will be given
in~\cite{KNT-VP}. 

The decisions to undress all data sets previously corrected
in~\cite{Hagiwara:2011af} are unchanged, with the exception of two
measurements of the inclusive hadronic $R$-ratio by the Crystal Ball
collaboration~\cite{Osterheld:1986hw,Edwards:1990pc}.  A
re-examination of the experimental analyses of these two measurements
has shown that not only do they include some treatment of the VP, but that they
also both account for this correction with sizeable systematic
uncertainties. Applying an extra VP correction here would be an
overestimate of the correction and its corresponding
uncertainty. Therefore, undressing of these data is not applied
here. In addition, it should be noted that in a separate
work~\cite{KLOEcombination}, two measurements by the KLOE
collaboration~\cite{Ambrosino:2008aa,Ambrosino:2010bv} in the
$\pi^+\pi^-$ channel are now undressed of VP effects using an updated
routine~\cite{FJ16VP} compared to the one used
previously~\cite{FJ03VP}. 

The undressing of narrow resonances in the $c\bar{c}$ and $b\bar{b}$
regions requires special attention. Importantly, the electronic width
of an individual resonance, $\Gamma_{ee}$, must be undressed of vacuum
polarisation effects using a parametrisation of the VP where the
correction {\em excludes} the contribution of that resonance, such
that 
\beq
\Gamma^0_{ee} = \frac{\Big(\alpha/\alpha_{\text{no res}}(M_{\rm res}^2)\Big)^2}{1+3\alpha/(4\pi)}\Gamma_{ee} \ .
\eeq
Here, $\alpha_{\text{no res}}$ is the effective QED coupling
neglecting the contribution of the resonance itself and is given by 
\beq
\alpha_{\text{no res}}(s) \equiv \frac{\alpha}{1-\Delta\alpha_{\text{no res}}(s)} \ ,
\eeq
where $\Delta\alpha_{\text{no res}}(s)$ is determined from equation
\eqref{eq:delAlpha} such that the input $R(s)$ does not include the
resonance that is being corrected. To include the resonance would 
lead to an inconsistent definition of the narrow resonance.

\subsubsection{Final state radiative corrections}\label{FSRcorrection}

Final state radiation (FSR) cannot be separated in an unambiguous way
in the measured hadronic cross sections. Therefore, while formally of
higher order in $\alpha$, FSR photons have to be taken into account
in the definition of the one particle irreducible hadronic blobs and
will already be included as part of the leading order hadronic
VP contributions. However, depending on the experimental analyses,
some amount of real photon FSR may have been removed during the event
selection. Adding back these missed contributions is model dependent
and not feasible for general hadronic final states. It is therefore
necessary to estimate the possible effects and their impact on the
accuracy of the data compilations. Similar to the case of VP corrections,
an extra radiative correction uncertainty due to FSR effects is then
estimated channel by channel. 

For the important $\pi^+\pi^-$ and $K^+ K^-$ channels, detailed studies
have been performed for this analysis. Here, and especially in the
limited energy range below $2$~GeV, it has been shown that scalar QED
provides a good description of photon FSR, see
e.g.~\cite{Hoefer:2001mx,Gluza:2002ui,Lees:2015qna}.  In the past, to
estimate possible FSR effects in $\pi^+ \pi^-$ and $K^+ K^-$
production, the fully inclusive, order $\alpha$ correction
to the cross section, 
\beq
\sigma^{(0)}_{\text{had,}(\gamma)}(s) =
\sigma^{(0)}_{\text{had}}(s)\biggr(1+\eta(s)\frac{\alpha}{\pi}\biggr)
\,, 
\eeq
has been used.\footnote{Here, the term `fully inclusive' means inclusive of effects from virtual and real (both soft and hard) one-photon emission.} Here, the function $\eta(s)$ is given e.g.\
in~\cite{Hoefer:2001mx} and the subscript $\gamma$ indicates the one
photon inclusive cross section. However, experimental cross section
measurements, by nature, include all virtual and soft real radiation
effects. Therefore, to estimate possibly missing photon FSR, ideally
only the effects from (hard) real radiation above/within
resolution/cut parameters which are specific for a given experiment
or analysis have to be estimated. Whereas in the calculation of the
inclusive correction a regularisation of the virtual and real soft
contributions is required to obtain the infrared finite result $\eta$,
the hard real radiation, $\eta^{\rm hard, real}$, can be estimated
numerically from 
\beq \label{etahardreal}
\eta^{\rm hard, real}(s)\ =\
\int^{s-2\sqrt{s}\Lambda}_{4m^2} ds' \rho_{\rm fin}(s,s^\prime)\,,
\eeq
where $m$ is the mass of the (scalar) particle, $\Lambda$ is a finite infrared
cut-off parameter on the invariant mass of the emitted photon and
$\rho_{\rm fin}$ is the radiator function (see appendix B
of~\cite{Gluza:2002ui}). 

\begin{figure}[!t]
\centering
    \includegraphics[width= 0.49\textwidth]{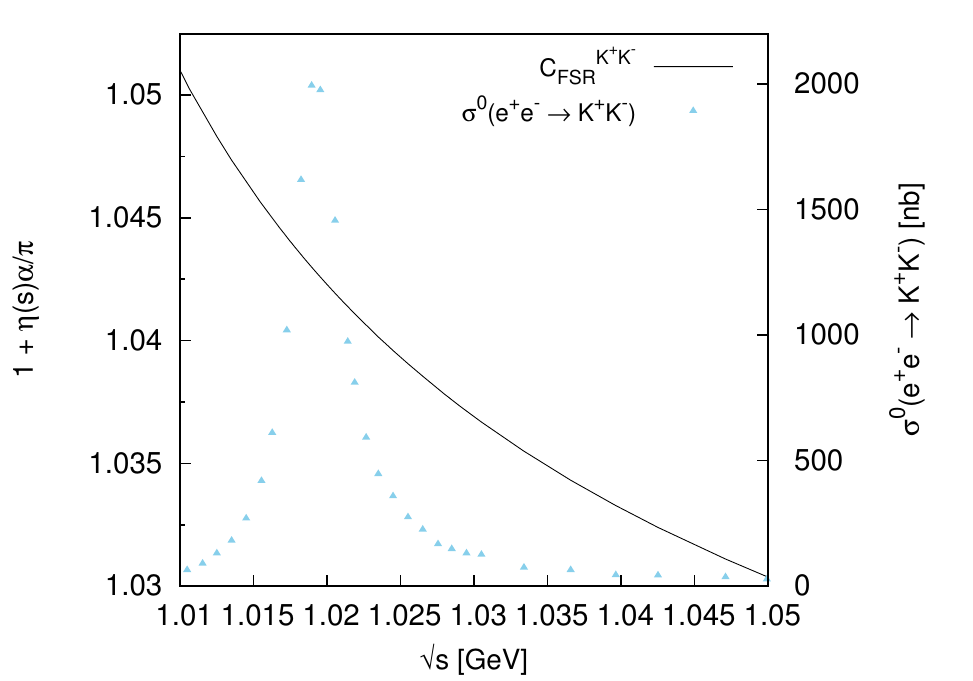}
    \includegraphics[width= 0.49\textwidth]{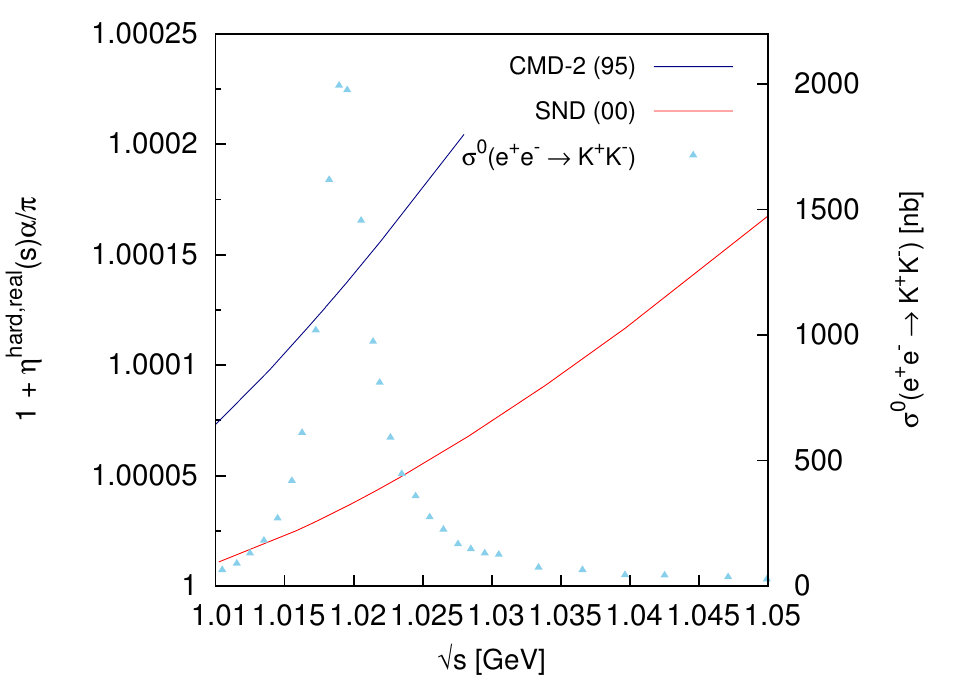}
  \caption{\small The effect of final state radiation in the $K^+K^-$
    channel in the $\phi$ resonance region. Left panel: the fully
    inclusive FSR correction $\eta(s)$. Right panel: hard real
    radiation $\eta^{\rm hard, real}(s)$, estimated with
    accolinearity cuts used in the two
    analyses~\cite{Akhmetshin:1995vz,Achasov:2000am}. The
    $e^+e^-\rightarrow K^+K^-$ cross section is also plotted for
    reference. }\label{fig:etaPhi} 
  \end{figure}
In the case of the $K^+K^-$ channel, by far the largest contribution
to $a_{\mu}$ and $\Delta\alpha$ (and their errors) comes from the energy
region of the $\phi$ peak, where the phase space for real radiation is
severely restricted. While the calculation of real radiation accounting 
for all experimental cuts would be very complicated and beyond the
scope of this work, an estimate can easily be made based on
equation~\eqref{etahardreal}, by relating $\Lambda$ to the cuts in the
photon accolinearity given in individual experimental analyses. In
Figure~\ref{fig:etaPhi}, the result for the fully inclusive correction
$\eta(s)$ (left panel) is compared to the estimates for the real hard
radiation, where $\eta^{\rm hard, real}(s)$ (right panel) now depends on the
accolinearity cuts as given by the two experimental analyses
depicted. Clearly, at and around the $\phi$ peak, phase space
restrictions strongly suppress any hard real radiation and using the
inclusive correction would lead to an overestimate of the possible
effects. Given the small size of both the possible correction in the $\phi$
region and the contribution of the $K^+K^-$ channel
above the $\phi$ to both mean value and error of $a_{\mu}$ and $\Delta\alpha$,
no correction or additional error estimate due to FSR is now applied
in the $K^+K^-$ channel. 
For the neutral kaon channel, hard photon radiation (which would
resolve the quark charges) is similarly suppressed and no FSR
correction or additional error are applied in this channel also. 

\begin{figure}[!t]
\centering
    \includegraphics[width= 0.55\textwidth]{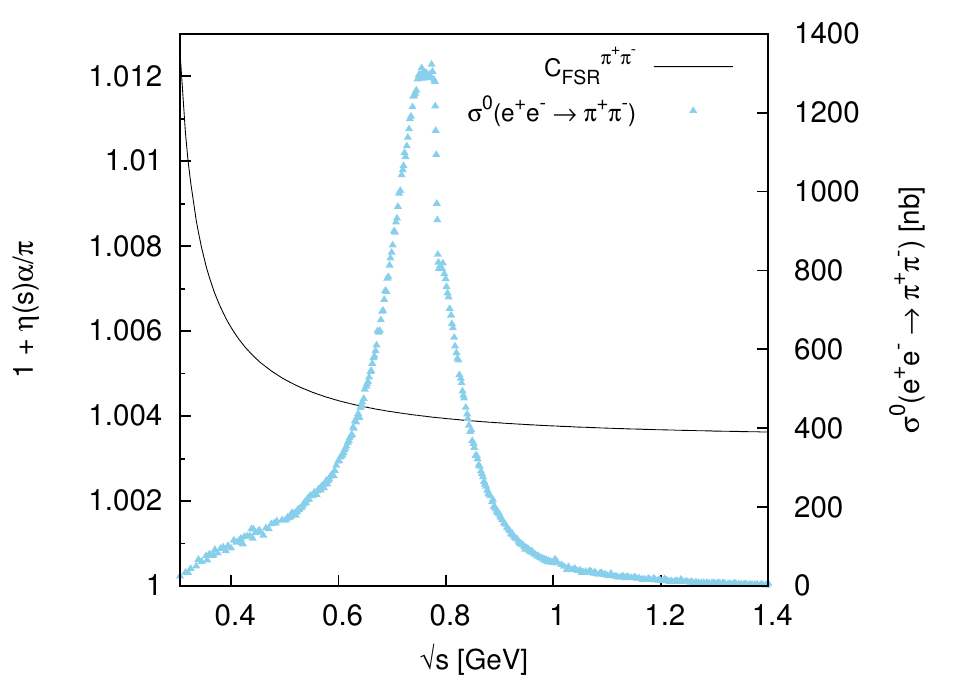}
  \caption{\small The behaviour of the inclusive FSR correction,
    $\eta(s)$, for the process $e^+e^-\rightarrow \pi^+\pi^-$. The
    $e^+e^-\rightarrow \pi^+\pi^-$ cross section is also plotted for
    reference.}\label{fig:etapipi} 
\end{figure}

The situation is different in the two pion channel. A study similar to
the two kaon channel showed that in principle larger contributions
from real radiation of the order of the inclusive correction can
arise. However, these contributions are strongly dependent on the cut
applied in equation~\eqref{etahardreal} and would require a more detailed,
measurement-by-measurement analysis, which is beyond the scope of this
work. In addition, for many of the data sets the required detailed
information is not available. Therefore, in sets which are understood 
to not include the full FSR corrections, the fully
inclusive correction (as shown in Figure~\ref{fig:etapipi}) is applied to
the respective measurements. 
In the $\pi^+ \pi^-$ data combination, recent sets from radiative
return, where additional photons are part of the leading order cross
section and are studied in detail as part of the analyses, have now
become dominant. Therefore, the impact of the fully inclusive FSR
correction to older sets is suppressed for both mean value and error
in comparison to previous analyses. The procedure to determine the
corresponding error estimate is discussed in the next chapter below,
whilst resulting numbers are given in Section~\ref{chap:pipi}, which
contains the detailed analysis of the $\pi^+\pi^-$ channel.

For the sub-leading, multi-hadron channels, there are, at present, no
equivalent FSR calculations. Depending on the experimental analysis,
they are (at least to some extent) estimated by Monte Carlo (MC) simulation and
contribute to the systematic uncertainties.\footnote{In the recent
  work~\cite{Jegerlehner:2017kke}, a version of the MC program {\tt
    Carlomat}~\cite{Kolodziej:2015eia} is discussed, which includes
  real photon radiation in multi-hadron channels. This could be used
  to further study the FSR effects.} 
However, for many data sets, it is far from clear to which extent FSR
effects are included in the systematic errors. Therefore, possible
effects are accounted for by applying an additional uncertainty
determined as a fraction of the respective contribution.

 \subsubsection{Estimating extra uncertainty due to radiative
   corrections}\label{radcorUncertainty} 

As in~\cite{Hagiwara:2003da,Hagiwara:2006jt,Hagiwara:2011af}, extra
uncertainties are estimated whenever additional radiative corrections
are applied. This is done, first and foremost, to account for any
under- or over-correction that may occur due to a lack of information
concerning the treatment of radiative corrections in the experimental
analyses. However, these radiative correction uncertainties also account
for any possibly incorrect treatment in the analyses,
e.g.\ missed FSR or inconsistent subtraction of VP contributions. This
is especially true for older data, where there is very little or even
no information at all regarding how the data have been treated. 

In each channel, the difference $\Delta a_{\mu}^{\rm
  vp}$ between the estimates of $a_\mu$ with and without additional VP
corrections is determined. For the uncertainty due to VP, one third of
the shift
\beq
\delta a_{\mu}^{\rm vp} = \frac{1}{3} \Delta a_{\mu}^{\rm vp} \,
\eeq
is taken. This is reduced from one half in~\cite{Hagiwara:2011af}, safe in the
knowledge that the KNT18 VP routine~\cite{KNT-VP}, which is
determined iteratively in a self-constistent way, is accurate to the
level of a permille when correcting the cross section and that newer
data sets are commonly undressed of VP effects by the experiment with
a modern routine. 

For the extra uncertainties due to FSR, there are now no contributions from the
$K^+K^-$ and $K^0_SK^0_L$ channels (see the discussion above).  For
the $\pi^+\pi^-$ channel, the full difference between the estimates of
$a_\mu^{\pi^+\pi^-}$ with and without additional FSR corrections is
taken as the FSR uncertainty. For all other channels, including the
inclusive data combination above 1.937 GeV, a fraction of 1\% of the
respective cross section is applied as the uncertainty.

The numerical estimates of all additional radiative correction
uncertainties are given in the respective sections for the individual
channels. The same procedures are applied in the calculation of the
contributions to $\Delta\alpha_{\rm had}$.

\subsection{Data combination}

The combination of $\sigma^{0}(e^+e^- \rightarrow$ hadrons) data and
its corresponding uncertainty has undergone a great deal of scrutiny
since~\cite{Hagiwara:2011af}. This has been due to the ever increasing
amount of data that have become available and due to a better
understanding of the treatment of correlated uncertainties that are
now a prominent feature of many cross section measurements. 
In~\cite{Hagiwara:2011af}, covariance matrices in the $\pi^+\pi^-$
channel from BaBar~\cite{Aubert:2009ad} and
KLOE~\cite{Ambrosino:2008aa,Ambrosino:2010bv} for statistical and
systematic uncertainties had already been included based on the method
of fitted renormalisation factors. In the following, a new method for
the combination of $\sigma^{0}_{\rm had}$ data is introduced, which
will allow to fully take into account all correlated uncertainties. 

\subsubsection{Clustering data points}

Within each hadronic channel, data points from different experiments
are assigned to {\em clusters}
(see~\cite{Hagiwara:2011af,Hagiwara:2003da,Hagiwara:2006jt} for
details).\footnote{The effect that different clusterings have on the
  $\chi^2_{\rm min}/{\rm d.o.f.}$ of the fit, the resulting
  $a_{\mu}^{\rm had, VP}$ value and its error has been discussed in
  detail in~\cite{Hagiwara:2011af}.} In this work, the clustering
algorithm has been improved. It is universal for all different
channels and only differs in the assigned maximum cluster size
$\delta$ (and $\delta_{\rm res}$, a maximum cluster size applicable at
individual narrow resonances). 
A scan over $\delta$ (and $\delta_{\rm res}$ if applicable) is
performed to determine a suitable clustering configuration which must
avoid both over- and under-clustering. Too wide or overpopulated
clusters would effectively lead to a re-binning of data points from
individual experiments and risk loss of information, while a too
narrow clustering would result, in the extreme, in an erratic
point-to-point representation of the cross section and no gain in the
accuracy. The preferred configuration is then chosen based on the
global $\chi^2_{\rm min}/{\rm d.o.f.}$ and the uncertainty on
$a_{\mu}^{\rm had, VP}$, combined with checks by eye of the resulting
spectral function.

\subsubsection{Fitting data} \label{ssSec: Fitting data}

The previous
analyses~\cite{Hagiwara:2011af,Hagiwara:2003da,Hagiwara:2006jt}
employed a non-linear $\chi^2$-minimisation utilising fitted
renormalisation factors as nuisance parameters that represented the
energy independent systematic uncertainties. Although a powerful
method, recent literature~\cite{Ball:2009qv} (see also~\cite{Benayoun:2015gxa}) have
highlighted the possibility that an incorrect treatment of multiplicative
normalisation uncertainties in a $\chi^2$ minimisation can incur a
systematic bias (see chapter 4 of~\cite{Ball:2009qv}).
In addition, although the non-linear $\chi^2$-minimisation used
in~\cite{Hagiwara:2011af} was adjusted to include covariance matrices,
the method's reliance on fitting energy independent renormalisation
factors prevented the use of correlated uncertainties to their full
capacity. As recent precise data, specifically radiative return
measurements in the $\pi^+\pi^-$ and $K^+K^-$ channels, have been
released with energy dependent uncertainties and non-trivial
bin-to-bin correlations for both the statistical and systematic
uncertainties, the fit procedure had to be modified to allow the full
use of all available correlations in a bias-free approach. 
Therefore, instead of fitting renormalisation factors (nuisance
parameters), an iterative fit procedure as advocated
in~\cite{Ball:2009qv} has been adopted, which re-initialises the full
covariance matrices at each iteration step and, consequently, avoids
bias.

Previously, in~\cite{Hagiwara:2011af,Hagiwara:2003da,Hagiwara:2006jt}, 
a constant cross section had been assumed 
across the width of each cluster. In this work, the fitted cross
section values at the cluster centres are obtained from the iterative 
$\chi^2$-minimisation where the cross section is taken to be linear
between adjacent cluster centres. This allows for a more stable
fit and is consistent with the trapezoidal rule integration utilised
for the $a_{\mu}^{\rm had, VP}$ and $\Delta\alpha_{\rm had}$ integrals. 
If data points at energies $\sqrt{s} = E_i^{(m)}$ are combined into
cluster $m$, then the weighted average of the cross section value
$R_m$ and energy $E_m$ for the {\em cluster centre} are 
\beq \label{R0RmI}
R_m =
\left[\sum^{N^{(m)}}_{i=1}\frac{R_{i}^{(m)}}{\left(\text{d}\tilde{R}_{i}^{(m)}\right)^2}\right]
\left[\sum^{N^{(m)}}_{i=1}\frac{1}{\left(\text{d}\tilde{R}_{i}^{(m)}\right)^2}\right]^{-1} 
\eeq 
and   
\beq 
E_m =
\left[\sum^{N^{(m)}}_{i=1}\frac{E_{i}^{(m)}}{\left(\text{d}\tilde{R}_{i}^{(m)}\right)^2}\right]
\left[\sum^{N^{(m)}}_{i=1}\frac{1}{\left(\text{d}\tilde{R}_{i}^{(m)}\right)^2}\right]^{-1}
\,, 
\eeq
where $R_i^{(m)}$ is the cross section value of data point $i$
contributing to cluster $m$, $N^{(m)}$ is the total number of data
points contributing to cluster $m$ and 
\beq \label{dtilde}
\text{d}\tilde{R}_{i}^{(m)} = \sqrt{(\text{d}R^{(m)}_{i; \
    \text{stat}})^2 + (\text{d}R^{(m)}_{i; \ \text{sys}})^2} \,. 
\eeq
$\text{d}R^{(m)}_{i; \ \text{stat}}$ and $\text{d}R^{(m)}_{i;
  \ \text{sys}}$ denote the absolute statistical and systematic
uncertainties, respectively. With a linear cross section now assumed,
if data point $i$ belongs to cluster $m$ and $E^{(m)}_{i} >
E_{m}$, then its interpolant cross section value $\mathcal{R}_{m}^{i}$
is given by 
\beq\label{int+}
\mathcal{R}_{m}^{i+} = R_m + \frac{(E_{i}^{(m)}-E_{m})}{(E_{m+1}-E_{m})}(R_{m+1}-R_{m}) \,,
\eeq
where the $+$ indicates that $E^{(m)}_{i} > E_{m}$. If, on the other hand, $E^{(m)}_{i} < E_{m}$ then
\beq\label{int-}
\mathcal{R}_{m}^{i-} = R_{m-1} + \frac{(E_{i}^{(m)}-E_{m-1})}{(E_{m}-E_{m-1})}(R_{m}-R_{m-1}) \,,
\eeq
where the $-$ indicates that $E_{i}^{(m)} <
E_{m}$. For data points at the borders where no interpolation is
possible, $\mathcal{R}_m^i$ is found by linear extrapolation.

A covariance matrix is constructed for the combination which contains
all the uncertainty and correlation information of all data
points. Using the covariance
matrix as defined by the data alone could result in bias
(see~\cite{D'Agostini:1993uj,Blobel:2003wa}). The covariance matrix is
therefore redefined at each step of the iteration using the fitted
$R_m$ values. Convergence of the iteration is observed in this work to
occur after only a few steps. 

The covariance matrix  ${\bf C}\big(i^{(m)},j^{(n)}\big)$ is given as
the sum of the statistical covariance matrix ${\rm C}^{\rm
  stat}\big(i^{(m)},j^{(n)}\big)$ and the systematic covariance matrix
${\rm C}^{\rm sys}\big(i^{(m)},j^{(n)}\big)$. At each stage of the
iteration, it is defined as 
\beq \label{ICk}
{\bf C}_I\big(i^{(m)},j^{(n)}\big) = {\text
  C}^{\text{stat}}\big(i^{(m)},j^{(n)}\big) + \frac{{\text
    C}^{\text{sys}}\big(i^{(m)},j^{(n)}\big)}{R_{i}^{(m)}R_{j}^{(n)}}
\mathcal{R}_{m}^{i,(I-1)}\mathcal{R}_{n}^{j,(I-1)} \,, 
\eeq
where the quantities $\mathcal{R}_{m}^{i, I}$ and
$\mathcal{R}_{n}^{j,I}$ are the interpolant cross sections given by
either equation \eqref{int+} or \eqref{int-} and $I$ denotes the
iteration number of the fit. This is then used as input into the now
{\em linear} $\chi^2$-function, 
\beq \label{IChi^2}
\chi^2_{I} =
\sum^{N_{\mathrm{tot}}}_{i=1}\sum^{N_{\mathrm{tot}}}_{j=1}\big(R_{i}^{(m)}
- \mathcal{R}_{m}^{i,I}\big) {\bf
  C}_{I}^{-1}\big(i^{(m)},j^{(n)}\big)\big(R_{j}^{(n)} -
\mathcal{R}_{n}^{j,I} \big) \,, 
\eeq
where $N_{\mathrm{tot}}$ is the total number of contributing data
points and ${\bf C}_{I}^{-1}\big(i^{(m)},j^{(n)}\big)$ is simply the
inverse of the covariance matrix defined in equation
\eqref{ICk}. Performing the minimisation yields a system of linear
equations 
\beq \label{gauss}
\sum^{N_{\mathrm{tot}}}_{j=1}\Biggr[\big(R_{j}^{(n)} -
\mathcal{R}_{n}^{j, I}\big) \frac{\del \mathcal{R}_{m}^{i}}{\del
  R_a}\Biggr] V_{I}^{-1}\big(m^{(i)},n^{(j)}\big) = 0 \ ,
\hspace{0.5cm} i = 1, ..., N_{\mathrm{tot}} \,, 
\eeq
where,
\beq 
V_{I}^{-1}\big(m^{(i)},n^{(j)}\big) =
\sum^{N^{(m)}}_{i=1}\sum^{N^{(n)}}_{j=1} {\bf
  C}_{I}^{-1}\big(i^{(m)},j^{(n)}\big) \,. 
\eeq
The solution to this yields the cluster centres $R_m$ and the
covariance matrix $V_{I}\big(m,n\big)$ which describes the correlation
between the errors ${\rm d}R_m$ and ${\rm d}R_n$. As in equations
\eqref{int+} and \eqref{int-}, the term $\mathcal{R}_{n}^{j}$ is to be
taken as either $\mathcal{R}_{n}^{j +}$, if $E_{j}^{(n)} > E_n$, or
$\mathcal{R}_{n}^{j -}$, if $E_{j}^{(n)} < E_n$. Subsequently, if
$E_{i}^{(m)} > E_m$, then 
\beq
\frac{\del \mathcal{R}_{m}^{i}}{\del R_a}\Biggr|_{E_{i}^{(m)} > E_m}\
=\ 
\frac{\del \mathcal{R}_{m}^{i+}}{\del R_a}\ =\ 
\biggr(1-\frac{(E_{i}^{(m)} - E_m)}{(E_{m+1} - E_m)}\biggr)\delta_{ma}
+ \frac{(E_{i}^{(m)} - E_m)}{(E_{m+1} - E_m)}\delta_{m+1,a} 
\eeq
and, if $E_{i}^{(m)} < E_m$, then
\beq
\frac{\del \mathcal{R}_{m}^{i}}{\del R_a}\Biggr|_{E_{i}^{(m)} < E_m}\
=\ 
\frac{\del \mathcal{R}_{m}^{i-}}{\del R_a}\ =\ 
\biggr(1-\frac{(E_{i}^{(m)} - E_{m-1})}{(E_{m} -
  E_{m-1})}\biggr)\delta_{m-1,a} + \frac{(E_{i}^{(m)} -
  E_{m-1})}{(E_{m} - E_{m-1})}\delta_{ma} \,,  
\eeq
where $\delta$ denotes the usual Kronecker delta.

In~\cite{Ball:2009qv}, it was stated that for the fit of parton
distribution functions, convergence is expected to occur after very few
iterations, which is also observed here. The final result includes the
total output covariance matrix $V(m,n)$ which, as in
\cite{Hagiwara:2011af}, is inflated according to the local
$\chi^2_{\rm min}/{\rm d.o.f.}$ for each cluster if $\chi^2_{\rm min}/{\rm
  d.o.f.} > 1$. This is done in order to account for any tensions
between the data. The use of the full covariance matrix allows for the inclusion of
any-and-all uncertainties and correlations that may exist between the
measurements. The flexibility to now make use of fully energy dependent
uncertainties ensures that the appropriate influence of the
correlations is incorporated into the determination of the cluster
centres, $R_m$, with the correct propagation of all experimental
errors to the $g-2$ uncertainty. 

Note that, for all channels, the differences between the old and the
new data combination procedures are small and lead to changes of
$a_{\mu}^{\rm had, VP}$ well within the quoted errors. Importantly,
combining data which have only global normalisation uncertainties
results in negligible differences between~\cite{Hagiwara:2011af} and this
work, indicating that previous results were largely unaffected by
the potential bias issue.

\subsubsection{Integration}

\begin{figure}[!t]
\centering
  \subfloat[The $\rho-\omega$ resonance region of the $\pi^+\pi^-$ channel.]{%
    \includegraphics[width= 0.5\textwidth]{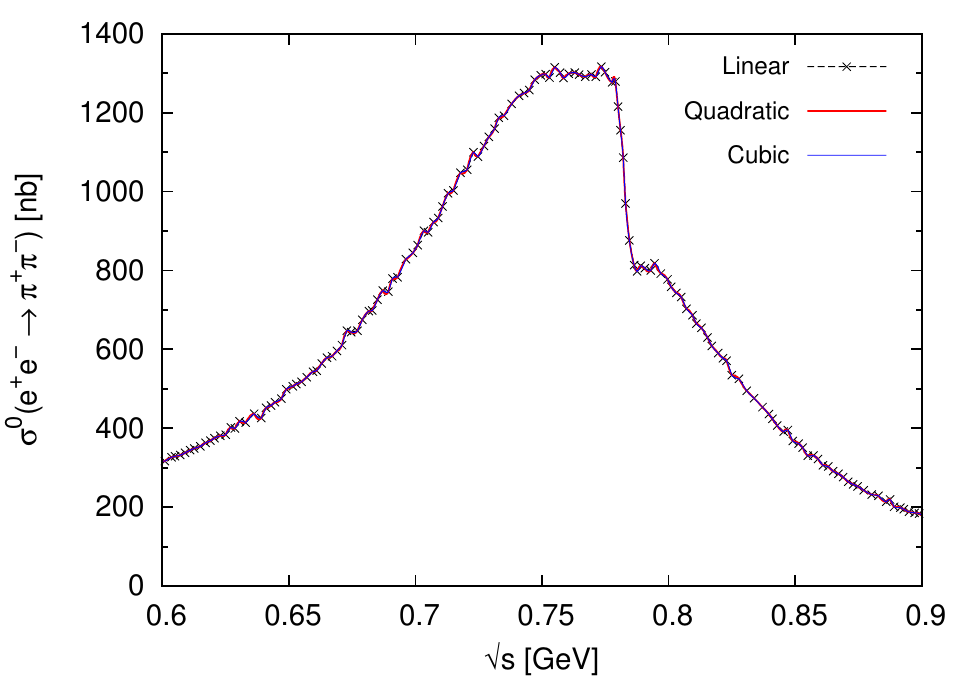}}\hfill
  \subfloat[The $\phi$ resonance region of the $K^+K^-$ channel.]{%
    \includegraphics[width= 0.5\textwidth]{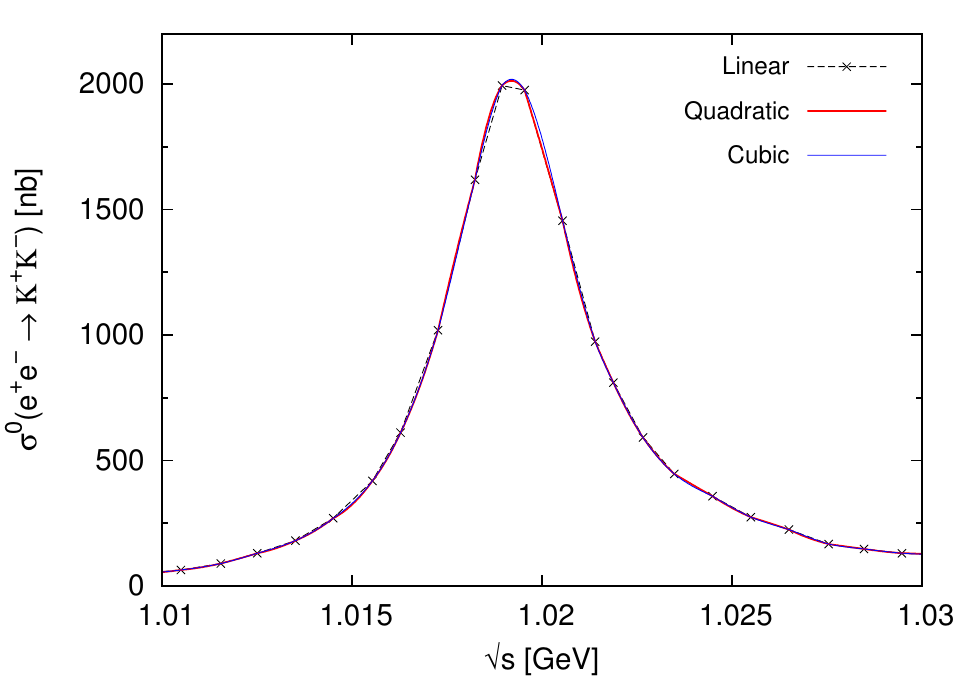}}\hfill
  \caption{\small The differences observed using linear, quadratic and
    cubic integration routines in prominent resonance regions in the
    $\pi^+\pi^-$ and $K^+K^-$
    channels.}\label{Fig:PolynomialIntegration} 
\end{figure}

As in~\cite{Hagiwara:2011af,Hagiwara:2003da,Hagiwara:2006jt}, the data
are integrated using a trapezoidal rule integral in order to obtain
$a_{\mu}^{\rm had, VP}$ and $\Delta \alpha_{\rm had}$. In principle, the use
of the trapezoidal rule integral could lead to unreliable results due
to the form of the kernel function or at narrow resonances if data are
sparse. However, with the current density of cross section
measurements, especially in the dominant hadronic channels, the
differences between trapezoidal rule integration and any higher order
polynomial approximation are consequently small and of no concern. This
can be seen in the plots in Figure~\ref{Fig:PolynomialIntegration}. 

The calculation of the uncertainty of $a_\mu^{\rm had, VP}$ and
$\Delta\alpha_{\rm had}$ has been modified to improve the
determination of the error contribution at the integral
boundaries. Should the upper or lower integral boundary, $E_a$, with
energy $E_m < E_a < E_{m+1}$ and cross section value $R_m < R_a <
R_{m+1}$, be found by linear interpolation (or extrapolation if it is
necessary to extend the integral boundaries), then the output covariance
matrix $V\big(m,n\big)$ is interpolated accordingly using the standard
error propagation formula  
\beq
V\big(a,b\big) = \sum_m\sum_n \frac{\del
  R_a}{\del R_m}V\big(m,n\big)\frac{\del R_b}{\del R_n} \,, 
\eeq
where $m$, $n$ run over all clusters, $b$ is a label for any energy
$E_b$ and $V\big(b,a\big)=V\big(a,b\big)$. 


\section{Determining $a_{\mu}^{\rm had, VP}$ and $\Delta\alpha_{\rm had}(M_Z^2)$}\label{amuanddeltaalpha}

The following section summarises the data combination and estimates of $a_{\mu}^{\rm had, \, LO \, VP}$ and \allowbreak $\Delta\alpha_{\rm had}(M_Z^2)$ from the leading and major sub-leading hadronic final states. All contributions from exclusive hadronic channels are evaluated up to $1.937$ GeV, which is the chosen transition point between the sum of exclusive channels and the inclusive $R$-ratio data in this work. This is discussed in detail in Section~\ref{exv.vs.inc.vs.pqcd}. Each contribution to $a_{\mu}^{\rm had, \, LO \, VP}$ is quoted with its respective statistical (stat) uncertainty, systematic (sys) uncertainty, VP (vp) correction uncertainty and FSR (fsr) correction uncertainty individually. This is followed by the contribution with a total (tot) uncertainty, determined from the individual sources added in quadrature,
\begin{align}
a_{\mu}^{\rm had, \, LO \, VP} & =  a_{\mu}^{\rm had, \, LO \, VP} \pm (\delta a_{\mu}^{\rm had, \, LO \, VP})_{\rm stat} \pm (\delta a_{\mu}^{\rm had, \, LO \, VP})_{\rm sys} \pm (\delta a_{\mu}^{\rm had, \, LO \, VP})_{\rm vp} \pm (\delta a_{\mu}^{\rm had, \, LO \, VP})_{\rm fsr} \nonumber
\\
\
& =  a_{\mu}^{\rm had, \, LO \, VP} \pm (\delta a_{\mu}^{\rm had, \, LO \, VP})_{\rm tot} \, .
\end{align}
All results for $a_{\mu}$ are given in units $10^{-10}$. For the contributions to $\Delta\alpha_{\rm had}(M_Z^2)$, only the mean value with the corresponding total uncertainty,
\beq
\Delta\alpha_{\rm had}(M_Z^2) = \Delta\alpha_{\rm had}(M_Z^2) \pm (\delta\Delta\alpha_{\rm had}(M_Z^2))_{\rm tot} \, ,
\eeq
is quoted and all results are given in units $10^{-4}$. In both cases, uncertainties include all available correlations and local $\chi^2$ inflation as discussed in Section~\ref{ssSec: Fitting data}. In the following, unless stated explicitly, only those data sets that are new additions since~\cite{Hagiwara:2011af}, or those that have undergone studies that are mentioned in the text, are discussed and referenced. All other contributing data sets are referenced in~\cite{Hagiwara:2003da,Hagiwara:2006jt,Hagiwara:2011af}. 

\subsection{$\pi^+\pi^-$ channel}\label{chap:pipi}
The  $\pi^+\pi^-$ channel is by far the most important contribution to $a_{\mu}^{\rm had, \, VP}$, dominating both its mean value and uncertainty. With twenty-six data sets totalling almost one thousand data points, it is also the largest individual data combination. Two new radiative return measurements from the KLOE collaboration~\cite{Babusci:2012rp} and the BESIII collaboration~\cite{Ablikim:2015orh} in the important $\rho$ region have greatly improved the estimate of this final state. The new data from the KLOE collaboration (KLOE12) agree well with both other measurements by KLOE (KLOE08~\cite{Ambrosino:2008aa} and KLOE10~\cite{Ambrosino:2010bv}). The three measurements are, in part, highly correlated, necessitating the construction of full statistical and systematic covariance matrices to be used in any combination of these data. This has been achieved in the separate work~\cite{KLOEcombination}, where all details regarding the correlations, the resulting combination of the KLOE data and a comparison with other experimental measurements of $\sigma_{\pi\pi(\gamma)}$ are presented. These covariance matrices are used here as input in the full $\pi^+\pi^-$ combination in order to fully incorporate the correlation information of the KLOE data as an influence on both the estimate of $a_{\mu}^{\pi^+\pi^-}$ and its uncertainty.

\begin{figure}[!t] 
  \centering
    \includegraphics[width=0.8\textwidth]{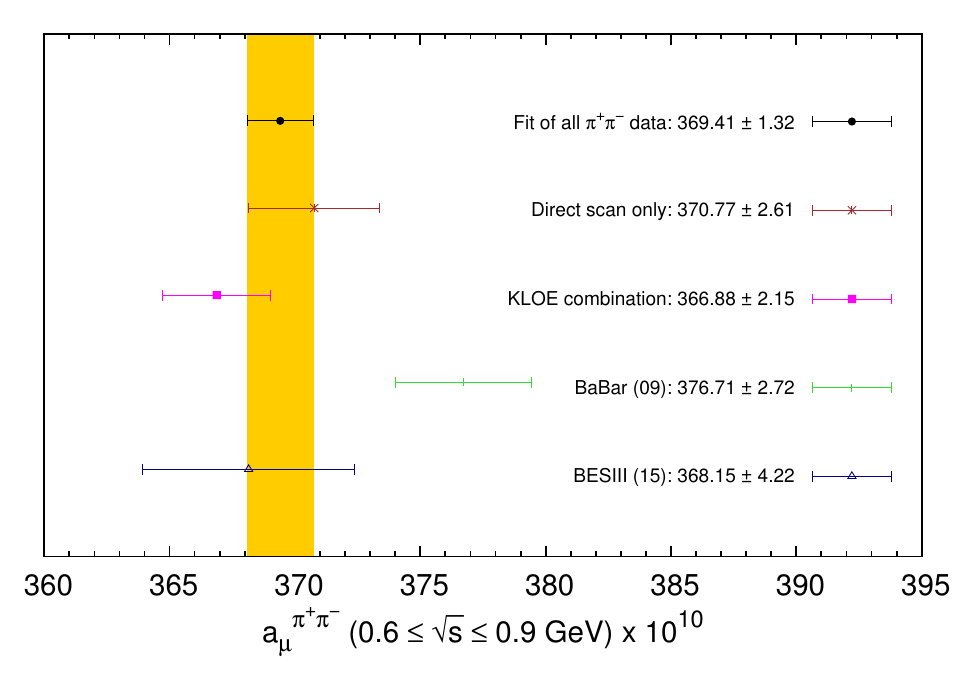}
     \caption{\small The comparison of the integration of the individual radiative return measurements and the combination of direct scan $\pi^+\pi^-$ measurements between $0.6 \leq \sqrt{s} \leq 0.9$ GeV.}     \label{fig:RadRetCompare}
\end{figure} 

\begin{figure}[!t]
\centering
  \subfloat{%
    \includegraphics[width= 0.5\textwidth]{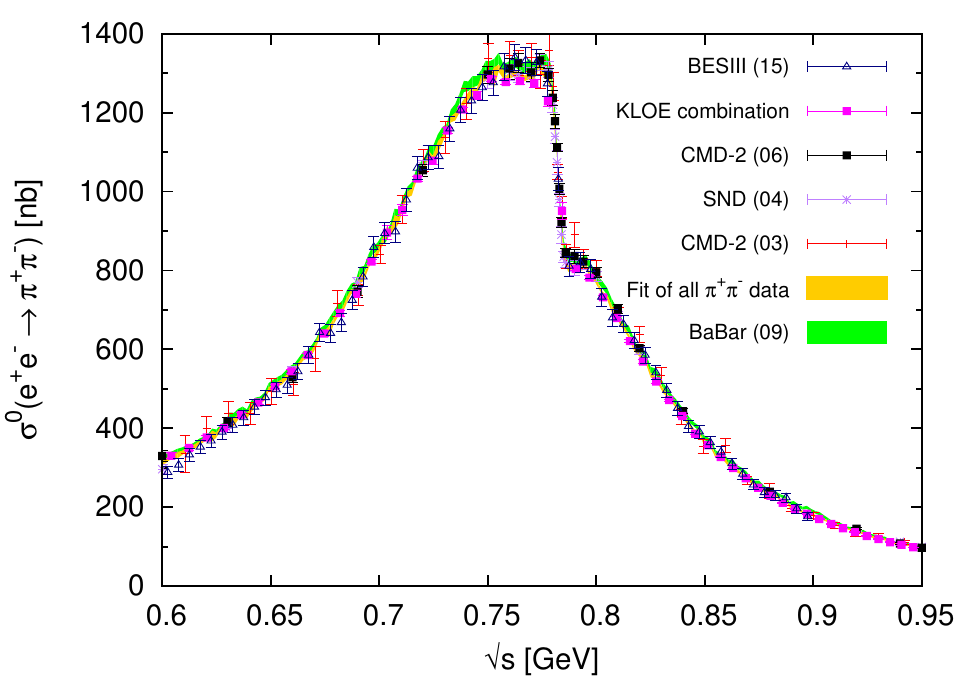}}\hfill
  \subfloat{%
    \includegraphics[width= 0.5\textwidth]{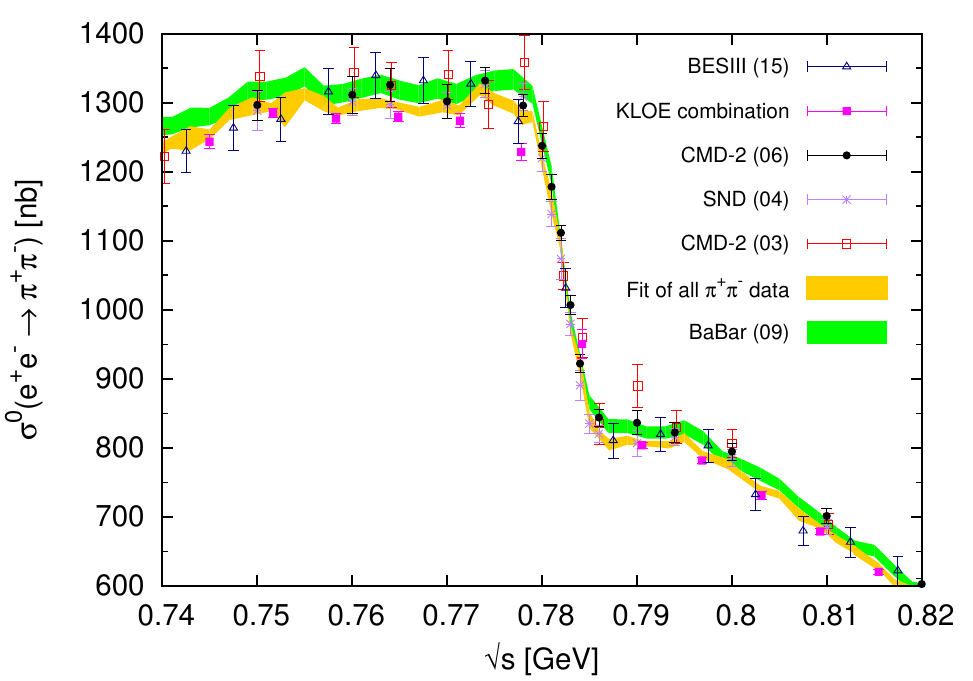}}\hfill
  \caption{\small Contributing data in the $\rho$ resonance region of the $\pi^+\pi^-$ channel plotted against the new fit of all data (left panel), with an enlargement of the $\rho-\omega$ interference region (right panel).}\label{fig:pipirho}
\end{figure}

\begin{figure}[!t] 
  \centering
    \includegraphics[width=0.9\textwidth]{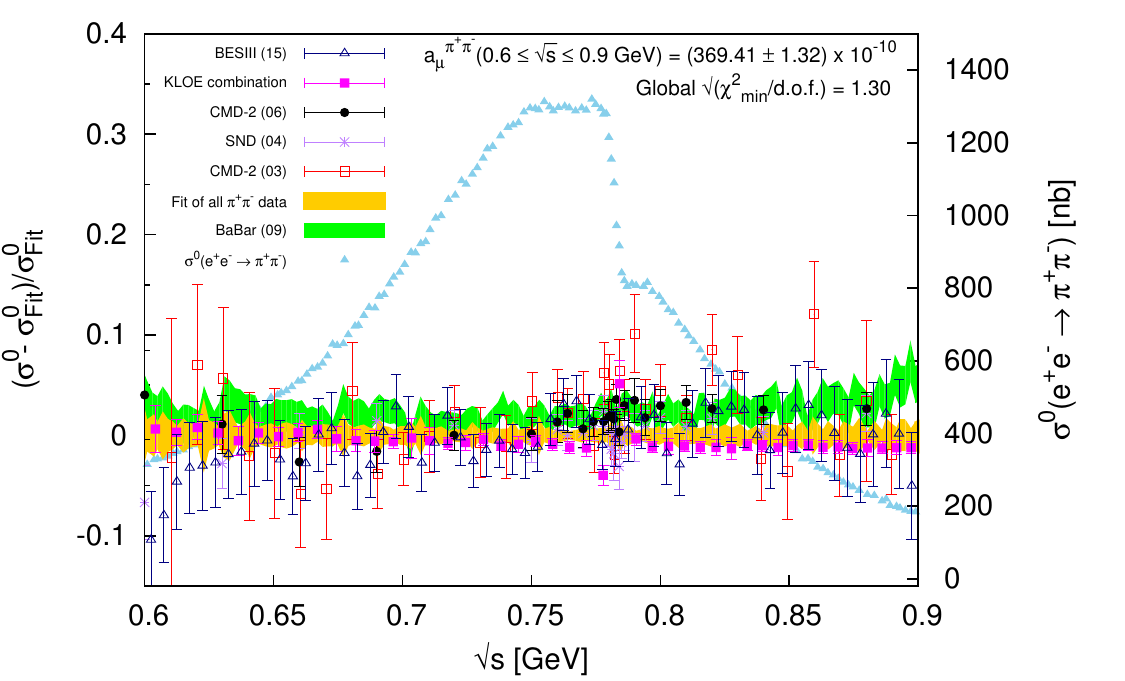}
     \caption{\small The relative difference of the radiative return and important direct scan data sets contributing to $a_{\mu}^{\pi^+\pi^-}$ and the fit of all data. For comparison, the individual sets have been normalised against the fit and have been plotted in the $\rho$ region. The light green band represents the BaBar data and their errors (statistical and systematic, added in quadrature). The yellow band represents the full data combination which incorporates all correlated statistical and systematic uncertainties. However, the width of the yellow band simply displays the square root of the diagonal elements of the total output covariance matrix of the fit. }     \label{fig:RadRetFit}
\end{figure}

\begin{figure}[!t] 
  \centering
    \includegraphics[width=0.9\textwidth]{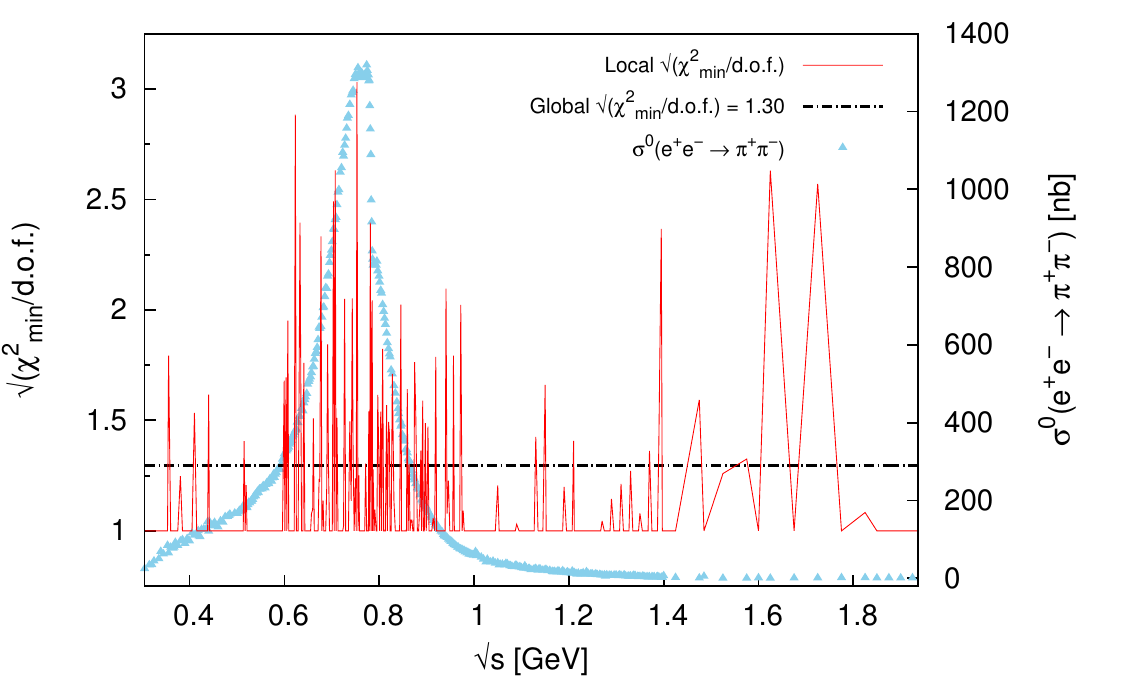}
     \caption{\small The effect of the local $\chi^2$ inflation and the overall global $\chi^2_{\rm min}/{\rm d.o.f.}$ in the $\pi^+\pi^-$ channel, which is plotted against the $e^+e^-\rightarrow \pi^+\pi^-$ cross section for reference.}     \label{fig:localvsglobal}
\end{figure}

The BESIII measurement in the $\rho$ resonance region (again with full statistical and systematic covariance matrices) allows for an in-depth comparison of the existing radiative return measurements already contributing to the $\pi^+\pi^-$ channel, namely the three measurements by the KLOE collaboration and the finely binned measurement from the BaBar collaboration~\cite{Aubert:2009ad}. In~\cite{Hagiwara:2011af}, details were given regarding tension between the KLOE and BaBar measurements, where the BaBar data were considerably higher. As is evident from Figure~\ref{fig:RadRetCompare}, tension exists between the BaBar data and all other contributing data in the dominant $\rho$ region. When considering this along with the plots of the resulting cross section in Figure~\ref{fig:pipirho}, it is clear that the new BESIII data agrees well with the KLOE data and the full $\pi^+\pi^-$ combination. Interestingly, however, it is in better agreement with the BaBar data at the peak of the resonance where the cross section is largest. Although BaBar still influences with an increase, the agreement between the other radiative return measurements and the direct scan data largely compensates for this effect. This is demonstrated by Figure~\ref{fig:RadRetFit}, with the combination clearly favouring the other measurements. Tension between data sets, however, still exists and is reflected in the local $\chi^2$ error inflation, which results in an $\sim15\%$ increase in the uncertainty of $a_{\mu}^{\pi^+\pi^-}$. The effect of this energy dependent error inflation is shown in Figure~\ref{fig:localvsglobal}, where the difference in using a local scaling of the error instead of a global one is clearly visible. Tensions arise in particular in the $\rho$ resonance region, where the cross section is large.

The full combination of all $\pi^+\pi^-$ data is found to give
\begin{align}
a_{\mu}^{\pi^+\pi^-}[0.305\leq \sqrt{s}\leq1.937\text{ GeV}] & = 502.97 \pm 1.14 \pm 1.59 \pm 0.06 \pm 0.14 \nonumber
\\
\  \label{pipiFull}
& = 502.97 \pm 1.97
\end{align}
and
\beq
\Delta\alpha_{\pi^+\pi^-}(M_Z^2)[0.305\leq \sqrt{s}\leq1.937 \text{ GeV}] = 34.26 \pm 0.12 \, .
\eeq
Although this value of $a_{\mu}^{\pi^+\pi^-}$ stays well within the error estimate of~\cite{Hagiwara:2011af}, it exhibits a substantial decrease of the mean value. This has been attributed to the new data combination routine which allows for the full use of correlations in the determination of the mean value as well as the uncertainty and the inclusion of the new, precise radiative return data which suppresses the influence of BaBar in the $\rho$ resonance region. 

In comparison with equation~\eqref{pipiFull}, the BaBar data alone in the same energy range give an estimate of $a_{\mu}^{\pi^+\pi^-}(\text{BaBar data only})  =  513.2 \pm 3.8$. Should all available $\pi^+\pi^-$ data be combined using a simple weighted average as in equation \eqref{R0RmI}, which only provides the error weighting to each cluster by its local uncertainty, the estimate for $a_{\mu}^{\pi^+\pi^-}$ would be $a_{\mu}^{\pi^+\pi^-}(\text{Naive weighted}$ ${\rm average})  =  509.1 \pm 2.9$. In this case, the estimate is strongly pulled up by the fine binning and high statistics of the BaBar data that dominate when no correlations are taken into account for the mean value. This difference of nearly $2\sigma$ when comparing to equation \eqref{pipiFull} indicates the importance of fully incorporating all available correlated uncertainties in any combination of the data.

The uncertainty has reduced by approximately one third. Again, this is due to the new, precise radiative return data which further dominate the $\pi^+\pi^-$ fit and the improvement of the overall combination which now fully incorporates the energy dependent correlations. In addition, the radiative corrections uncertainties have reduced since~\cite{Hagiwara:2011af}, as discussed in detail in Section~\ref{radcorr}.

\subsection{$\pi^+\pi^-\pi^0$ channel}

\begin{figure}[!t] 
  \centering
    \includegraphics[width=0.6\textwidth]{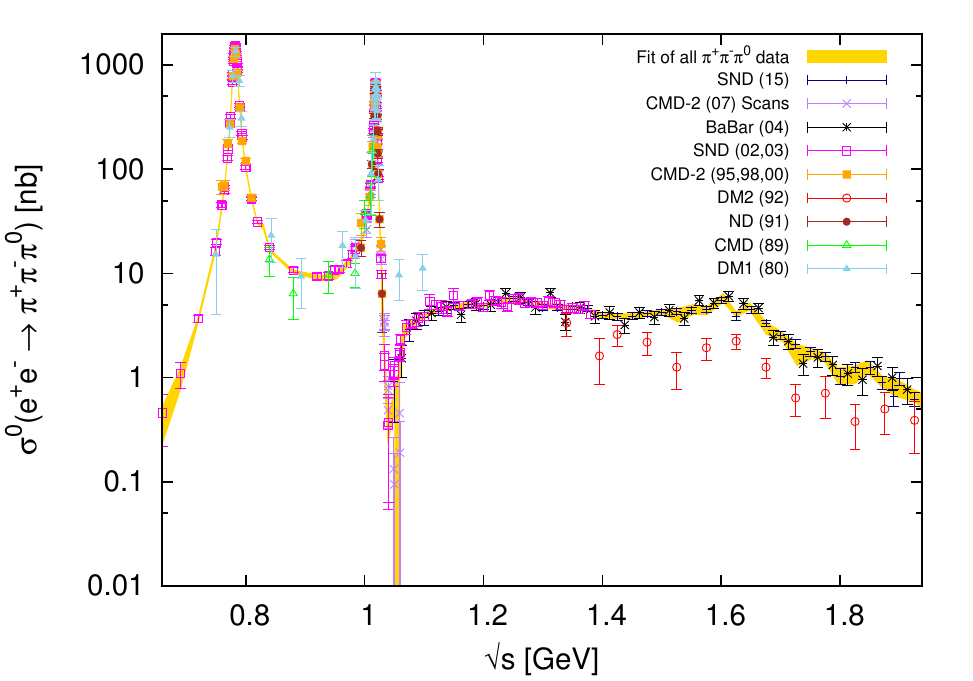}
     \caption{\small The cross section $\sigma^{0}(e^+e^-\rightarrow\pi^+\pi^-\pi^0)$ in the range $0.66\leq \sqrt{s}\leq1.937$ GeV, where the prominent $\omega$ and $\phi$ resonances are clearly visible.} \label{fig:pi+pi-pi0full}
\end{figure}

\begin{figure}[!t]
\centering
  \subfloat[The $\omega$ resonance.]{%
    \includegraphics[width= 0.5\textwidth]{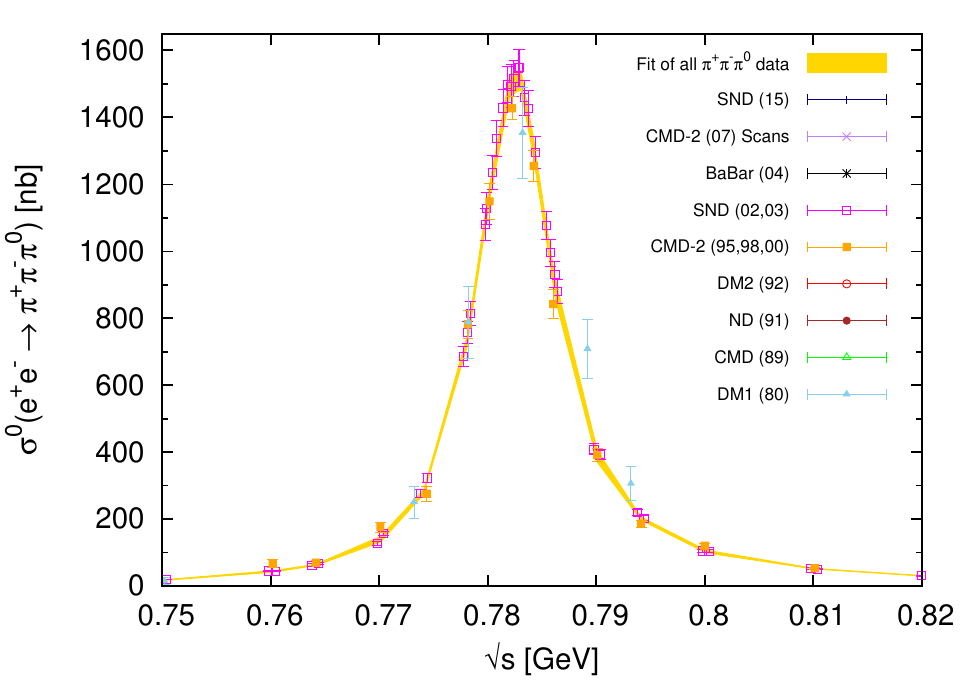}}\hfill
  \subfloat[The $\phi$ resonance.]{%
    \includegraphics[width= 0.5\textwidth]{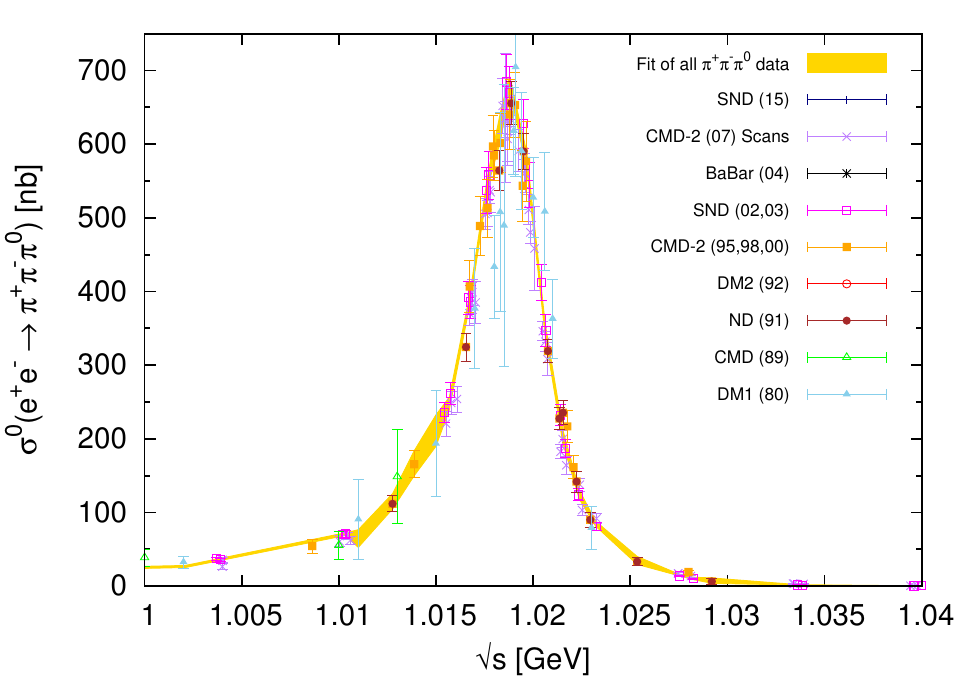}}\hfill
  \caption{\small Enlargements of the resonance regions in the $\pi^+\pi^-\pi^0$ final state.}\label{fig:pi+pi-pi0res}
\end{figure}

Since~\cite{Hagiwara:2011af}, there has only been one new addition to the $\pi^+\pi^-\pi^0$ channel~\cite{Aulchenko:2015mwt}. This new data set improves this channel away from resonance, where previously only BaBar data~\cite{Aubert:2004kj} had provided a contribution of notable precision. Compared to~\cite{Hagiwara:2011af}, an additional change is applied to three separate data scans over the $\phi$ resonance in a measurement by CMD-2~\cite{Akhmetshin:2006sc}, where the systematic uncertainties between the three scans are now taken to be fully correlated~\cite{FedorCorrelations}. These changes, along with the new data combination routine, have resulted in an improved estimate of 
\begin{align}
a_{\mu}^{\pi^+\pi^-\pi^0}[0.66\leq \sqrt{s}\leq1.937 \text{ GeV}] & =47.79 \pm 0.22 \pm 0.71 \pm 0.13 \pm 0.48 \nonumber
\\
\
& = 47.79 \pm 0.89
\end{align}
and
\beq
\Delta\alpha_{\pi^+\pi^-\pi^0}(M_Z^2)[0.66\leq \sqrt{s}\leq1.937 \text{ GeV}] = 4.77 \pm 0.08 \, .
\eeq
Figure~\ref{fig:pi+pi-pi0full} shows the full integral range of the data for the $\pi^+\pi^-\pi^0$ cross section. Figure~\ref{fig:pi+pi-pi0res} shows an enlargement of the $\omega$ and $\phi$ resonance regions in this channel.

\subsection{$4\pi$ channels}

The $\pi^+\pi^-\pi^+\pi^-$ channel now includes two new additions since~\cite{Hagiwara:2011af}. First, an improved statistics measurement by the BaBar collaboration in the range $0.6125\leq \sqrt{s}\leq4.4875$ GeV~\cite{Lees:2012cr} supersedes their previous measurement in this channel~\cite{Aubert:2005eg}. More recently, a data set by the CMD-3 collaboration in the range $0.92\leq \sqrt{s}\leq1.06$ GeV~\cite{Akhmetshin:2016dtr} has been completed, which better resolves the interference pattern of the $\phi \rightarrow \pi^+\pi^-\pi^+\pi^-$ transition that is clearly evident in the non-resonant cross section. In addition, the M3N thesis data~\cite{Paulot:1979thesis} that are not published, but were previously included in this channel, are now discarded. With these changes, 
\begin{figure}[t] 
  \centering
    \includegraphics[width=0.6\textwidth]{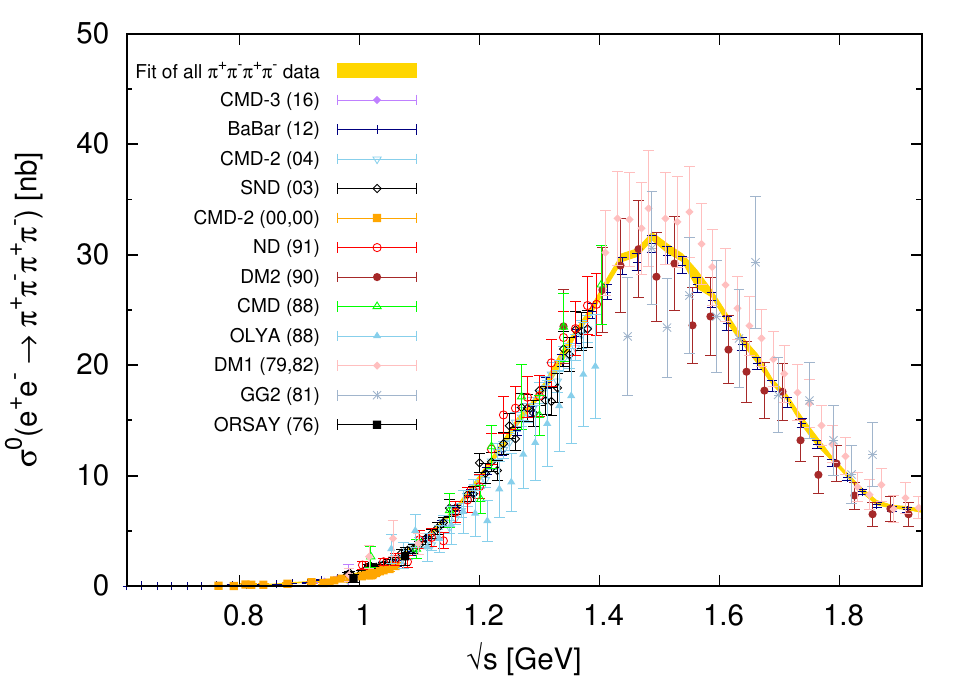}
     \caption{\small The cross section $\sigma^{0}(e^+e^-\rightarrow\pi^+\pi^-\pi^+\pi^-)$ in the range $0.6125\leq \sqrt{s}\leq1.937$ GeV.} \label{fig:pi+pi-pi+pi-full}
\end{figure}
\begin{figure}[!t] 
  \centering
    \includegraphics[width=0.6\textwidth]{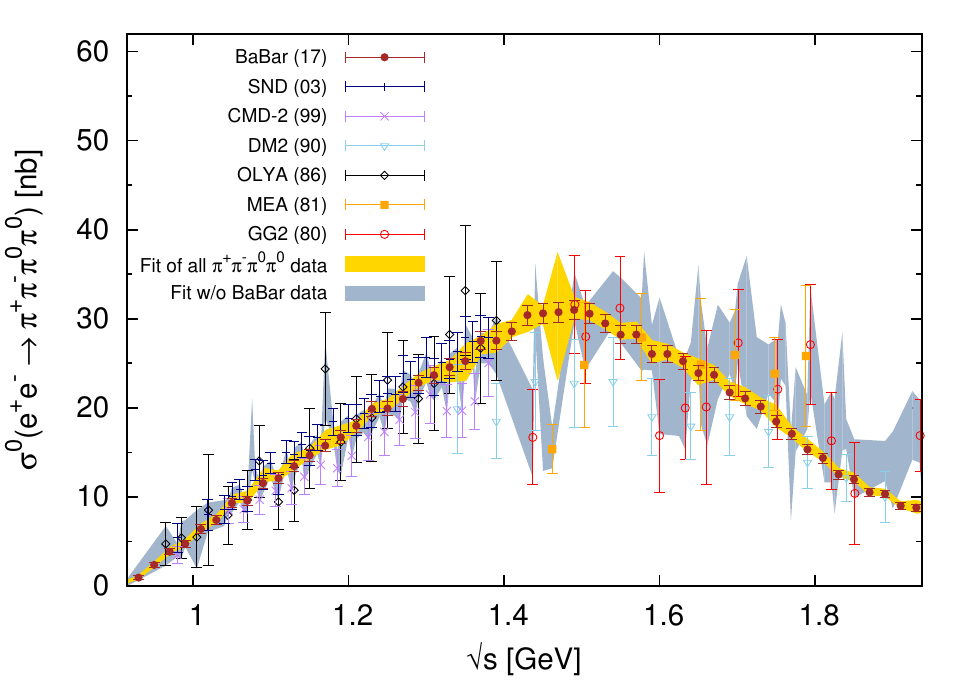}
     \caption{\small The cross section $\sigma^{0}(e^+e^-\rightarrow\pi^+\pi^-\pi^0\pi^0)$ in the range $0.850\leq \sqrt{s}\leq1.937$ GeV.} \label{fig:pi+pi-pi0pi0full}
\end{figure} 
\begin{align}
a_{\mu}^{\pi^+\pi^-\pi^+\pi^-}[0.6125\leq \sqrt{s}\leq1.937 \text{ GeV}] & =14.87 \pm 0.02 \pm 0.13 \pm 0.03 \pm 0.15 \nonumber
\\
\
& = 14.87 \pm 0.20
\end{align}
and
\beq
\Delta\alpha_{\pi^+\pi^-\pi^+\pi^-}(M_Z^2)[0.6125\leq \sqrt{s}\leq1.937 \text{ GeV}] = 4.02 \pm 0.05 \, .
\eeq
Here, the mean value has increased since~\cite{Hagiwara:2011af} largely due to the new BaBar data. The fitted cross section and data are displayed in Figure~\ref{fig:pi+pi-pi+pi-full}.

The picture for the $\pi^+\pi^-\pi^0\pi^0$ final state has also improved, with a new measurement of this channel by BaBar~\cite{TheBaBar:2017vzo} providing the only new data in this channel since 2003. As with the $\pi^+\pi^-\pi^+\pi^-$ channel, the M3N data~\cite{Paulot:1979thesis} in this channel have been omitted. The estimate for this channel is
\begin{align}
a_{\mu}^{\pi^+\pi^-\pi^0\pi^0}[0.850\leq \sqrt{s}\leq1.937 \text{ GeV}] & =19.39 \pm 0.09 \pm 0.74 \pm 0.04 \pm 0.19 \nonumber
\\
\
& = 19.39 \pm 0.78
\end{align}
and
\beq
\Delta\alpha_{\pi^+\pi^-\pi^0\pi^0}(M_Z^2)[0.850\leq \sqrt{s}\leq1.937 \text{ GeV}] = 5.00 \pm 0.20\, ,
\eeq
where there is clear improvement since~\cite{Hagiwara:2011af}. The uncertainty contribution from $\pi^+\pi^-\pi^0\pi^0$ is, however, still relatively large in comparison with its contribution to $a_{\mu}^{\rm had, \, LO \, VP}$ and $\Delta\alpha_{\rm had}$ and requires better new data. The new fit of the data for the bare cross section $e^+e^-\rightarrow\pi^+\pi^-\pi^0\pi^0$ is shown in Figure~\ref{fig:pi+pi-pi0pi0full}. The fit without the new BaBar data is shown for comparison to highlight the large improvement this new data set provides. 

\subsection{$K\bar{K}$ channels}

 \begin{figure}[!t]
\centering
  \subfloat[Full range of the cross section $\sigma^{0}(e^+e^-\rightarrow K^+K^-)$.]{%
    \includegraphics[width= 0.6\textwidth]{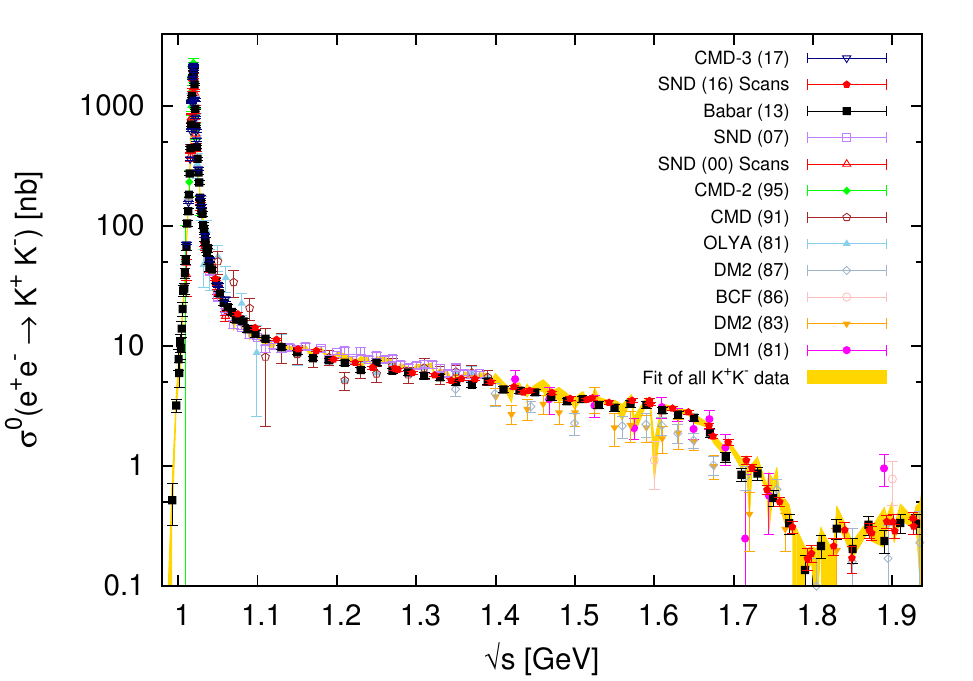}}\hfill
  \subfloat[The $\phi$ resonance.]{%
    \includegraphics[width= 0.6\textwidth]{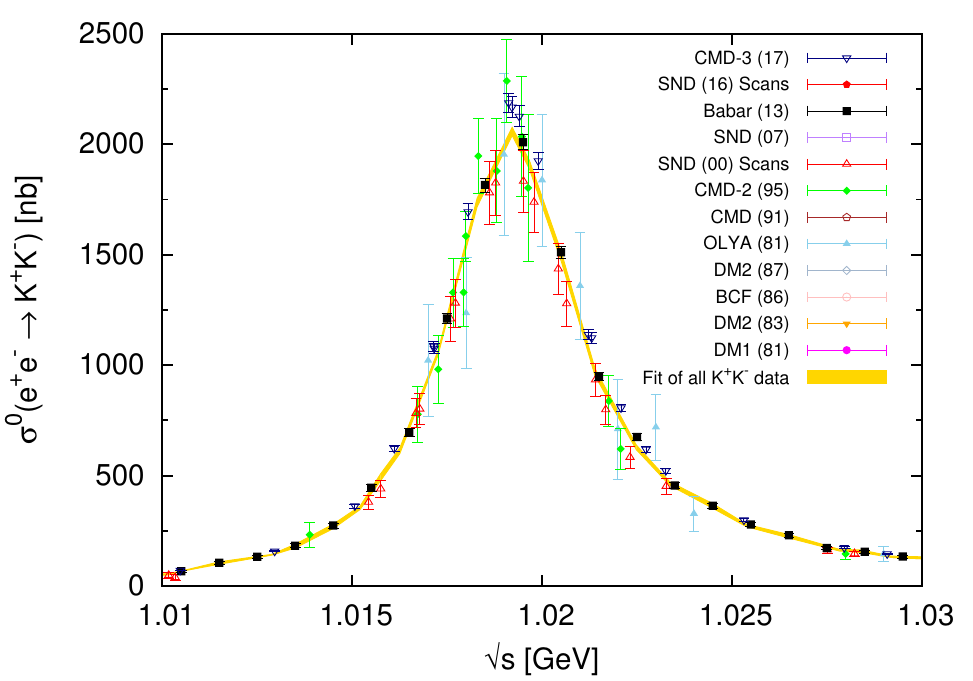}}\hfill
     \caption{\small The cross section $\sigma^{0}(e^+e^-\rightarrow K^+K^-)$ in the range $0.9875\leq \sqrt{s}\leq1.937$ GeV and an enlargement of the $\phi$ resonance. The large influence of the BaBar data (black squares) overwhelms the older data.} \label{fig:K+K-full}
\end{figure}

The $K^+K^-$ channel now includes a precise and finely binned measurement by the BaBar collaboration,  supplemented with full statistical and systematic covariance matrices~\cite{Lees:2013gzt}. This is the first and only example to date of the release of energy dependent, correlated uncertainties outside of the $\pi^+\pi^-$ channel and has an overwhelming influence on the data combination. There is also a new measurement in this channel of the $\phi$ resonance by the CMD-3 collaboration~\cite{Kozyrev:2017agm}. The existing CMD-2 scans in the same region~\cite{Akhmetshin:2008gz} are omitted from this work as they suffer from an overestimation of the trigger efficiency for slow kaons~\cite{Kozyrev:2017agm} and are awaiting reanalysis\cite{CMD-2trigger}. In addition, two new scans by the SND collaboration measured at the tail of the $\phi$ and into the continuum are included~\cite{Achasov:2016lbc}. The systematic uncertainties of these two scans, along with the existing two scans by SND~\cite{Achasov:2000am}, are considered to be fully correlated~\cite{FedorCorrelations}. The combination of the available $K^+K^-$ data now gives
\begin{align}
a_{\mu}^{K^+K^-}[0.9875\leq \sqrt{s}\leq1.937 \text{ GeV}] & =23.03 \pm 0.08 \pm 0.20 \pm 0.03 \pm 0.00 \nonumber
\\
\
& = 23.03 \pm 0.22
\end{align}
and
\beq
\Delta\alpha_{K^+K^-}(M_Z^2)[0.9875\leq \sqrt{s}\leq1.937 \text{ GeV}] = 3.37 \pm 0.03 \, ,
\eeq
which exhibits an increase of the mean value of more than 1$\sigma$ from the estimate in~\cite{Hagiwara:2011af} attributed to the inclusion of the new BaBar and CMD-3 data. The cross section of the process $e^+e^-\rightarrow K^+K^-$ is displayed in Figure~\ref{fig:K+K-full}.

 \begin{figure}[!t]
\centering
  \subfloat[Full range of the cross section $\sigma^{0}(e^+e^-\rightarrow K^0_S K^0_L)$.]{%
    \includegraphics[width= 0.6\textwidth]{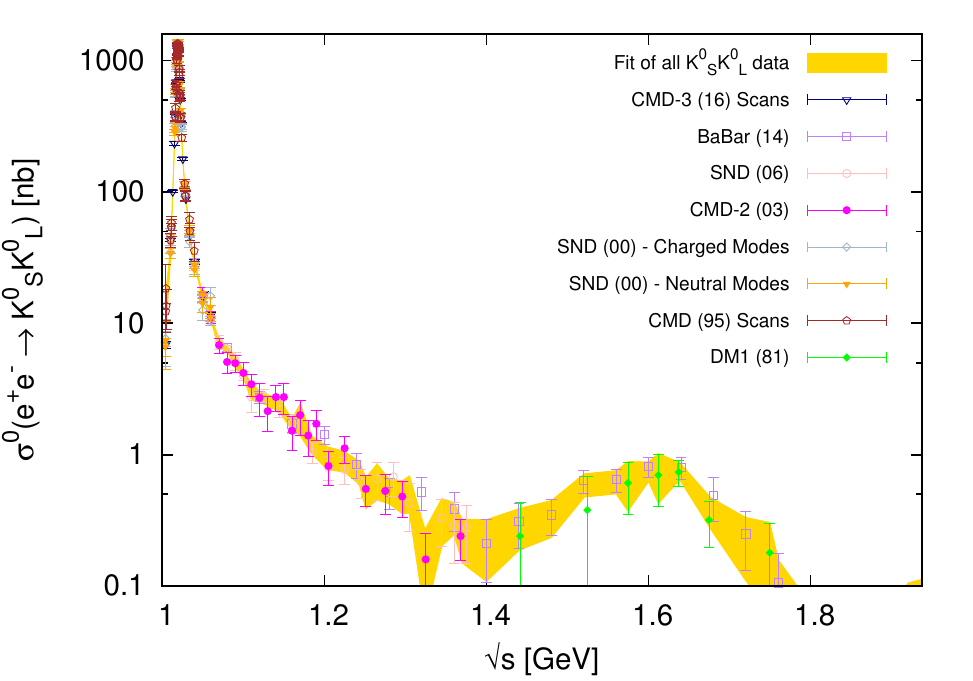}}\hfill
  \subfloat[The $\phi$ resonance.]{%
    \includegraphics[width= 0.6\textwidth]{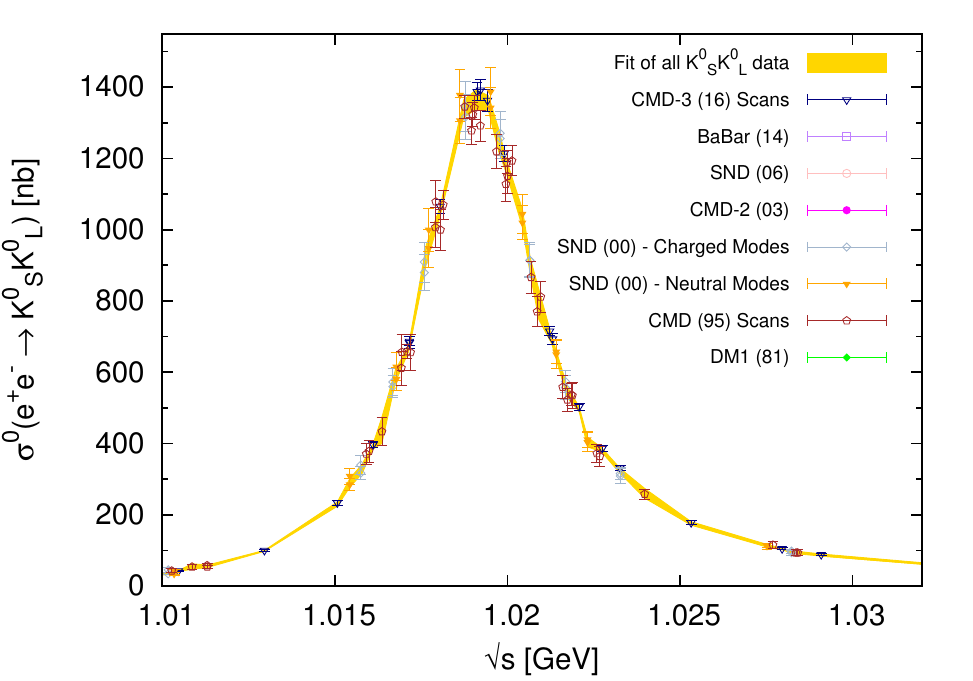}}\hfill
     \caption{\small The cross section $\sigma^{0}(e^+e^-\rightarrow K^0_S K^0_L)$ with an enlargement of the $\phi$ resonance.} \label{fig:K0SK0Lfull}
\end{figure}
The uncertainty has drastically improved since~\cite{Hagiwara:2011af} with much of the change being due to a finer clustering over the $\phi$ resonance after the inclusion of the new high statistics BaBar data. Following from the discussion in Section~\ref{FSRcorrection}, that there is now no FSR correction applied to this channel and, therefore, there is no extra radiative correction uncertainty due to FSR. It should also be noted that any FSR correction would result in an increase of $a_{\mu}^{K^+K^-}$, showing again the strong influence the new data have had in this channel to increase the mean value since~\cite{Hagiwara:2011af}, where previously an FSR correction was applied.

New data for the $K^0_S K^0_L$ final state is included from the BaBar collaboration above the $\phi$ resonance~\cite{Lees:2014xsh} and from the CMD-3 collaboration on the $\phi$~\cite{Kozyrev:2016raz}. Two existing measurements in this channel~\cite{Achasov:2000am,Akhmetshin:1999ym} have multiple data scans of which the systematic uncertainties are now taken to be fully correlated~\cite{FedorCorrelations}. This combination results in a contribution of
\begin{align}
a_{\mu}^{K^0_S K^0_L}[1.00371\leq \sqrt{s}\leq1.937 \text{ GeV}] & =13.04 \pm 0.05 \pm 0.16 \pm 0.10 \pm 0.00 \nonumber
\\
\
& = 13.04 \pm 0.19
\end{align}
and
\beq
\Delta\alpha_{K^0_S K^0_L}(M_Z^2)[1.00371\leq \sqrt{s}\leq1.937 \text{ GeV}] = 1.77 \pm 0.03 \, .
\eeq
Again, there is also no additional FSR correction uncertainty applied to this channel, with the reasoning in Section~\ref{FSRcorrection} enforced by the probability of photon emission being highly suppressed for a neutral final state and given the limited phase space. The cross section of the process $e^+e^-\rightarrow K^0_S K^0_L$ is displayed in Figure~\ref{fig:K0SK0Lfull}.

\subsection{$K\bar{K}\pi$ and $K\bar{K}2\pi$ channels} \label{Sec:KKnpi}

\begin{figure}[!t]
\centering
 \includegraphics[width= 0.6\textwidth]{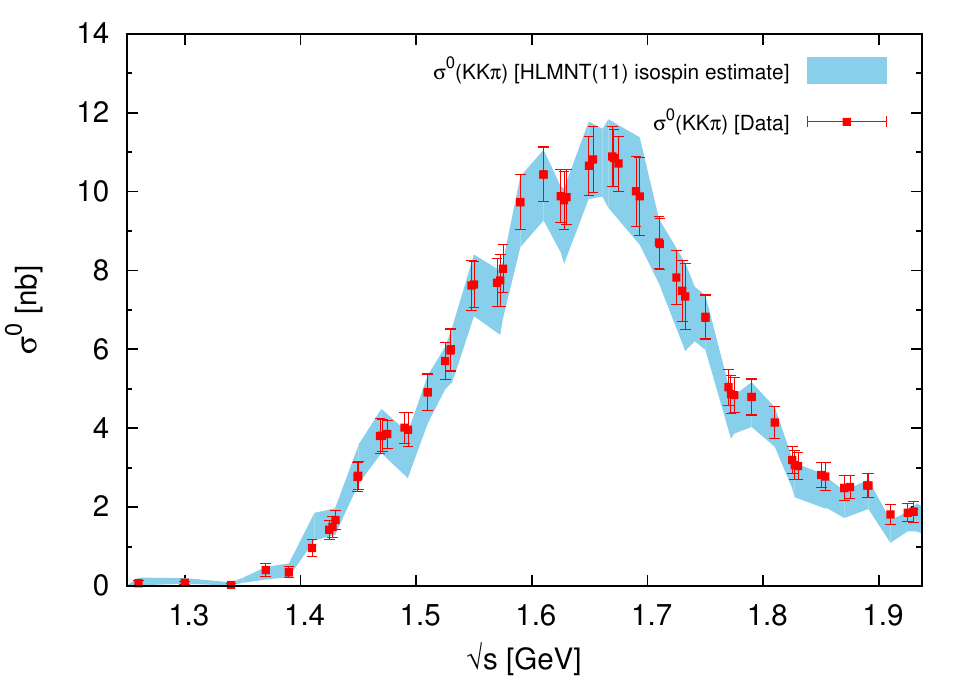}
 \caption{\small The measured cross section $\sigma^{0}(KK\pi)$ compared to the estimate from the previously used isospin relation.} \label{fig:KKpi}
\end{figure}

Since~\cite{Hagiwara:2011af}, the neutral final state $K^0_S K^0_L\pi^0$ has been measured by BaBar~\cite{TheBABAR:2017vgl} and SND~\cite{Achasov:2017vaq}, completing all modes that contribute to the $KK\pi$ final state and removing the reliance on isospin for this channel (other than $K^0_S \cong K^0_L$). Therefore, the $KK\pi$ cross section is now calculated using
\begin{align}
\sigma(KK\pi) & =  \sigma(K^0_S K^\pm \pi^\mp) + \sigma(K^0_L K^\pm \pi^\mp) + \sigma(K^+ K^- \pi^0) + \sigma(K^0_S K^0_L \pi^0) \nonumber
\\
\
& \simeq  2\sigma(K^0_S K^\pm \pi^\mp) + \sigma(K^+ K^- \pi^0) + \sigma(K^0_S K^0_L \pi^0) \, ,
\end{align}
resulting in a contribution of
\begin{align}
a_{\mu}^{KK\pi}[1.260\leq \sqrt{s}\leq1.937 \text{ GeV}]   & =2.71 \pm 0.05 \pm 0.11 \pm 0.01 \pm 0.01 \nonumber
\\
\
& = 2.71 \pm 0.12
\end{align}
and
\beq
\Delta\alpha_{KK\pi}(M_Z^2)[1.260\leq \sqrt{s}\leq1.937 \text{ GeV}] = 0.89 \pm 0.04 \, .
\eeq
In~\cite{Hagiwara:2011af}, the isospin estimate in the same energy range yielded,
\beq
a_{\mu}^{KK\pi}(\text{HLMNT11 isospin estimate})   =  2.65 \pm 0.14 .
\eeq
This good agreement between the HLMNT11 isospin estimate and the data-based approach in this analysis is also demonstrated in Figure~\ref{fig:KKpi}.

 \begin{figure}[!t]
\centering
 \includegraphics[width= 0.6\textwidth]{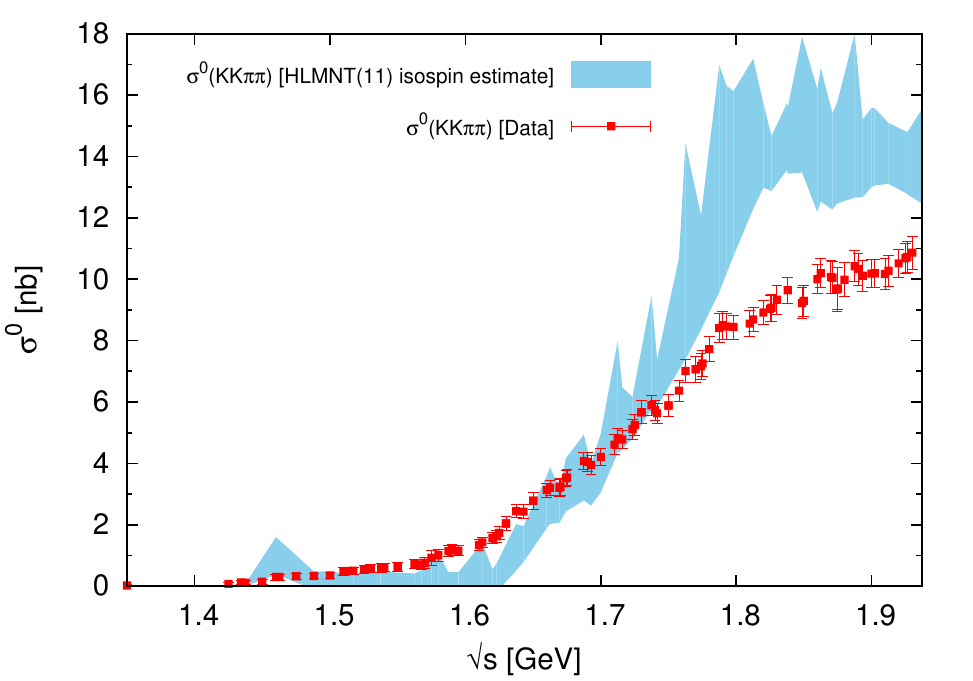}
 \caption{\small The cross section $\sigma^{0}(KK\pi\pi)$ compared to the previous estimate using isospin relations.} \label{fig:KKpipi}
\end{figure}

For $KK2\pi$, BaBar have measured the previously missing modes $K^0_S K^0_L \pi^+\pi^-$, $K^0_S K^0_S\pi^+\pi^-$~\cite{Lees:2014xsh}, $K^0_S K^0_L 2\pi^0$~\cite{TheBABAR:2017vgl} and $K^0_S K^\pm \pi^\mp \pi^0$~\cite{TheBABAR:2017aph}, such that this contribution is now determined using
\begin{align}
\sigma(KK2\pi) & =  \sigma(K^+ K^- \pi^0\pi^0) + \sigma(K^+ K^- \pi^+\pi^-) + \sigma(K^0 \bar{K^0}\pi^+\pi^-) + \sigma(K^0_S K^0_L 2 \pi^0)   \nonumber
\\
\
& \ \ \ + \ \sigma(K^0_S K^\pm\pi^\mp\pi^0) + \sigma(K^0_L K^\pm\pi^\mp\pi^0)   \nonumber
\\
\
& \simeq   \sigma(K^+ K^- \pi^0\pi^0) + \sigma(K^+ K^- \pi^+\pi^-) + \sigma(K^0 \bar{K^0}\pi^+\pi^-) + \sigma(K^0_S K^0_L 2 \pi^0)   \nonumber
\\
\
& \ \ \ + \ 2\sigma(K^0_S K^\pm\pi^\mp\pi^0) \, .
\end{align}
Here, again, it is assumed that $K^0_S \cong K^0_L$ and, hence, 
\begin{align}
\sigma(K^0 \bar{K^0}\pi^+\pi^-) & = \sigma(K^0_S K^0_L \pi^+\pi^-) + \sigma(K^0_S K^0_S \pi^+\pi^-) + \sigma(K^0_L K^0_L \pi^+\pi^-) \nonumber
\\
\
& \simeq  \sigma(K^0_S K^0_L \pi^+\pi^-) + 2\sigma(K^0_S K^0_S \pi^+\pi^-) \, .
\end{align}
Therefore, the estimate in this channel is now found to be
\begin{align} \label{eq:KK2pi}
a_{\mu}^{KK2\pi}[1.350\leq \sqrt{s}\leq1.937 \text{ GeV}]   & =1.93 \pm 0.03 \pm 0.07 \pm 0.01 \pm 0.01 \nonumber
\\
\
& = 1.93 \pm 0.08
\end{align}
and
\beq
\Delta\alpha_{KK2\pi}(M_Z^2)[1.350\leq \sqrt{s}\leq1.937 \text{ GeV}] = 0.75 \pm 0.03 \, .
\eeq
Comparing equation~\eqref{eq:KK2pi} with the HLMNT11 isospin estimate in the same energy range of
\beq
a_{\mu}^{KK2\pi}(\text{HLMNT11 isospin estimate})   =  2.51 \pm 0.35 
\eeq
and examining Figure~\ref{fig:KKpipi}, it is evident that the isospin relations provided a poor estimate of this final state. Using the data, $KK2\pi$ contributes a much smaller mean value with a greatly reduced uncertainty.

\subsection{Inclusive $R$-ratio data}

\begin{figure}[!t] 
  \centering
    \includegraphics[width=0.6\textwidth]{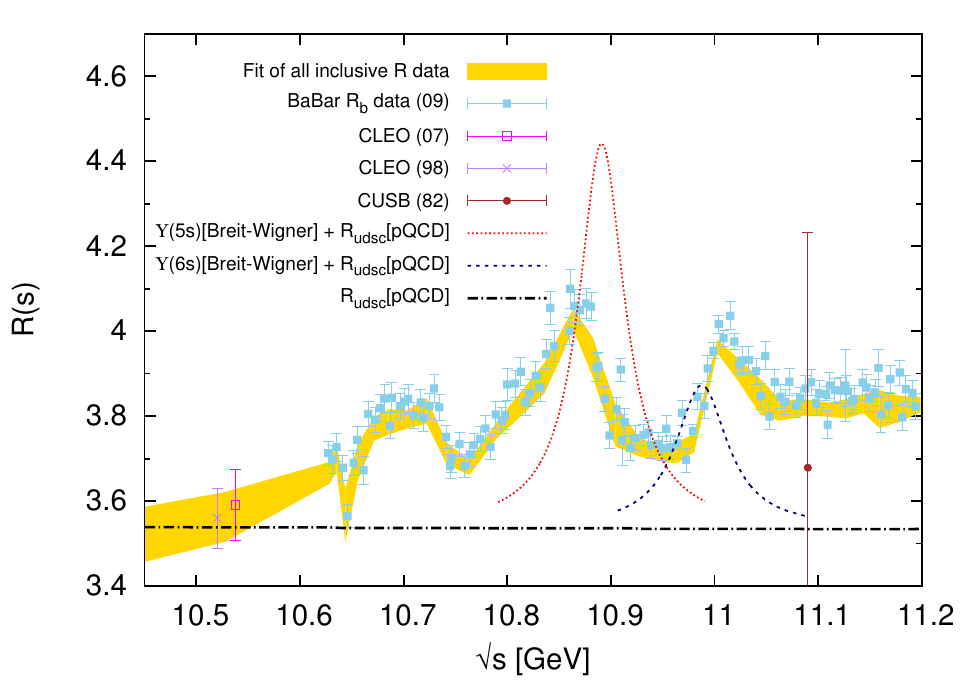}
     \caption{\small The fit of inclusive $R$ data in the region of the newly included BaBar $R_{b}$ data, with the BaBar $R_{b}$ is shown as light blue markers. The resonance structures of the $\Upsilon(5S)$ and $\Upsilon(6S)$ states are clearly visible.} \label{fig:BaBarRb}
\end{figure}

The combination of inclusive hadronic $R$-ratio data between $1.937\leq\sqrt{s}\leq 11.2$ GeV has three new data additions since~\cite{Hagiwara:2011af}. The first of these are the precise BaBar $R_{b}$ data between $10.54\leq\sqrt{s}\leq 11.20$ GeV~\cite{Aubert:2008ab}, which must be deconvoluted of the effects from initial state radiation (ISR) and must have the radiative tails of the resonances from the $\Upsilon(1S-4S)$ states removed (see~\cite{R.Liao:thesis}). These data are then added to the perturbative QCD (pQCD) estimate of $R_{udsc}$~\cite{Harlander:2002ur} to be included as an accurate data set in the inclusive channel in this region. The inclusion of the BaBar $R_{b}$ data is particularly beneficial as it resolves the resonances of the $\Upsilon(5S)$ and $\Upsilon(6S)$ states, removing the need to estimate these structures using a Breit-Wigner parametrisation as was done in~\cite{Hagiwara:2003da,Hagiwara:2006jt,Hagiwara:2011af}. These data are shown in Figure~\ref{fig:BaBarRb}, together with the previously used incoherent sum of the resonance parametrisations which are clearly very different from the $b\bar{b}$ cross section as measured by BaBar. Note that apart from the CLEO(98) data point~\cite{Ammar:1997sk} at $10.52$ GeV, the CLEO(07) data point~\cite{Besson:2007aa} at $10.538$ GeV and the CUSB data point at $11.09$ GeV~\cite{Rice:1982br}, there are no other data in this $b\bar{b}$ resonance region.

\begin{figure}[!t] 
  \centering
    \includegraphics[width=0.6\textwidth]{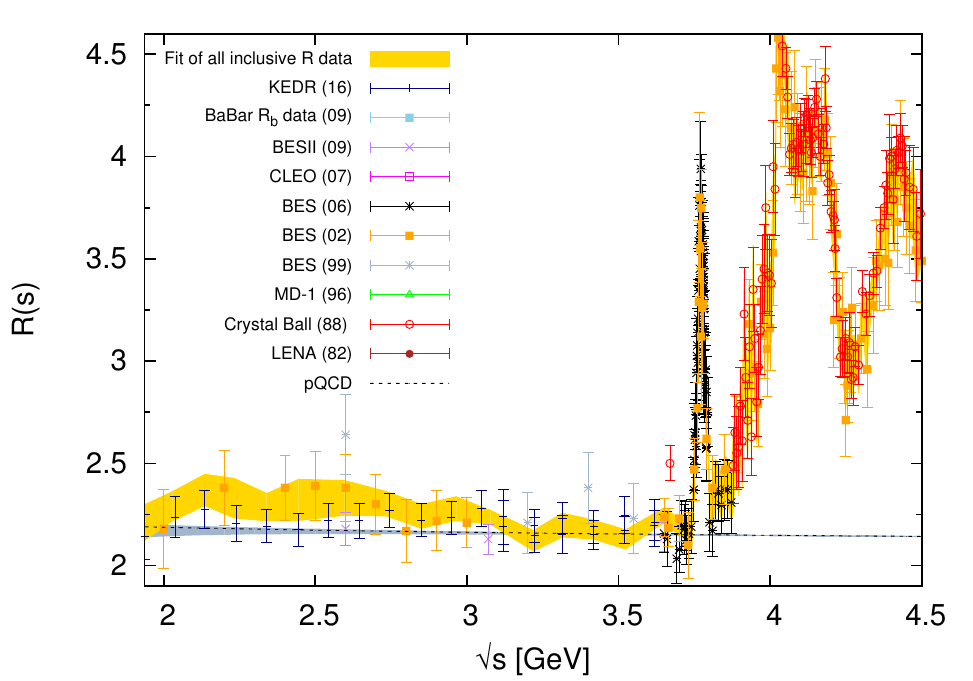}
     \caption{\small The combination of inclusive $R$ data in the region $1.937\leq\sqrt{s}\leq 4.50$ GeV. For comparison, the fit and the contributing data are plotted against the estimate of pQCD, represented by the dashed line and grey band.} \label{fig:inc2-4}
\end{figure}

\begin{figure}[!t] 
  \centering
    \includegraphics[width=0.9\textwidth]{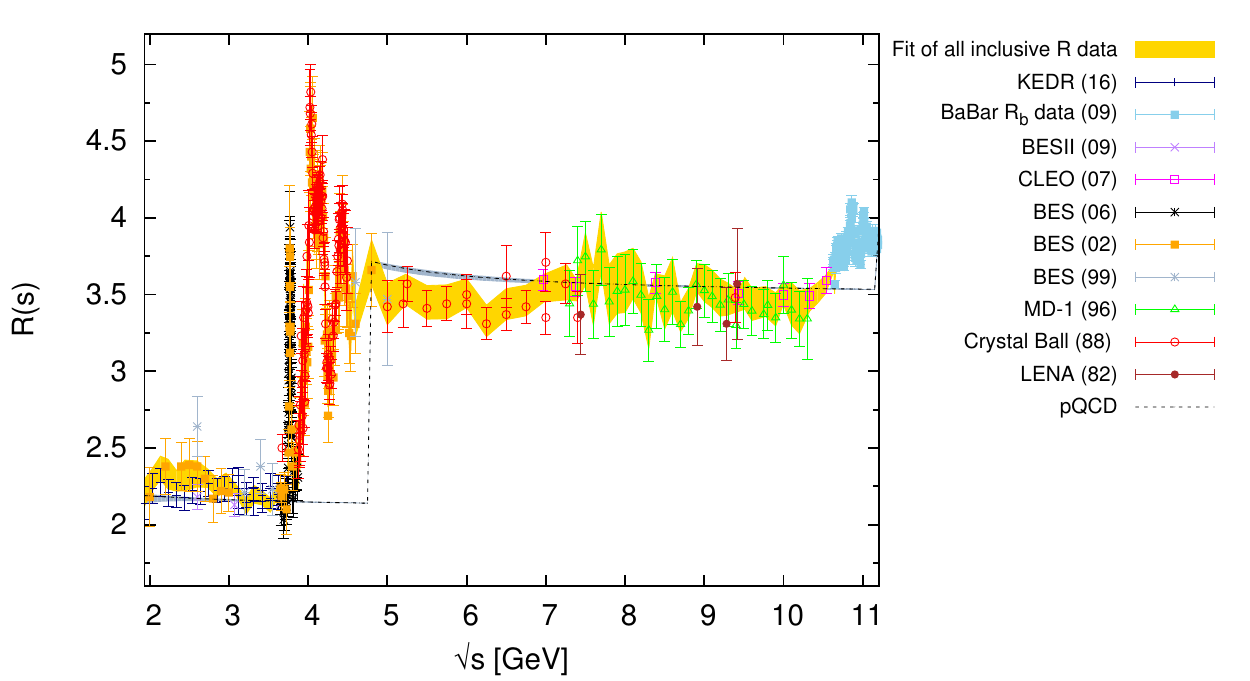}
     \caption{\small Compilation of inclusive data in range $1.937\leq \sqrt{s}\leq11.2\text{ GeV}$. The dashed line and surrounding grey band shows the estimate from pQCD for comparison. The yellow band represents the total uncertainty of the inclusive data combination.} \label{fig:incFull}
\end{figure}

The other new data are recent precise measurements by the KEDR collaboration: one set between $1.84\leq\sqrt{s}\leq 3.05$ GeV~\cite{Anashin:2016hmv} and two scans in the energy range $3.12\leq\sqrt{s}\leq 3.72$ GeV~\cite{Anashin:2015woa}. For the latter, the systematic uncertainties are taken to be fully correlated~\cite{FedorCorrelations}. The fit of the inclusive data in the range $1.937\leq\sqrt{s}\leq 3.80$ GeV is shown in Figure~\ref{fig:inc2-4}, which demonstrates the good agreement between KEDR and pQCD. In~\cite{Hagiwara:2011af}, the decision was made to use pQCD in the range $2.6\leq\sqrt{s}\leq 3.73$ GeV, where the quality of inclusive data was poor, with an error inflated according to the percentage errors of the inclusive BES data in this region~\cite{Ablikim:2009ad}. With the new KEDR data~\cite{Anashin:2016hmv,Anashin:2015woa}, the inclusive data combination is much improved, as shown in Figure~\ref{fig:inc2-4}. In this range, the data combination results in
\beq
a_{\mu}^{\rm had, \, LO \, VP}[\text{inc.}, 2.60\leq \sqrt{s}\leq3.73\text{ GeV}]  = 11.19 \pm 0.17 \ ,
\eeq
whereas from pQCD (with an inflated uncertainty~\cite{Hagiwara:2011af}), the estimate would be
\beq
a_{\mu}^{\rm had, \, LO \, VP}[\text{pQCD}, 2.60\leq \sqrt{s}\leq3.73\text{ GeV}]  = 10.82 \pm 0.35 \ .
\eeq
For the larger energy range $1.937\leq\sqrt{s}\leq 11.2$ GeV, the resulting data combination is displayed in Figure~\ref{fig:incFull}. As well as the differences observed between the data and pQCD below the charm threshold, the data above it (unchanged since\cite{Hagiwara:2011af}) also show a slight variation from the prediction of pQCD. Considering that with the new, precise KEDR data, the differences between the inclusive data and pQCD are not as large as previously and that this work is aiming at a predominantly data-driven analysis, the contributions in the entire inclusive data region are now estimated using the inclusive data alone. Hence, for this analysis, the contribution from the inclusive data is found to be
\begin{align}
a_{\mu}^{\rm had, \, LO \, VP}[\text{inc.}, 1.937\leq \sqrt{s}\leq11.2\text{ GeV}] & =43.67 \pm 0.17 \pm 0.48 \pm 0.01 \pm 0.44 \nonumber
\\
\
& = 43.67 \pm 0.67
\end{align}
and
\beq
\Delta\alpha_{\text{had}}(M_Z^2)[\text{inc.}, 1.937\leq \sqrt{s}\leq11.2\text{ GeV}] =  82.82 \pm 1.05\ .
\eeq

\subsubsection{Transition region between exclusive and inclusive data}\label{exv.vs.inc.vs.pqcd}

The transition region between the sum of exclusive states and the inclusive $R$-ratio data is of interest and deserves re-examination. For the sum of exclusive channels, whilst many measurements extend to 2 GeV or beyond, with increasing energy the inclusion of more and more multi-hadronic final states is required to achieve a reliable estimate of the total hadronic cross section. Previously, in~\cite{Hagiwara:2011af}, the sum of exclusive data was used up to 2 GeV, which defined the transition point between the exclusive sum and the inclusive data combination. In this analysis, the new KEDR data~\cite{Anashin:2016hmv} contribute two data points below 2 GeV, extending the lower boundary of the inclusive data down to $1.841$ GeV (compared to 2 GeV in~\cite{Hagiwara:2011af}) and providing an opportunity to reconsider the previous choice concerning the data input in this region. 

\begin{figure}[!t]
  \centering
    \includegraphics[width=0.7\textwidth]{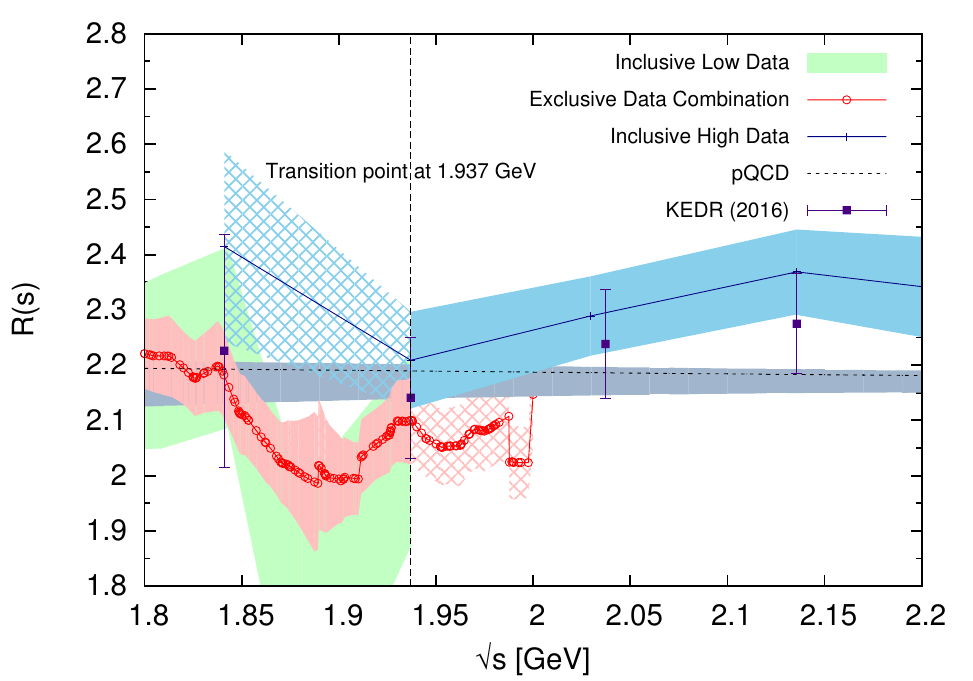}
     \caption{\small The energy region between $1.8\leq \sqrt{s} \leq 2.2$ GeV where the fit of inclusive {\em R} ratio measurements (light blue band) replaces the sum of exclusive hadronic final states (red band) from 1.937 GeV to 2 GeV. The patterned light blue band and patterned red band show the continuation of the inclusive data combination below 1.937 GeV and the continuation of the exclusive sum above 1.937 GeV, respectively. The recent KEDR data are individually marked and included in the inclusive data fit. The light green band shows the data combination of old inclusive hadronic cross section data that exist between $1.43\leq \sqrt{s} \leq 2.00$ GeV, which were previously discussed in~\cite{Hagiwara:2011af} and are not used due to their lack of precision. The estimate from pQCD is included for comparison as a dashed line with an error band which is dominated by the variation of the renormalisation scale $\mu$ in the range $\frac{1}{2}\sqrt{s} < \mu < 2\sqrt{s}$.} \label{fig:exc.vs.inc.vs.pQCD}
\end{figure}

From the lower boundary of the KEDR measurement up to 2 GeV,  the resulting contributions to $a_{\mu}^{\rm had, \, LO \, VP}$ from the sum of exclusive states, the inclusive data combination and pQCD are given in Table~\ref{Tab:excinc}. The integrated values of the inclusive data and pQCD agree within errors. However, the contribution from the sum of exclusive states disagrees with the estimates from both the inclusive data and pQCD, where the sum of exclusive provides a smaller contribution. This is particularly visible in Figure~\ref{fig:exc.vs.inc.vs.pQCD}, where although the sum of exclusive states agrees with the two inclusive data points below 2 GeV at their respective energies, the combined sum of exclusive states is lower in general. This is largely attributed to the new data for the $\pi^+\pi^-\pi^0\pi^0$ final state, where Figure~\ref{fig:pi+pi-pi0pi0full} shows that these new data result in a clear reduction of the fitted cross section below 2 GeV.\footnote{Interestingly, as can be seen from Figure~\ref{fig:exc.vs.inc.vs.pQCD}, the sum of exclusive states is in good agreement with the imprecise and, therefore, unused inclusive hadronic cross section data that exist between $1.43\leq \sqrt{s} \leq 2.00$ GeV. This is in contrast with the findings in the previous analyses~\cite{Hagiwara:2003da,Hagiwara:2006jt,Hagiwara:2011af}, which observed that the inclusive data in this range were lower than the sum of exclusive states.} Due to this effect, the previous transition point in~\cite{Hagiwara:2011af} between the sum of exclusive states and the inclusive data combination at 2 GeV is no longer the preferred choice in this work, where it is clear from Figure~\ref{fig:exc.vs.inc.vs.pQCD} that these two different choices for the data input are largely incompatible at this point. A more natural choice for this transition point is now 1.937 GeV, where it can be seen from Figure~\ref{fig:exc.vs.inc.vs.pQCD} that all available data choices at this energy are in agreement within errors. This is further substantiated by Table~\ref{Tab:excinc}, where the value for $a_{\mu}^{\rm had, \, LO \, VP}$ from the contribution from exclusive states below 1.937 GeV summed with the contribution from the inclusive data combination above 1.937 GeV is, within errors, in agreement with the integrated values of all other choices for the data input. Consequently, in this work, this is chosen to be the transition point between the sum of exclusive states and the inclusive $R$-ratio data.
\begin{table}[t]
\hspace{-0.2cm}
\scalebox{0.9}{
 {\renewcommand{\arraystretch}{0.9}
\setlength{\extrarowheight}{3pt}
  \begin{tabular}{|l|c|}
  \hline 
Input  & $a_{\mu}^{\rm had, \, LO \, VP} [1.841 \leq \sqrt{s} \leq 2.00 \ {\rm GeV}] \times 10^{10}$  \\ 
  \hline
  Exclusive sum & $6.06 \pm 0.17$  \\
  Inclusive data & $6.67 \pm 0.26$   \\
  pQCD & $6.38 \pm 0.11$   \\
\hline
  Exclusive ($<1.937$ GeV) + inclusive ($> 1.937$ GeV) & $6.23 \pm 0.13$   \\
\hline
  \end{tabular}
  } 
  }  \caption{\small Comparison of results for $a_{\mu}^{\rm had, \, LO \, VP} [1.841 \leq \sqrt{s} \leq 2.00 \ {\rm GeV}]$ from the different available inputs in this region.}\label{Tab:excinc}
    \end{table} 

\subsection{Total contribution of $a_{\mu}^{\rm had, \, LO \, VP}$ and $\Delta\alpha_{\rm had}^{(5)}(M_Z^2)$}\label{sec:amuanddelalpha}

\begin{table}[htbp]
\hspace{-0.2cm}
\scalebox{0.9}{
 {\renewcommand{\arraystretch}{0.9}
 \begin{tabular}{|l|c|c|c|c|}
\hline																							
{\bf Channel}	&	{\bf Energy range (GeV)}							&	$a_{\mu}^{\rm had, \, LO \, VP} \times 10^{10}$					&	$\Delta\alpha_{\rm had}^{(5)}(M_Z^2) \times 10^{4}$					&	{\bf New data}	\\
\hline																							
\multicolumn{5}{|c|}{Chiral perturbation theory (ChPT) threshold contributions} \\																							
\hline																							
$\pi^0\gamma$	&	$	\hphantom{00}m_{\pi}	\leq	\sqrt{s}	\leq	0.600	$	&	$	\hphantom{0}0.12	\pm	0.01	$	&	$	\hphantom{0}0.00	\pm	0.00	$	&	-	\\
$\pi^+\pi^-$	&	$	\hphantom{0}2m_{\pi}	\leq	\sqrt{s}	\leq	0.305	$	&	$	\hphantom{0}0.87	\pm	0.02	$	&	$	\hphantom{0}0.01	\pm	0.00	$	&	-	\\
$\pi^+\pi^-\pi^0$	&	$	\hphantom{0}3m_{\pi}	\leq	\sqrt{s}	\leq	0.660	$	&	$	\hphantom{0}0.01	\pm	0.00	$	&	$	\hphantom{0}0.00	\pm	0.00	$	&	-	\\
$\eta\gamma$	&	$	\hphantom{00}m_{\eta}	\leq	\sqrt{s}	\leq	0.660	$	&	$	\hphantom{0}0.00	\pm	0.00	$	&	$	\hphantom{0}0.00	\pm	0.00	$	&	-	\\
\hline																							
\multicolumn{5}{|c|}{Data based channels ($\sqrt{s} \leq 1.937$ GeV)}																							\\
\hline																							
$\pi^0\gamma$	&	$	0.600	\leq	\sqrt{s}	\leq	1.350	$	&	$	\hphantom{0}4.46	\pm	0.10	$	&	$	\hphantom{0}0.36	\pm	0.01	$	&	\cite{Achasov:2016bfr}	\\
$\pi^+\pi^-$	&	$	0.305	\leq	\sqrt{s}	\leq	1.937	$	&	$	502.97	\pm	1.97\hphantom{0}	$	&	$	34.26	\pm	0.12	$	&	\cite{Babusci:2012rp,Ablikim:2015orh}	\\
$\pi^+\pi^-\pi^0$	&	$	0.660	\leq	\sqrt{s}	\leq	1.937	$	&	$	47.79	\pm	0.89	$	&	$	\hphantom{0}4.77	\pm	0.08	$	&	\cite{Aulchenko:2015mwt}	\\
$\pi^+\pi^-\pi^+\pi^-$	&	$	0.613	\leq	\sqrt{s}	\leq	1.937	$	&	$	14.87	\pm	0.20	$	&	$	\hphantom{0}4.02	\pm	0.05	$	&	\cite{Lees:2012cr,Akhmetshin:2016dtr}	\\
$\pi^+\pi^-\pi^0\pi^0$	&	$	0.850	\leq	\sqrt{s}	\leq	1.937	$	&	$	19.39	\pm	0.78	$	&	$	\hphantom{0}5.00	\pm	0.20	$	&	\cite{TheBaBar:2017vzo}	\\
$(2\pi^+2\pi^-\pi^0)_{\text{no }\eta}$	&	$	1.013	\leq	\sqrt{s}	\leq	1.937	$	&	$	\hphantom{0}0.99	\pm	0.09	$	&	$	\hphantom{0}0.33	\pm	0.03	$	&	-	\\
$3\pi^+3\pi^-$	&	$	1.313	\leq	\sqrt{s}	\leq	1.937	$	&	$	\hphantom{0}0.23	\pm	0.01	$	&	$	\hphantom{0}0.09	\pm	0.01	$	&	\cite{Akhmetshin:2013xc}	\\
$(2\pi^+2\pi^-2\pi^0)_{\text{no }\eta\omega}$	&	$	1.322	\leq	\sqrt{s}	\leq	1.937	$	&	$	\hphantom{0}1.35	\pm	0.17	$	&	$	\hphantom{0}0.51	\pm	0.06	$	&	-	\\
$K^+K^-$	&	$	0.988	\leq	\sqrt{s}	\leq	1.937	$	&	$	23.03	\pm	0.22	$	&	$\hphantom{0}	3.37	\pm	0.03	$	&	\cite{Lees:2013gzt,Kozyrev:2017agm,Achasov:2016lbc}	\\
$K^0_S K^0_L$	&	$	1.004	\leq	\sqrt{s}	\leq	1.937	$	&	$	13.04	\pm	0.19	$	&	$	\hphantom{0}1.77	\pm	0.03	$	&	\cite{Lees:2014xsh,Kozyrev:2016raz}	\\
$KK\pi$	&	$	1.260	\leq	\sqrt{s}	\leq	1.937	$	&	$	\hphantom{0}2.71	\pm	0.12	$	&	$	\hphantom{0}0.89	\pm	0.04	$	&	\cite{TheBABAR:2017vgl,Achasov:2017vaq}	\\
$KK2\pi$	&	$	1.350	\leq	\sqrt{s}	\leq	1.937	$	&	$	\hphantom{0}1.93	\pm	0.08	$	&	$	\hphantom{0}0.75	\pm	0.03	$	&	\cite{Lees:2014xsh,TheBABAR:2017vgl,TheBABAR:2017aph}	\\
$\eta\gamma$	&	$	0.660	\leq	\sqrt{s}	\leq	1.760	$	&	$	\hphantom{0}0.70	\pm	0.02	$	&	$	\hphantom{0}0.09	\pm	0.00	$	&	\cite{Achasov:2013eli}	\\
$\eta\pi^+\pi^-$	&	$	1.091	\leq	\sqrt{s}	\leq	1.937	$	&	$	\hphantom{0}1.29	\pm	0.06	$	&	$	\hphantom{0}0.39	\pm	0.02	$	&	\cite{Achasov:2017kqm,TheBABAR:2018vvb}	\\
$(\eta\pi^+\pi^-\pi^0)_{\text{no }\omega}$	&	$	1.333	\leq	\sqrt{s}	\leq	1.937	$	&	$	\hphantom{0}0.60	\pm	0.15	$	&	$	\hphantom{0}0.21	\pm	0.05	$	&	\cite{CMD-3:2017tgb}	\\
$\eta2\pi^+2\pi^-$	&	$	1.338	\leq	\sqrt{s}	\leq	1.937	$	&	$	\hphantom{0}0.08	\pm	0.01	$	&	$	\hphantom{0}0.03	\pm	0.00	$	&	-	\\
$\eta\omega$	&	$	1.333	\leq	\sqrt{s}	\leq	1.937	$	&	$	\hphantom{0}0.31	\pm	0.03	$	&	$	\hphantom{0}0.10	\pm	0.01	$	&	\cite{CMD-3:2017tgb,Achasov:2016qvd}	\\
$\om(\to\pi^0\gamma)\pi^0$	&	$	0.920	\leq	\sqrt{s}	\leq	1.937	$	&	$	\hphantom{0}0.88	\pm	0.02	$	&	$	\hphantom{0}0.19	\pm	0.00	$	&	\cite{Achasov:2012zza,Achasov:2016zvn}	\\
$\eta\phi$	&	$	1.569	\leq	\sqrt{s}	\leq	1.937	$	&	$	\hphantom{0}0.42	\pm	0.03	$	&	$	\hphantom{0}0.15	\pm	0.01	$	&	-	\\
$\phi\to$ unaccounted	&	$	0.988	\leq	\sqrt{s}	\leq	1.029	$	&	$	\hphantom{0}0.04	\pm	0.04	$	&	$	\hphantom{0}0.01	\pm	0.01	$	&	-	\\
$\eta\omega\pi^0$	&	$	1.550	\leq	\sqrt{s}	\leq	1.937	$	&	$	\hphantom{0}0.35	\pm	0.09	$	&	$	\hphantom{0}0.14	\pm	0.04	$	&	\cite{Achasov:2016eyg}	\\
$\eta(\rightarrow{\rm npp})K\bar{K}_{\text{no }\phi \rightarrow K\bar{K}}$	&	$	1.569	\leq	\sqrt{s}	\leq	1.937	$	&	$	\hphantom{0}0.01	\pm	0.02	$	&	$	\hphantom{0}0.00	\pm	0.01	$	&	\cite{Aubert:2007ym,TheBABAR:2017vgl}	\\
$p\bar{p}$	&	$	1.890	\leq	\sqrt{s}	\leq	1.937	$	&	$	\hphantom{0}0.03	\pm	0.00	$	&	$	\hphantom{0}0.01	\pm	0.00	$	&	\cite{Akhmetshin:2015ifg}	\\
$n\bar{n}$	&	$	1.912	\leq	\sqrt{s}	\leq	1.937	$	&	$	\hphantom{0}0.03	\pm	0.01	$	&	$	\hphantom{0}0.01	\pm	0.00	$	&	\cite{Achasov:2014ncd}	\\
\hline																							
\multicolumn{5}{|c|}{Estimated contributions ($\sqrt{s} \leq 1.937$ GeV) }																							\\
\hline																							
$(\pi^+\pi^-3\pi^0)_{\text{no }\eta}$	&	$	1.013	\leq	\sqrt{s}	\leq	1.937	$	&	$	\hphantom{0}0.50	\pm	0.04	$	&	$	\hphantom{0}0.16	\pm	0.01	$	&	-	\\
$(\pi^+\pi^-4\pi^0)_{\text{no }\eta}$	&	$	1.313	\leq	\sqrt{s}	\leq	1.937	$	&	$	\hphantom{0}0.21	\pm	0.21	$	&	$	\hphantom{0}0.08	\pm	0.08	$	&	-	\\
$KK3\pi$	&	$	1.569	\leq	\sqrt{s}	\leq	1.937	$	&	$	\hphantom{0}0.03	\pm	0.02	$	&	$	\hphantom{0}0.02	\pm	0.01	$	&	-	\\
$\omega(\rightarrow{\rm npp})2\pi$	&	$	1.285	\leq	\sqrt{s}	\leq	1.937	$	&	$	\hphantom{0}0.10	\pm	0.02	$	&	$	\hphantom{0}0.03	\pm	0.01	$	&	-	\\
$\omega(\rightarrow{\rm npp})3\pi$	&	$	1.322	\leq	\sqrt{s}	\leq	1.937	$	&	$	\hphantom{0}0.17	\pm	0.03	$	&	$	\hphantom{0}0.06	\pm	0.01	$	&	-	\\
$\omega(\rightarrow{\rm npp})KK$	&	$	1.569	\leq	\sqrt{s}	\leq	1.937	$	&	$	\hphantom{0}0.00	\pm	0.00	$	&	$	\hphantom{0}0.00	\pm	0.00	$	&	-	\\
$\eta\pi^+\pi^-2\pi^0$	&	$	1.338	\leq	\sqrt{s}	\leq	1.937	$	&	$	\hphantom{0}0.08	\pm	0.04	$	&	$	\hphantom{0}0.03	\pm	0.02	$	&	-	\\
\hline																							
\multicolumn{5}{|c|}{Other contributions ($\sqrt{s} >1.937$ GeV)}																							\\
\hline																							
Inclusive channel	&	$	1.937	\leq	\sqrt{s}	\leq	11.199	$	&	$	43.67	\pm	0.67	$	&	$	82.82	\pm	1.05	$	&	\cite{Aubert:2008ab,Anashin:2016hmv,Anashin:2015woa}	\\
$J/\psi$	&				-				&	$	\hphantom{0}6.26	\pm	0.19	$	&	$	\hphantom{0}7.07	\pm	0.22	$	&	-	\\
$\psi'$	&				-				&	$	\hphantom{0}1.58	\pm	0.04	$	&	$	\hphantom{0}2.51	\pm	0.06	$	&	-	\\
$\Upsilon(1S-4S)$	&				-				&	$	\hphantom{0}0.09	\pm	0.00	$	&	$	\hphantom{0}1.06	\pm	0.02	$	&	-	\\
pQCD	&	$	11.199	\leq	\sqrt{s}	\leq	\infty	$\hphantom{000}	&	$	\hphantom{0}2.07	\pm	0.00	$	&	$	124.79	\pm	0.10	\hphantom{0}$	&	-	\\
\hline																							
\hline																							
Total	&	$	m_{\pi}	\leq	\sqrt{s}	\leq	\infty	$\hphantom{0}	&	$	693.26	\pm	2.46\hphantom{0}	$	&	$	276.11	\pm	1.11\hphantom{0}	$	&	-	\\
\hline																																													

\end{tabular} 
} 
}\caption{Summary of the contributions to $a_{\mu}^{\rm had, \, LO \, VP}$ and $\Delta\alpha_{\rm had}^{(5)}(M_Z^2)$ calculated in this analysis. The first column indicates the hadronic final state or individual contribution, the second column gives the respective energy range of the contribution, the third column states the determined value of $a_{\mu}^{\rm had, \, LO \, VP}$, the fourth column states the determined value of $\Delta\alpha_{\rm had}^{(5)}(M_Z^2)$ and the last column indicates any new data that have been included since~\cite{Hagiwara:2011af}. The last row describes the total contribution obtained from the sum of the individual final states, with the uncertainties added in quadrature.} \label{tab:amuhad}
\end{table}
 \begin{figure}[!t]
\centering
  \subfloat[Fractional contributions to $a_{\mu}^{\rm had, \, LO \, VP}$.]{%
    \includegraphics[width= 0.8\textwidth]{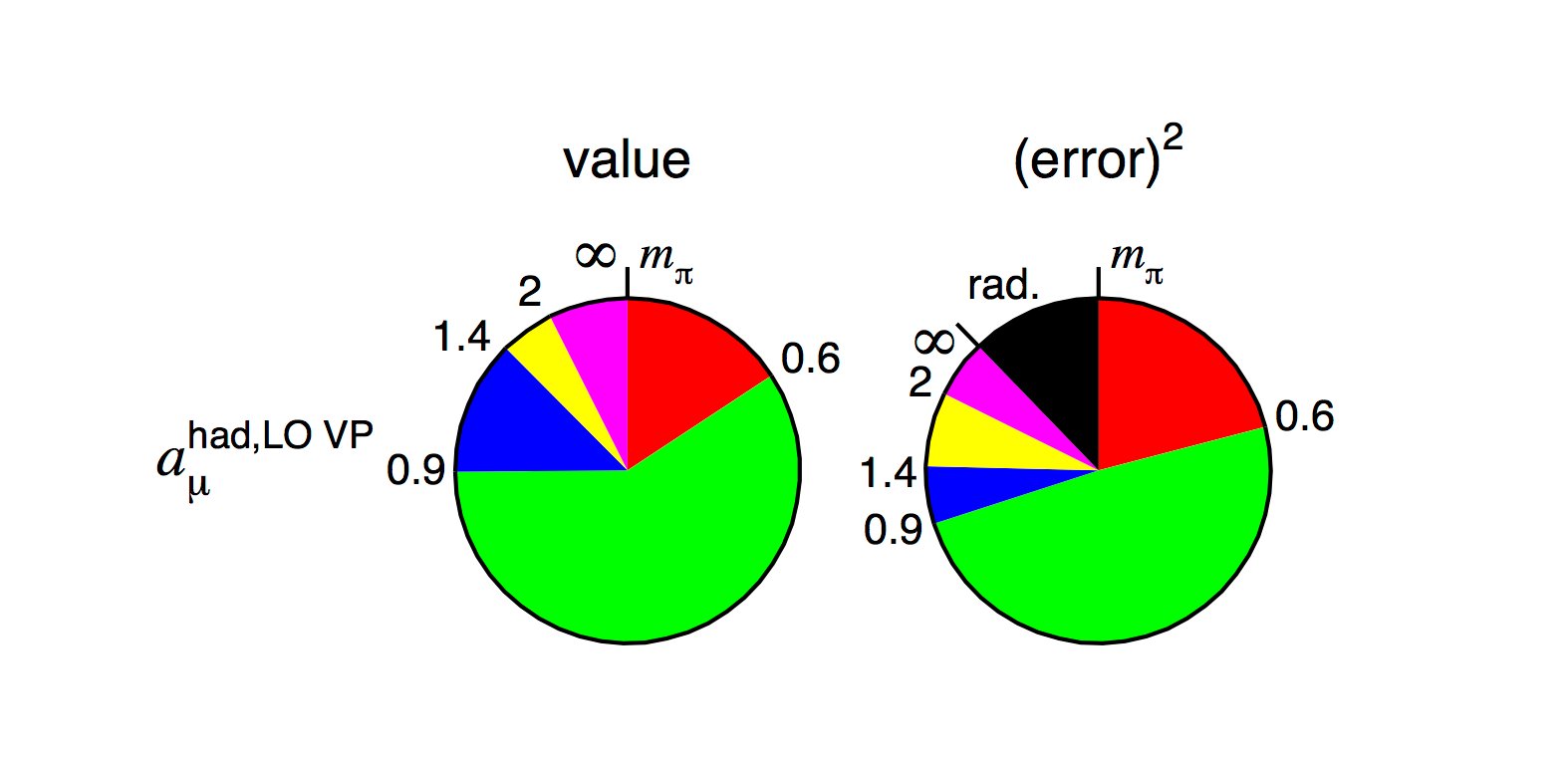}}\hfill
  \subfloat[Fractional contributions to $\Delta\alpha_{\rm had}^{(5)}(M_Z^2)$]{%
    \includegraphics[width= 0.8\textwidth]{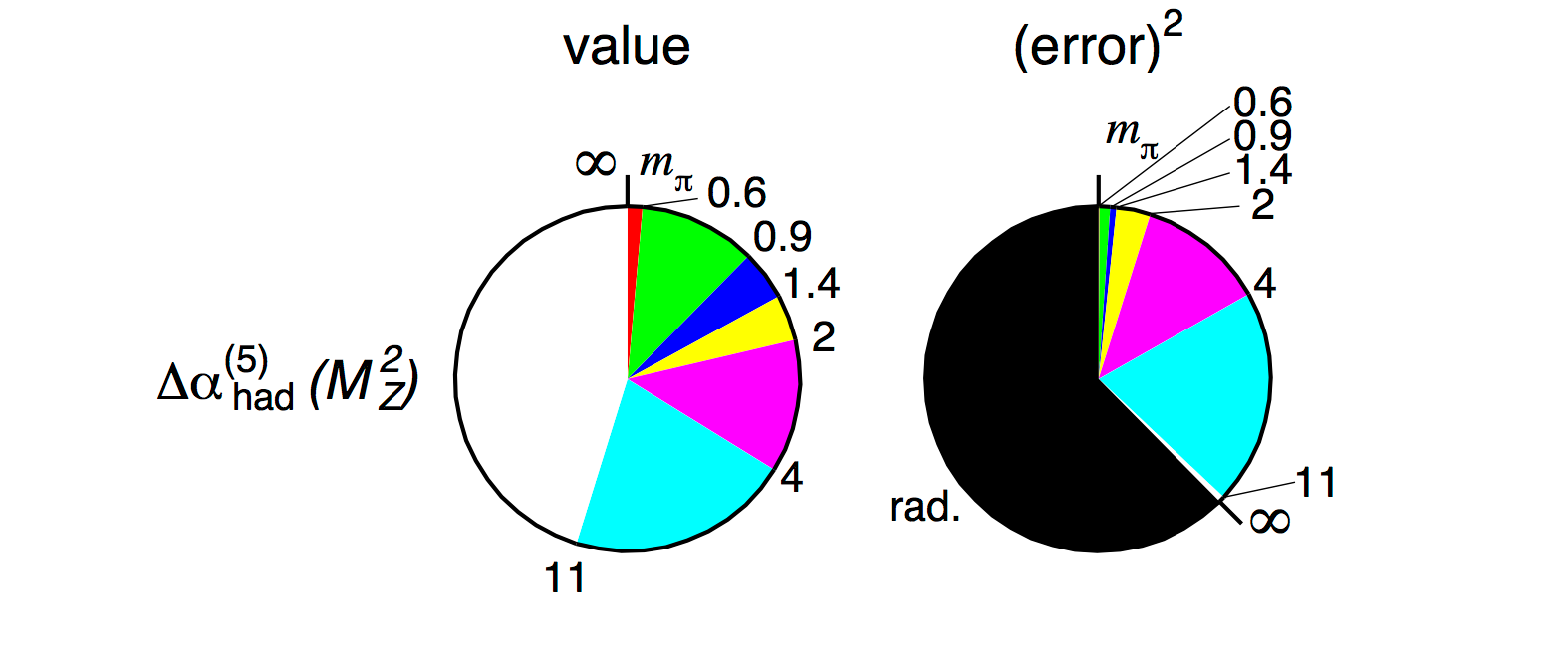}}\hfill
  \caption{\small Pie charts showing the fractional contributions to the total mean value (left pie chart) and (error)$^2$ (right pie chart) of both $a_{\mu}^{\rm had, \, LO \, VP}$ (upper panel) and $\Delta\alpha_{\rm had}^{(5)}(M_Z^2)$ (lower panel) from various energy intervals. The energy intervals for $a_{\mu}^{\rm had, \, LO \, VP}$ are defined by the boundaries $m_{\pi}$, 0.6, 0.9, 1.43, 2.0 and $\infty$ GeV. For $\Delta\alpha_{\rm had}^{(5)}(M_Z^2)$, the intervals are defined by the energy boundaries $m_{\pi}$, 0.6, 0.9, 1.43, 2.0, 4.0, 11.2 and $\infty$ GeV. In both cases, the (error)$^2$ includes all experimental uncertainties (including all available correlations) and local $\chi^2_{\rm min}/{\rm d.o.f.}$ inflation. The fractional contribution to the (error)$^2$ from the radiative correction uncertainties are shown in black and indicated by `rad.'.}  \label{piechart}
\end{figure} 
 \begin{figure}[htbp]
\centering
    \vspace{2cm}
  \subfloat[The hadronic $R$-ratio.]{%
    \includegraphics[width= 1.0\textwidth]{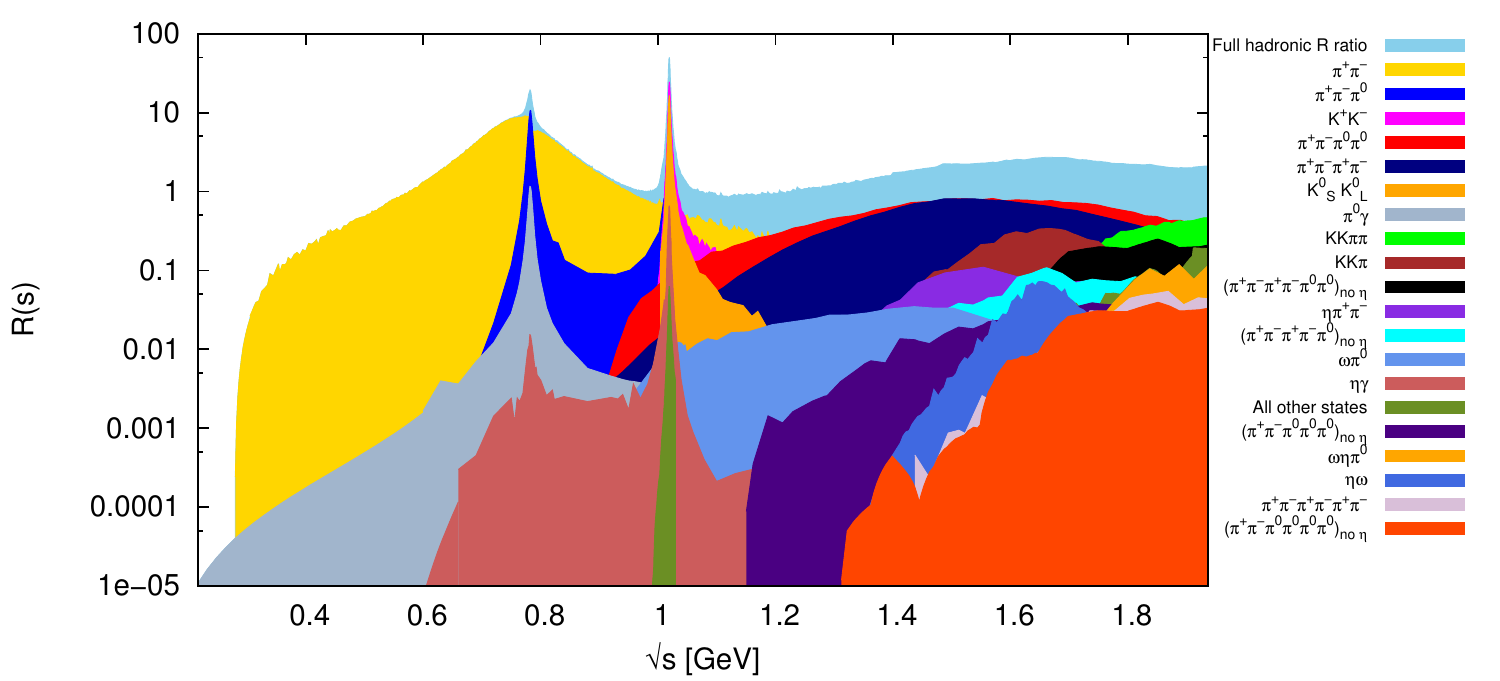}}\hfill
    \vspace{0.5cm}
  \subfloat[The uncertainty of the hadronic $R$-ratio.]{%
    \includegraphics[width= 0.97\textwidth]{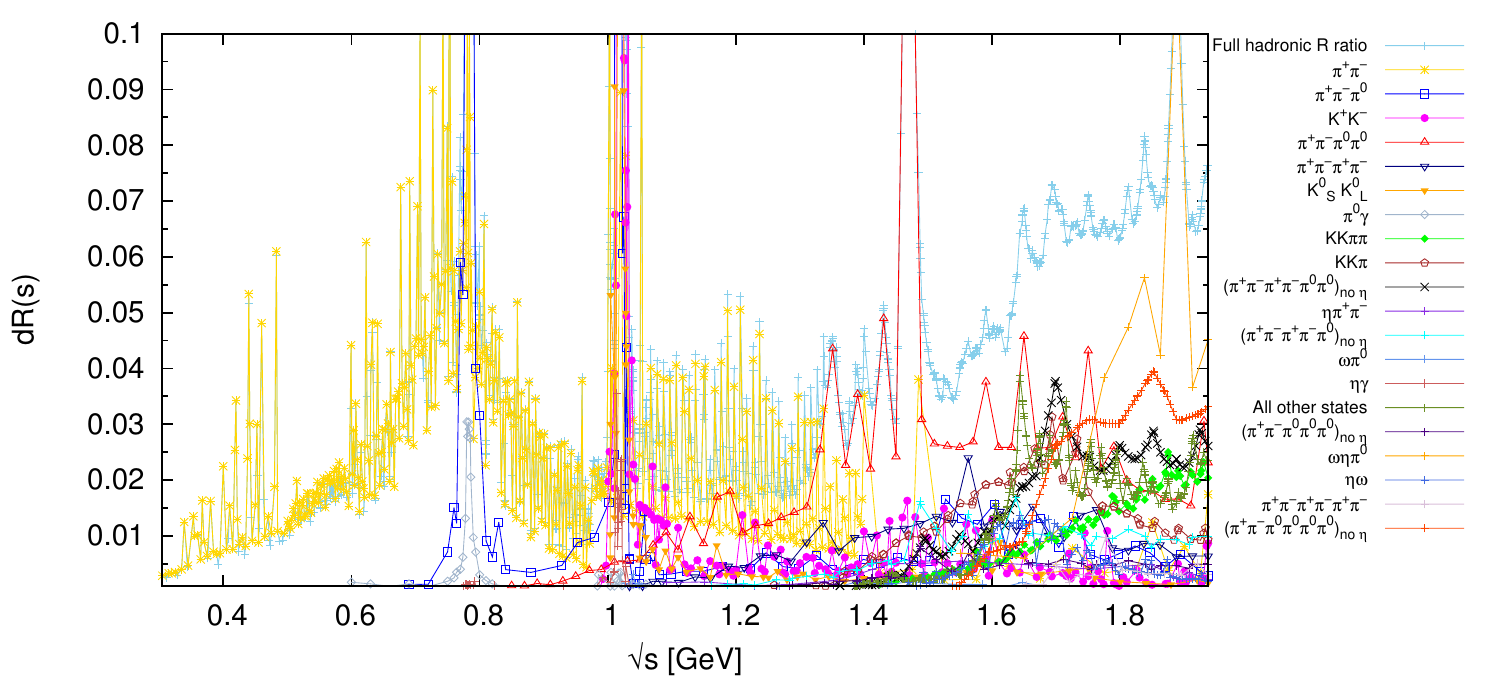}}\hfill
  \caption{\small Contributions to the total hadronic $R$-ratio from the different final states (upper panel) and their uncertainties (lower panel) below 1.937 GeV. The full $R$-ratio and its uncertainty is shown in light blue in each plot, respectively. Each final state is included as a new layer on top in decreasing order of the size of its contribution to $a_{\mu}^{\rm had, \, LO \, VP}$.}  \label{hadxSec}
\end{figure} 
\begin{figure}[!htbp] 
  \centering
    \includegraphics[width=1.0\textwidth]{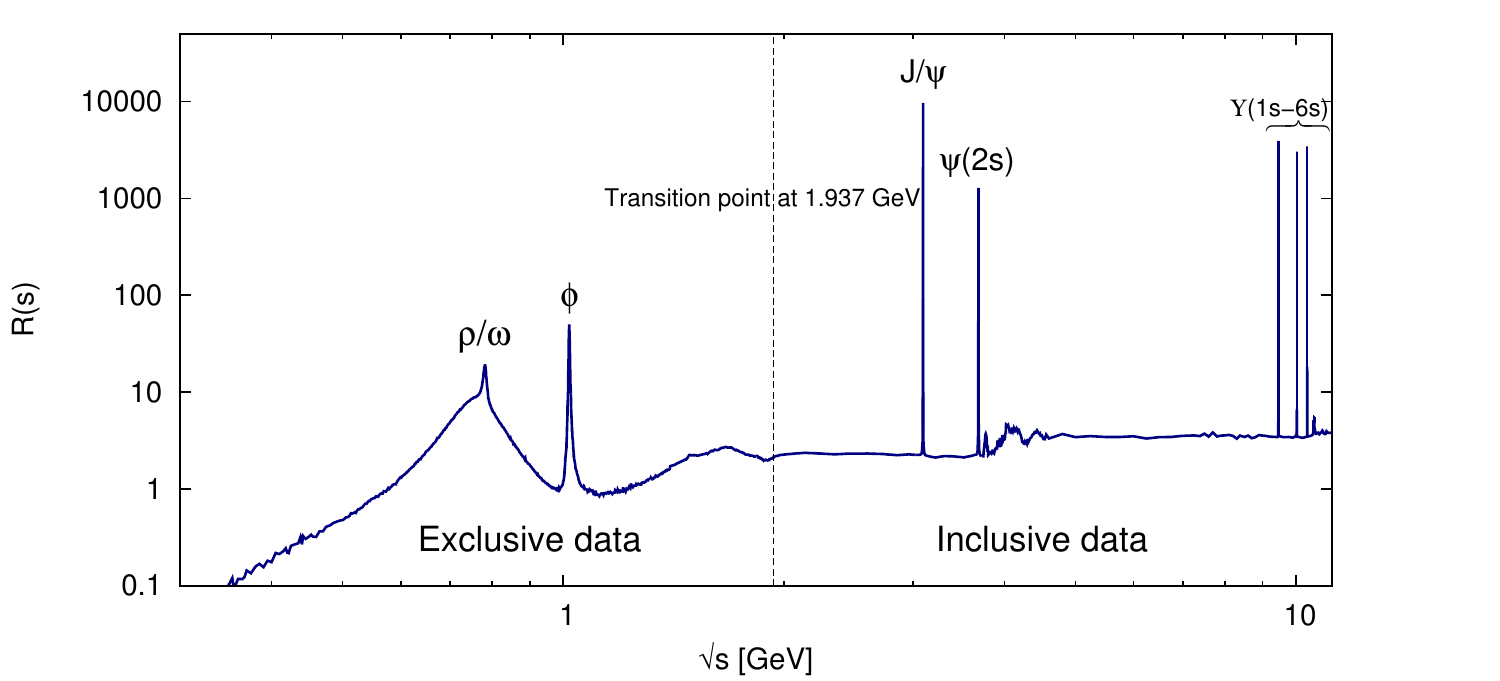}
     \caption{\small The resulting hadronic $R$-ratio shown in the range $m_{\pi}\leq\sqrt{s}\leq 11.1985$ GeV, where the prominent resonances are labelled.}     \label{rHad}
\end{figure} 

Table~\ref{tab:amuhad} lists all contributions from individual channels contributing to $a_{\mu}^{\rm had, \, LO \, VP}$, with the corresponding total. From the sum of these contributions, the estimate for $a_{\mu}^{\rm had, \, LO \, VP}$ from this analysis is
\begin{align} \label{LOHVP_KNT18}
a_{\mu}^{\rm had, \, LO \, VP} & = (693.26  \pm 1.19_{\rm stat} \pm 2.01_{\rm sys} \pm 0.22_{\rm vp} \pm 0.71_{\rm fsr}) \times 10^{-10}  \nonumber
\\
\
& =  (693.26 \pm 2.46_{\rm tot}) \times 10^{-10} \ , 
\end{align}
where the uncertainties include all available correlations and local $\chi^2$ inflation as discussed in Section~\ref{ssSec: Fitting data}.
Using the same data compilation as described for the calculation of $a_{\mu}^{\rm had, \, LO \, VP}$, the next-to-leading order (NLO) contribution to $a_{\mu}^{\rm had, VP}$ is determined here to be
\begin{align}\label{NLOHVP_KNT18}
a_{\mu}^{\rm had,\, NLO\, VP} & = (-9.82  \pm 0.02_{\rm stat} \pm 0.03_{\rm sys} \pm 0.01_{\rm vp} \pm 0.02_{\rm fsr}) \times 10^{-10}  \nonumber
\\
\
& =  (-9.82 \pm 0.04_{\rm tot}) \times 10^{-10} \ .
\end{align}
The corresponding result for $\Delta\alpha_{\rm had}^{(5)}(M_Z^2)$ is
\begin{align} \label{delalphahad_KNT18}
\Delta\alpha_{\rm had}^{(5)}(M_Z^2) & = (276.11  \pm  0.26_{\rm stat} \pm 0.68_{\rm sys} \pm 0.14_{\rm vp} \pm 0.83_{\rm fsr}) \times 10^{-4}  \nonumber
\\
\
& = (276.11 \pm 1.11_{\rm tot}) \times 10^{-4} \ ,
\end{align}
where the superscript (5) indicates the contributions from all quark flavours except the top quark, which is added separately. In each case, the errors from the individual channels and sources of uncertainty are added in quadrature to determine the total error. The fractional contributions to the total mean value and uncertainty of both $a_{\mu}^{\rm had, \, LO \, VP}$ and $\Delta\alpha_{\rm had}^{(5)}(M_Z^2)$ from various energy intervals is shown in Figure~\ref{piechart}. Here, the dominance of the energy region below $0.9$ GeV for $a_{\mu}^{\rm had, \, LO \, VP}$ and its uncertainty is clearly evident, with this being predominantly due to the contributions from the $\pi^+\pi^-$ channel. Notably, the pie chart depicting the fractional contributions to the (error)$^2$ of $\Delta\alpha_{\rm had}^{(5)}(M_Z^2)$ reveals how the uncertainty on this quantity is dominated by the contributions from the radiative correction uncertainties. Mostly, this large error contribution comes from the uncertainty due to possible FSR applied to the combination of inclusive data above 1.937 GeV discussed in Section~\ref{radcorUncertainty}. This, in particular, highlights the differences in the kernel functions of the respective dispersion integrals for $a_{\mu}^{\rm had, \, LO \, VP}$ and $\Delta\alpha_{\rm had}^{(5)}(M_Z^2)$, where contributions from higher energies have a larger influence on $\Delta\alpha_{\rm had}^{(5)}(M_Z^2)$ than on $a_{\mu}^{\rm had, \, LO \, VP}$. If, instead of a data driven analysis, the region above 1.937 GeV was estimated using pQCD, it would effectively eliminate the impacting radiative correction uncertainties in this region. Figure~\ref{hadxSec} shows the contributions from all hadronic final states to the hadronic $R$-ratio and its uncertainty below 1.937 GeV. Here, the individual final states are displayed separately as well as with the resulting total hadronic $R$-ratio. The full compilation for the hadronic $R$-ratio is shown in Figure~\ref{rHad}. The data vector and corresponding covariance matrix of the hadronic $R$-ratio in the range $m_{\pi}\leq\sqrt{s}\leq 11.1985$ GeV determined in this work are available upon request from the authors.

\subsection{Comparison with the HLMNT11 evaluation}

\begin{table}[htbp]
\centering
\scalebox{0.975}{
  {\renewcommand{\arraystretch}{0.9}
  \begin{tabular}{|l|c|c|c|}
\hline																						
{\bf Channel}	&	{\bf This work (KNT18)}							&	{\bf HLMNT11}~\cite{Hagiwara:2011af}							&	{\bf Difference}				\\
\hline																						
\multicolumn{4}{|c|}{Chiral perturbation theory (ChPT) threshold contributions} \\																						
\hline																						
$\pi^0\gamma$	&	$	\hphantom{0}	0.12	\pm	0.01		$	&	$	\hphantom{0}	0.12	\pm	0.01		$	&	$	0.00		$	\\
$\pi^+\pi^-$	&	$	\hphantom{0}	0.87	\pm	0.02		$	&	$	\hphantom{0}	0.87	\pm	0.02		$	&	$	0.00		$	\\
$\pi^+\pi^-\pi^0$	&	$	\hphantom{0}	0.01	\pm	0.00		$	&	$	\hphantom{0}	0.01	\pm	0.00		$	&	$	0.00		$	\\
$\eta\gamma$	&	$	\hphantom{0}	0.00	\pm	0.00		$	&	$	\hphantom{0}	0.00	\pm	0.00		$	&	$	0.00		$	\\
\hline																						
\multicolumn{4}{|c|}{Data based channels ($\sqrt{s} \leq 2$ GeV)}																						\\
\hline																						
$\pi^0\gamma$	&	$	\hphantom{0}	4.46	\pm	0.10		$	&	$	\hphantom{0}	4.54	\pm	0.14		$	&	$	-0.08	\hphantom{-}	$	\\
$\pi^+\pi^-$	&	$		502.99	\pm	1.97	\hphantom{0}	$	&	$		505.77	\pm	3.09	\hphantom{0}	$	&	$	-2.78	\hphantom{-}	$	\\
$\pi^+\pi^-\pi^0$	&	$		47.82	\pm	0.89		$	&	$		47.51	\pm	0.99		$	&	$	0.31		$	\\
$\pi^+\pi^-\pi^+\pi^-$	&	$		15.17	\pm	0.21		$	&	$		14.65	\pm	0.47		$	&	$	0.52		$	\\
$\pi^+\pi^-\pi^0\pi^0$	&	$		19.80	\pm	0.79		$	&	$		20.37	\pm	1.26		$	&	$	-0.57	\hphantom{-}	$	\\
$(2\pi^+2\pi^-\pi^0)_{\text{no }\eta}$	&	$	\hphantom{0}	1.08	\pm	0.10		$	&	$	\hphantom{0}	1.20	\pm	0.10		$	&	$	-0.12	\hphantom{-}	$	\\
$3\pi^+3\pi^-$	&	$	\hphantom{0}	0.28	\pm	0.02		$	&	$	\hphantom{0}	0.28	\pm	0.02		$	&	$	0.00		$	\\
$(2\pi^+2\pi^-2\pi^0)_{\text{no }\eta\omega}$	&	$	\hphantom{0}	1.60	\pm	0.20		$	&	$	\hphantom{0}	1.80	\pm	0.24		$	&	$	-0.20	\hphantom{-}	$	\\
$K^+K^-$	&	$		23.05	\pm	0.22		$	&	$		22.15	\pm	0.46		$	&	$	0.90		$	\\
$K^0_S K^0_L$	&	$		13.05	\pm	0.19		$	&	$		13.33	\pm	0.16		$	&	$	-0.28	\hphantom{-}	$	\\
$KK\pi$	&	$	\hphantom{0}	2.80	\pm	0.12		$	&	$	\hphantom{0}	2.77	\pm	0.15		$	&	$	0.03		$	\\
$KK2\pi$	&	$	\hphantom{0}	2.42	\pm	0.09		$	&	$	\hphantom{0}	3.31	\pm	0.58		$	&	$	-0.89	\hphantom{-}	$	\\
$\eta\gamma$	&	$	\hphantom{0}	0.70	\pm	0.02		$	&	$	\hphantom{0}	0.69	\pm	0.02		$	&	$	0.01		$	\\
$\eta\pi^+\pi^-$	&	$	\hphantom{0}	1.32	\pm	0.06		$	&	$	\hphantom{0}	0.98	\pm	0.24		$	&	$	0.34		$	\\
$(\eta\pi^+\pi^-\pi^0)_{\text{no }\omega}$	&	$	\hphantom{0}	0.63	\pm	0.15		$	&				-				&	$	0.63		$	\\
$\eta2\pi^+2\pi^-$	&	$	\hphantom{0}	0.11	\pm	0.02		$	&	$	\hphantom{0}	0.11	\pm	0.02		$	&	$	0.00		$	\\
$\eta\omega$	&	$	\hphantom{0}	0.31	\pm	0.03		$	&	$	\hphantom{0}	0.42	\pm	0.07		$	&	$	-0.11	\hphantom{-}	$	\\
$\om(\to\pi^0\gamma)\pi^0$	&	$	\hphantom{0}	0.88	\pm	0.02		$	&	$	\hphantom{0}	0.77	\pm	0.03		$	&	$	0.11		$	\\
$\eta\phi$	&	$	\hphantom{0}	0.45	\pm	0.04		$	&	$	\hphantom{0}	0.46	\pm	0.03		$	&	$	-0.01	\hphantom{-}	$	\\
$\phi\to$ unaccounted	&	$	\hphantom{0}	0.04	\pm	0.04		$	&	$	\hphantom{0}	0.04	\pm	0.04		$	&	$	0.00		$	\\
$\eta\om\pi^0$	&	$	\hphantom{0}	0.42	\pm	0.10		$	&				-				&	$	0.42		$	\\
﻿$\eta(\rightarrow{\rm npp})K\bar{K}_{\text{no }\phi \rightarrow K\bar{K}}$	&	$	\hphantom{0}	0.01	\pm	0.01		$	&				-				&	$	0.01		$	\\
$p\bar{p}$	&	$	\hphantom{0}	0.07	\pm	0.00		$	&	$	\hphantom{0}	0.06	\pm	0.00		$	&	$	0.01		$	\\
$n\bar{n}$	&	$	\hphantom{0}	0.06	\pm	0.01		$	&	$	\hphantom{0}	0.07	\pm	0.02		$	&	$	-0.01	\hphantom{-}	$	\\
\hline																						
\multicolumn{4}{|c|}{Estimated contributions ($\sqrt{s} \leq 2$ GeV) }																						\\
\hline																						
$(\pi^+\pi^-3\pi^0)_{\text{no }\eta}$	&	$	\hphantom{0}	0.53	\pm	0.05		$	&	$	\hphantom{0}	0.60	\pm	0.05		$	&	$	-0.07	\hphantom{-}	$	\\
$(\pi^+\pi^-4\pi^0)_{\text{no }\eta}$	&	$	\hphantom{0}	0.25	\pm	0.25		$	&	$	\hphantom{0}	0.28	\pm	0.28		$	&	$	-0.03	\hphantom{-}	$	\\
$KK3\pi$	&	$	\hphantom{0}	0.08	\pm	0.03		$	&	$	\hphantom{0}	0.08	\pm	0.04		$	&	$	0.00		$	\\
$\omega(\rightarrow{\rm npp})2\pi$	&	$	\hphantom{0}	0.10	\pm	0.02		$	&	$	\hphantom{0}	0.11	\pm	0.02		$	&	$	-0.01	\hphantom{-}	$	\\
$\omega(\rightarrow{\rm npp})3\pi$	&	$	\hphantom{0}	0.20	\pm	0.04		$	&	$	\hphantom{0}	0.22	\pm	0.04		$	&	$	-0.02	\hphantom{-}	$	\\
$\omega(\rightarrow{\rm npp})KK$	&	$	\hphantom{0}	0.01	\pm	0.00		$	&	$	\hphantom{0}	0.01	\pm	0.00		$	&	$	0.00		$	\\
$\eta\pi^+\pi^-2\pi^0$	&	$	\hphantom{0}	0.11	\pm	0.05		$	&	$	\hphantom{0}	0.11	\pm	0.06		$	&	$	0.00		$	\\
\hline																						
\multicolumn{4}{|c|}{Other contributions ($\sqrt{s} >2$ GeV)}																						\\
\hline																						
Inclusive channel	&	$		41.27	\pm	0.62		$	&	$		41.40	\pm	0.87		$	&	$	-0.13	\hphantom{-}	$	\\
$J/\psi$	&	$	\hphantom{0}	6.26	\pm	0.19		$	&	$	\hphantom{0}	6.24	\pm	0.16		$	&	$	0.02		$	\\
$\psi'$	&	$	\hphantom{0}	1.58	\pm	0.04		$	&	$	\hphantom{0}	1.56	\pm	0.05		$	&	$	0.02		$	\\
$\Upsilon(1S-4S)$	&	$	\hphantom{0}	0.09	\pm	0.00		$	&	$	\hphantom{0}	0.10	\pm	0.00		$	&	$	-0.01	\hphantom{-}	$	\\
pQCD	&	$	\hphantom{0}	2.07	\pm	0.00		$	&	$	\hphantom{0}	2.06	\pm	0.00		$	&	$	0.01		$	\\
\hline																						
\hline																						
Total	&	$		693.26	\pm	2.46	\hphantom{0}	$	&	$		694.91	\pm	4.27	\hphantom{0}	$	&	$	-1.64	\hphantom{-}	$	\\
\hline																																											
  \end{tabular} 
  }
}\caption{Comparison of the contributions to $a_{\mu}^{\rm had, \, LO \, VP}$ calculated in the HLMNT11 analysis~\cite{Hagiwara:2011af} and in this work (KNT18), where all results are given in units of $a_{\mu}^{\rm had, \, LO \, VP} \times 10^{10}$. The first column indicates the final state or individual contribution, the second column gives the KNT18 estimate, the third column states the HLMNT11 estimate and the last column gives the difference between the two evaluations.}\label{tab:HLMNTvsKNT} 
\end{table}
\begin{table}[htbp]
\vspace{-0.5cm}
\centering
\scalebox{0.975}{
  {\renewcommand{\arraystretch}{0.9}
  \begin{tabular}{|l|c|c|}
\hline													
{\bf Channel}	&	{\bf This work (KNT18)}			&	{\bf HLMNT11}~\cite{Hagiwara:2011af}			\\
\hline													
$\pi^+\pi^-$	&	$	1.3	$	&	$	1.4	$	\\
$\pi^+\pi^-\pi^0$	&	$	2.1	$	&	$	3.0	$	\\
$\pi^+\pi^-\pi^+\pi^-$	&	$	1.8	$	&	$	1.7	$	\\
$\pi^+\pi^-\pi^0\pi^0$	&	$	2.0	$	&	$	1.3	$	\\
$(2\pi^+2\pi^-\pi^0)_{\text{no }\eta}$	&	$	1.0	$	&	$	1.2	$	\\
$(2\pi^+2\pi^-2\pi^0)_{\text{no }\eta\omega}$	&	$	3.5	$	&	$	4.0	$	\\
$K^+K^-$	&	$	2.1	$	&	$	1.9	$	\\
$K^0_S K^0_L$	&	$	0.8	$	&	$	0.8	$	\\
\hline													
  \end{tabular} 
  }
}\caption{Comparison of the global $\sqrt{\chi^2_{\rm min}/{\rm d.o.f.}}$ for the leading and major sub-leading channels determined in the HLMNT11 analysis~\cite{Hagiwara:2011af} and in this work (KNT18). The first column indicates the final state or individual contribution, the second column gives the KNT18 value, the third column states the HLMNT11 value.}\label{tab:HLMNTvsKNTchi2} 
\end{table}
\begin{figure}[!t] 
  \centering
    \includegraphics[width=0.8\textwidth]{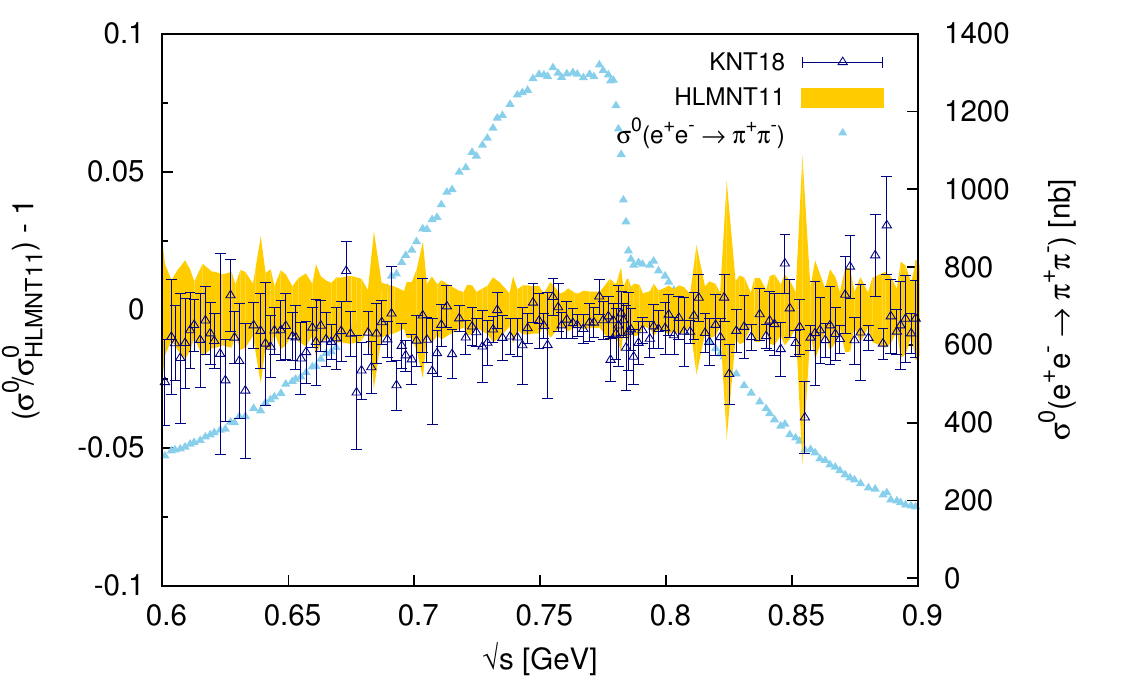}
     \caption{\small The normalised difference of the clusters of the $\pi^+\pi^-$ data fit from the KNT18 analysis with respect to those from the HLMNT11 analysis in the range $0.6\leq\sqrt{s}\leq0.9$ GeV. The width of the yellow band represents the total uncertainties of the clusters of the HLMNT11 analysis. The $\pi^+\pi^-$ cross section is displayed for reference.}     \label{pipiHLMNTcompare}
\end{figure} 

To understand further how the changes in the data combination/input have altered the estimate of $a_{\mu}^{\rm had, \, LO \, VP}$ and its uncertainty, a comparison of the results from this analysis and the previous HLMNT11 evaluation~\cite{Hagiwara:2011af} is particularly interesting. Table~\ref{tab:HLMNTvsKNT} gives a channel-by-channel comparison of the two works, highlighting the differences in the individual contributions for each channel and the total sum over their respective energy ranges.\footnote{Note that the results for individual contributions to $a_{\mu}^{\rm had, \, LO \, VP}$ from this work that are listed in Table~\ref{tab:HLMNTvsKNT} differ from those given earlier in Section~\ref{amuanddeltaalpha} and in Table~\ref{tab:amuhad}, as for a comparison with HLMNT11~\cite{Hagiwara:2011af}, contributions to $a_{\mu}^{\rm had, \, LO \, VP}$ from exclusive channels are evaluated up to 2 GeV. However, to consistently compare the final results for $a_{\mu}^{\rm had, \, LO \, VP}$ between the two works, the total KNT18 result given in Table~\ref{tab:HLMNTvsKNT} is not determined as the sum of the individual contributions listed above it, but is the final result for $a_{\mu}^{\rm had, \, LO \, VP}$ calculated in this work using the exclusive channels evaluated up to 1.937 GeV. Summing the KNT18 values listed in Table~\ref{tab:HLMNTvsKNT} (i.e.~choosing to evaluate the sum of exclusive states from this work up to 2 GeV), results in $a_{\mu}^{\rm had, \, LO \, VP} = (693.06 \pm 2.45)\times 10^{-10}$.} The largest difference occurs in the $\pi^+\pi^-$ channel, where the mean value in this work is lower by almost $1\sigma$ of the HLMNT11 analysis and the uncertainty has reduced by approximately one third. As described in the in-depth discussion of the $2\pi$ contribution in Section~\ref{chap:pipi}, this is largely due to the new, precise and highly correlated radiative return data from KLOE and BESIII and the capability of the new data combination method to utilise the correlations to their full capacity. The global $\sqrt{\chi^2_{\rm min}/{\rm d.o.f.}}$ of the leading and major sub-leading channels in this work are compared to those from the HLMNT11 analysis~\cite{Hagiwara:2011af} in Table~\ref{tab:HLMNTvsKNTchi2}. The reduction of the global $\sqrt{\chi^2_{\rm min}/{\rm d.o.f.}}$ for the $\pi^+\pi^-$ channel further highlights that the data combination for this channel has improved. The energy dependent changes in the resonance region are shown in Figure~\ref{pipiHLMNTcompare}, where it can be seen that, as expected from the comparison of the  $\pi^+\pi^-$ results in Table~\ref{tab:HLMNTvsKNT}, the KNT18 data combination is in good agreement with the HLMNT11 analysis, but sits lower overall.

The $K^+K^-$ channel shows tension with the HLMNT11 analysis, where the new data in this channel from BaBar~\cite{Lees:2013gzt} and CMD-3~\cite{Kozyrev:2017agm} have incurred a large increase in the mean value, whilst also improving the uncertainty despite the small increase in global $\sqrt{\chi^2_{\rm min}/{\rm d.o.f}}$. This is also the case for the $\pi^+\pi^-\pi^+\pi^-$ channel. Other tensions include the $K^0_S K^0_L$, $\eta\pi^+\pi^-$, $\eta\omega$ and $\omega(\rightarrow\pi^0\gamma)\pi^0$ channels, where again, the new, more precise data have resulted in changes outside the quoted HLMNT11 uncertainties. The $KK2\pi$ channel exhibits a similar change as discussed in Section~\ref{Sec:KKnpi}. All other channels are in good agreement between the different analyses. It it important to note that this work includes three channels that were not included as part of the HLMNT11 analysis: $(\eta\pi^+\pi^-\pi^0)_{\text{no }\omega}$, $\eta\omega\pi^0$ and $\eta\big(\rightarrow$ non-purely pionic (npp)$\big)K\bar{K}_{\text{no }\phi \rightarrow K\bar{K}}$, where these final states were previously unmeasured by experiment and not estimated through isospin relations. Overall, due to the large reduction in the $\pi^+\pi^-$ channel, it is found that the estimate of $a_{\mu}^{\rm had, \, LO \, VP}$ has decreased between the HLMNT11 analysis and this work, although this decrease is well within the uncertainty. In total, the uncertainty has been reduced by $\sim 42\%$ with respect to the HLMNT11 analysis. 

\subsection{Comparison with other similar works}\label{sec:DHMZcompare}

\begin{table}[htbp]
\vspace{-1cm}
\scalebox{0.88}{
 {\renewcommand{\arraystretch}{0.95}
  \begin{tabular}{|l|c|c|c|}
\hline																							
{\bf Channel}	&	{\bf This work (KNT18)}							&	{\bf DHMZ17}﻿~\cite{Davier:2017zfy}							&	{\bf Difference}					\\
\hline																							
\multicolumn{4}{|c|}{Data based channels ($\sqrt{s} \leq 1.8$ GeV)}																							\\
\hline																							
$\pi^0\gamma$ (data + ChPT)	&	$	\hphantom{0}	4.58	\pm	0.10		$	&	$	\hphantom{0}	4.29	\pm	0.10		$	&	$		0.29		$	\\
$\pi^+\pi^-$ (data + ChPT)	&	$		503.74	\pm	1.96	\hphantom{0}	$	&	$		507.14	\pm	2.58	\hphantom{0}	$	&	$		-3.40	\hphantom{-}	$	\\
$\pi^+\pi^-\pi^0$ (data + ChPT)	&	$		47.70	\pm	0.89		$	&	$		46.20	\pm	1.45		$	&	$		1.50		$	\\
$\pi^+\pi^-\pi^+\pi^-$	&	$		13.99	\pm	0.19		$	&	$		13.68	\pm	0.31		$	&	$		0.31		$	\\
$\pi^+\pi^-\pi^0\pi^0$	&	$		18.15	\pm	0.74		$	&	$		18.03	\pm	0.54		$	&	$		0.12		$	\\
$(2\pi^+2\pi^-\pi^0)_{\text{no }\eta}$	&	$	\hphantom{0}	0.79	\pm	0.08		$	&	$	\hphantom{0}	0.69	\pm	0.08		$	&	$		0.10		$	\\
$3\pi^+3\pi^-$	&	$	\hphantom{0}	0.10	\pm	0.01		$	&	$	\hphantom{0}	0.11	\pm	0.01		$	&	$		-0.01	\hphantom{-}	$	\\
$(2\pi^+2\pi^-2\pi^0)_{\text{no }\eta\omega}$	&	$	\hphantom{0}	0.77	\pm	0.11		$	&	$	\hphantom{0}	0.72	\pm	0.17		$	&	$		0.05		$	\\
$K^+K^-$	&	$		23.00	\pm	0.22		$	&	$		22.81	\pm	0.41		$	&	$		0.19		$	\\
$K^0_S K^0_L$	&	$		13.04	\pm	0.19		$	&	$		12.82	\pm	0.24		$	&	$		0.22		$	\\
$KK\pi$	&	$	\hphantom{0}	2.44	\pm	0.11		$	&	$	\hphantom{0}	2.45	\pm	0.15		$	&	$		-0.01	\hphantom{-}	$	\\
$KK2\pi$	&	$	\hphantom{0}	0.86	\pm	0.05		$	&	$	\hphantom{0}	0.85	\pm	0.05		$	&	$		0.01		$	\\
$\eta\gamma$ (data + ChPT)	&	$	\hphantom{0}	0.70	\pm	0.02		$	&	$	\hphantom{0}	0.65	\pm	0.02		$	&	$		0.05		$	\\
$\eta\pi^+\pi^-$	&	$	\hphantom{0}	1.18	\pm	0.05		$	&	$	\hphantom{0}	1.18	\pm	0.07		$	&	$		0.00		$	\\
$(\eta\pi^+\pi^-\pi^0)_{\text{no }\omega}$	&	$	\hphantom{0}	0.48	\pm	0.12		$	&	$	\hphantom{0}	0.39	\pm	0.12		$	&	$		0.09		$	\\
$\eta2\pi^+2\pi^-$	&	$	\hphantom{0}	0.03	\pm	0.01		$	&	$	\hphantom{0}	0.03	\pm	0.01		$	&	$		0.00		$	\\
$\eta\omega$	&	$	\hphantom{0}	0.29	\pm	0.02		$	&	$	\hphantom{0}	0.32	\pm	0.03		$	&	$		-0.03	\hphantom{-}	$	\\
$\om(\to\pi^0\gamma)\pi^0$	&	$	\hphantom{0}	0.87	\pm	0.02		$	&	$	\hphantom{0}	0.94	\pm	0.03		$	&	$		-0.07	\hphantom{-}	$	\\
$\eta\phi$	&	$	\hphantom{0}	0.33	\pm	0.03		$	&	$	\hphantom{0}	0.36	\pm	0.03		$	&	$		-0.03	\hphantom{-}	$	\\
$\phi\to$ unaccounted	&	$	\hphantom{0}	0.04	\pm	0.04		$	&	$	\hphantom{0}	0.05	\pm	0.00		$	&	$		-0.01	\hphantom{-}	$	\\
$\eta\om\pi^0$	&	$	\hphantom{0}	0.10	\pm	0.05		$	&	$	\hphantom{0}	0.06	\pm	0.04		$	&	$		0.04		$	\\
﻿$\eta(\rightarrow{\rm npp})K\bar{K}_{\text{no }\phi \rightarrow K\bar{K}}$	&	$	\hphantom{0}	0.00	\pm	0.01		$	&	$	\hphantom{0}	0.01	\pm	0.01		$	&	$	\hphantom{*}	-0.01$*	$\hphantom{-}	$	\\
\hline																							
\multicolumn{4}{|c|}{Estimated contributions ($\sqrt{s} \leq 1.8$ GeV) }																							\\
\hline																							
$(\pi^+\pi^-3\pi^0)_{\text{no }\eta}$	&	$	\hphantom{0}	0.40	\pm	0.04		$	&	$	\hphantom{0}	0.35	\pm	0.04		$	&	$		0.05		$	\\
$(\pi^+\pi^-4\pi^0)_{\text{no }\eta}$	&	$	\hphantom{0}	0.12	\pm	0.12		$	&	$	\hphantom{0}	0.11	\pm	0.11		$	&	$		0.01		$	\\
$KK3\pi$	&	$	\hphantom{0}	-0.02	\pm	0.01	\hphantom{-}	$	&	$	\hphantom{0}	-0.03	\pm	0.02	\hphantom{-}	$	&	$		0.01		$	\\
$\omega(\rightarrow{\rm npp})2\pi$	&	$	\hphantom{0}	0.08	\pm	0.01		$	&	$	\hphantom{0}	0.08	\pm	0.01		$	&	$		0.00		$	\\
$\omega(\rightarrow{\rm npp})3\pi$	&	$	\hphantom{0}	0.10	\pm	0.02		$	&	$	\hphantom{0}	0.36	\pm	0.01		$	&	$		-0.26	\hphantom{-}	$	\\
$\omega(\rightarrow{\rm npp})KK$	&	$	\hphantom{0}	0.00	\pm	0.00		$	&	$	\hphantom{0}	0.01	\pm	0.00		$	&	$		-0.01	\hphantom{-}	$	\\
$\eta\pi^+\pi^-2\pi^0$	&	$	\hphantom{0}	0.03	\pm	0.01		$	&	$	\hphantom{0}	0.03	\pm	0.01		$	&	$		0.00		$	\\
\hline																							
\multicolumn{4}{|c|}{Other contributions}																							\\
\hline																							
$J/\psi$	&	$	\hphantom{0}	6.26	\pm	0.19		$	&	$	\hphantom{0}	6.28	\pm	0.07		$	&	$		-0.02	\hphantom{-}	$	\\
$\psi'$	&	$	\hphantom{0}	1.58	\pm	0.04		$	&	$	\hphantom{0}	1.57	\pm	0.03		$	&	$		0.01		$	\\
$\Upsilon(1S-4S)$	&	$	\hphantom{0}	0.09	\pm	0.00		$	&				-				&	$	\hphantom{**}	0.09		$**	\\
\hline																							
\multicolumn{4}{|c|}{Contributions by energy region}																							\\
\hline																							
$1.8 \leq \sqrt{s} \leq 3.7$ GeV	&	$		34.54	\pm	0.56		$ (data)	&	$	\hphantom{***}	33.45	\pm	0.65		$ (pQCD)***	&	$		1.09		$	\\
$3.7 \leq \sqrt{s} \leq 5.0$ GeV	&	$	\hphantom{0}	7.33	\pm	0.11		$ (data)	&	$	\hphantom{0}	7.29	\pm	0.03		$ (data)	&	$		0.04		$	\\
$5.0 \leq \sqrt{s} \leq 9.3$ GeV	&	$	\hphantom{0}	6.62	\pm	0.10		$ (data)	&	$	\hphantom{0}	6.86	\pm	0.04		$ (pQCD)	&	$		-0.24	\hphantom{-}	$	\\
$9.3 \leq \sqrt{s} \leq 12.0$ GeV	&	$	\hphantom{0}	1.12	\pm	0.01		$ (data+pQCD)	&	$	\hphantom{0}	1.21	\pm	0.01		$ (pQCD)	&	$		-0.09	\hphantom{-}	$	\\
$12.0 \leq \sqrt{s} \leq 40.0$ GeV	&	$	\hphantom{0}	1.64	\pm	0.00		$ (pQCD)	&	$	\hphantom{0}	1.64	\pm	0.00		$ (pQCD)	&	$		0.00		$	\\
$> 40.0$ GeV	&	$	\hphantom{0}	0.16	\pm	0.00		$ (pQCD)	&	$	\hphantom{0}	0.16	\pm	0.00		$ (pQCD)	&	$		0.00		$	\\
\hline																							
\hline																							
Total	&	$		693.3	\pm	2.5	\hphantom{0}	$	&	$		693.1	\pm	3.4	\hphantom{0}	$	&	$		0.2	\hphantom{0}	$	\\
\hline																																																																																										
  \multicolumn{4}{l}{ }  \\																			\multicolumn{4}{l}{*DHMZ have not removed the decay of $\eta$ to pionic states which incurs a double counting of this }  \\
  \multicolumn{4}{l}{ \  contribution with the $KKn\pi$ channels.}  \\
 \multicolumn{4}{l}{**DHMZ include the contributions from the $\Upsilon$ resonances in the energy region $9.3 \leq \sqrt{s} \leq 12.0$ GeV.}  \\
\multicolumn{4}{l}{***DHMZ have inflated errors to account for differences between data and pQCD.}  \\
  \end{tabular} 
  }
}\caption{Comparison of the contributions to $a_{\mu}^{\rm had, \, LO \, VP}$ calculated by DHMZ17 and in this work (KNT18), where all results are given in units $a_{\mu}^{\rm had, \, LO \, VP} \times 10^{10}$. The first column indicates the final state or individual contribution, the second column gives the KNT18 estimate, the third column states the DHMZ17 estimate and the last column gives the difference between the two evaluations. For the final states in this work that have low energy contributions estimated from chiral perturbation theory (see~\cite{Hagiwara:2003da}), the contributions from these regions have been added to the contributions from the respective data.}\label{tab:DHMZvsKNT}
\end{table}

The DHMZ group have recently released a new estimate of $a_{\mu}^{\rm had, \, LO \, VP}$~\cite{Davier:2017zfy} which, due to a similar data input, is directly comparable with this work and provides insight into how choices with regards to the data combination can affect results. In particular, with the uncertainties of $a_{\mu}^{\rm had, \, LO \, VP}$ from both the KNT18 and DHMZ17 analyses now less than $0.5\%$, it is important that these differences are understood in order to quantify the reliability of different approaches and results. In~\cite{Davier:2017zfy}, the authors provide a channel-by-channel breakdown of their estimates for the different finals states, which are compared to the respective estimates from this work in Table~\ref{tab:DHMZvsKNT}. For the exclusive data channels, the DHMZ group choose to take the contributions from these data up to 1.8 GeV, relying on estimates from pQCD above this (with inflated errors for the pQCD data below the $c\bar{c}$ threshold). As such, the estimates from this work in Table~\ref{tab:DHMZvsKNT} have been recalculated to mimic the chosen energy regions of the DHMZ analysis and allow for a consistent comparison.

When comparing the total estimate of $a_{\mu}^{\rm had, \, LO \, VP}$ from the two analyses, the results seem to be in very good agreement. However, as can be seen from Table~\ref{tab:DHMZvsKNT}, this masks much larger differences in the estimates from individual channels. The most striking of these is the estimate for the $\pi^+\pi^-$ channel, where there is a tension of slightly more than $1\sigma$ between the KNT18 and DHMZ17 results. This is unexpected when considering the data input for both analyses are likely to be similar and, therefore, points to marked differences in the way the data are combined. The higher value of the DHMZ17 estimate seems to suggest that their data combination favours the data from the BaBar measurement, with this data set being the only single set that could influence the mean value of the $\pi^+\pi^-$ channel to be as high. This behaviour is similar to the result obtained from combining the $\pi^+\pi^-$ data using only a simple weighted average as discussed in Section~\ref{chap:pipi}. In turn, this effect is compensated by other major sub-leading final states having larger estimates in this work compared to the DHMZ17 analysis. Specifically, the $\pi^+\pi^-\pi^0$, $\pi^+\pi^-\pi^+\pi^-$ and $K^0_SK^0_L$ estimates are noticeably lower in the DHMZ17 analysis. In addition, there is tension in the region between $1.8 \leq \sqrt{s} \leq 3.7$ GeV, where the choice to use data in this region has a higher integrated contribution to $a_{\mu}^{\rm had, \, LO \, VP}$ than the DHMZ17 estimate from pQCD. This is particularly significant when reconsidering Figure~\ref{fig:exc.vs.inc.vs.pQCD}, where it was observed that the sum of exclusive states from in the range $1.8 \leq \sqrt{s} \leq 2.0$ GeV has a cross section that is lower than the estimate from pQCD. The differences seen in Table~\ref{tab:DHMZvsKNT} above 1.8 GeV are then caused by cross section data below the charm production threshold being higher than pQCD (see Figure~\ref{fig:inc2-4}) and lower than pQCD above it (see Figure~\ref{fig:incFull}). It should be noted that the estimate for the $\omega(\rightarrow{\rm npp})3\pi$ final state from isospin relations, although only a small contribution to $a_{\mu}^{\rm had, \, LO \, VP}$, exhibits a significant difference between the two analyses, suggesting a different relation has been used in the DHMZ17 analysis than in this work.

\begin{figure}[!t] 
  \centering
    \includegraphics[width=0.82\textwidth]{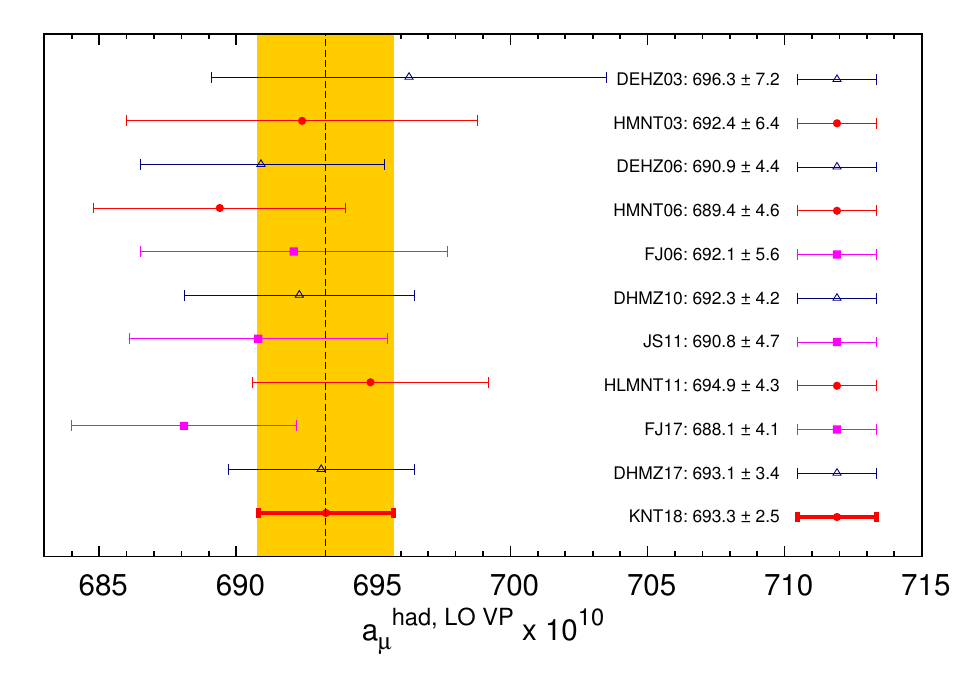}
     \caption{\small Comparison of recent and previous evaluations of $a_{\mu}^{\rm had, \, LO \, VP}$ determined from $e^+e^-\rightarrow {\rm hadrons}$ cross section data. The analyses listed in chronological order are: DEHZ03~\cite{Davier:2003pw}, HMNT03~\cite{Hagiwara:2003da}, DEHZ06~\cite{Davier:2007ua}, HMNT06~\cite{Hagiwara:2006jt}, FJ06~\cite{Jegerlehner:2006ju}, DHMZ10~\cite{Davier:2010nc}, JS11~\cite{Jegerlehner:2011ti}, HLMNT11~\cite{Hagiwara:2011af}, FJ17~\cite{Jegerlehner:2017lbd} and DHMZ17~\cite{Davier:2017zfy}. The prediction from this work is listed as KNT18, which defines the uncertainty band that the other analyses are compared to.}     \label{amuHVPCompare}
\end{figure}

As well as the DHMZ analysis, an updated work by F.~Jegerlehner (FJ17)~\cite{Jegerlehner:2017lbd} resulted in an estimate of $a_{\mu, {\rm FJ17}}^{\rm had, \, LO \, VP} = (688.07 \pm 4.14) \times 10^{-10}$ based on the available $e^+e^-$ data. Within errors, this result is in agreement with both this work and the DHMZ17 analysis. Interestingly, unlike the comparison with DHMZ17, the two-pion contribution in the energy range $0.316 \leq\sqrt{s}\leq 2.0$ GeV is found in the FJ17 analysis to be $a_{\mu, {\rm FJ17}}^{\pi^+\pi^-} = (502.16 \pm 2.44) \times 10^{-10}$~\cite{Jegerlehner:2017gek}, which is in good agreement with the KNT18 estimate in the same energy range of  $a_{\mu, {\rm KNT18}}^{\pi^+\pi^-} = (501.68 \pm 1.71) \times 10^{-10}$. However, a more detailed comparison with the estimates of other channels determined in~\cite{Jegerlehner:2017lbd,Jegerlehner:2017gek} is not possible as the FJ17 analysis chooses to estimate certain resonance contributions using available parametrisations~\cite{PDG2016} instead of using the available data. A comparison of recent and previous evaluations of $a_{\mu}^{\rm had, \, LO \, VP}$ determined from $e^+e^-\rightarrow {\rm hadrons}$ cross section data is shown in Figure~\ref{amuHVPCompare},\footnote{In addition, an evaluation using a global fit based on the Broken Hidden Local Symmetry model~\cite{Benayoun:2015gxa} prefers lower values for $a_{\mu}^{\rm had, \, LO \, VP}$ than those displayed in Figure~\ref{amuHVPCompare}.} which highlights the agreement between the different works and the improvement in the precision of the respective analyses.

\subsection{SM prediction of $g-2$ of the muon} \label{g-2muon}

The Standard Model (SM) prediction of the anomalous magnetic moment of the muon is determined by summing the contributions from all sectors of the SM, such that
\beq \label{amuSMeq}
a_{\mu}^{\rm SM}  = a_{\mu}^{\rm QED} + a_{\mu}^{\rm EW} + a_{\mu}^{\rm had} \, ,
\eeq
where $a_{\mu}^{\rm QED}$ is the QED contribution, $a_{\mu}^{\rm EW}$ is the electro-weak contribution and $a_{\mu}^{\rm had}$ are the hadronic contributions due to the hadronic vacuum polarisation and hadronic light-by-light scattering (see equation~\eqref{eq:amu_had}). The QED contributions are known up to and including five-loop accuracy. The five-loop calculation has recently been completed numerically 
by Kinoshita {\em et al.}~\cite{Aoyama:2012wk,Aoyama:2017uqe} to evaluate all 12,762 five-loop diagrams. This calculation includes all contributions that are due to photons and leptons alone. They are found to be
\begin{align}
a_{\mu}^{\rm QED} & = 11 \ 658 \ 471.8971 \ (0.0007) \ (0.0017) \ (0.0006) \ (0.0072) \times 10^{-10} \nonumber
\\
\
& = (11 \ 658 \ 471.8971 \pm 0.007)  \times 10^{-10}  \, ,
\end{align}
where the uncertainties are owing to the uncertainty on the lepton masses, the four-loop contributions, the five-loop contributions and the determination of $\alpha$ using measurements of $^{87}{\rm Rb}$, respectively. With such a precise determination of $a_{\mu}^{\rm QED}$ resulting from a perturbative series that converges extremely well, the QED result seems stable. It should be noted, however, that the four-loop and five-loop contributions rely heavily on numerical integrations and independent checks of these results are crucial. This has been recently accomplished through several different analyses~\cite{Kataev:2012kn,Baikov:2013ula,Kurz:2013exa, Kurz:2015bia,Kurz:2016bau,Laporta:2017okg,Volkov:2017xaq}, which corroborate the results from Kinoshita and collaborators. Therefore, it is safe to assume that the estimate for the QED contribution is well under control.

The contribution from the EW sector is well known to two-loop accuracy~\cite{Czarnecki:1995wq,Czarnecki:1995sz,Peris:1995bb,Czarnecki:2002nt,Czarnecki:2017rlm}. With the mass of the Higgs now known, the updated estimate~\cite{Gnendiger:2013pva} gives
\beq
a_{\mu}^{\rm EW}  = (15.36 \pm 0.10) \times 10^{-10} \, ,
\eeq
where the knowledge of the Higgs mass has halved the uncertainty of this contribution compared to the estimate used in~\cite{Hagiwara:2011af}. Although a relatively small contribution when compared to $a_{\mu}^{\rm QED}$, the uncertainty is not negligible considering the projected experimental accuracy, but is small when compared to the hadronic uncertainties. However, with this contribution known safely to two-loop accuracy, the electroweak estimate is also very well under control.  

For the hadronic vacuum polarisation contributions, the leading order and next-to-leading order contributions have been calculated in this work. The LO contribution, from equation~\eqref{LOHVP_KNT18}, was found to be $a_{\mu}^{\rm had, \, LO \, VP}  =  (693.26 \pm 2.46) \times 10^{-10}$ and the NLO was given in equation~\eqref{NLOHVP_KNT18} as $a_{\mu}^{\rm had, \, NLO \, VP}  =  (-9.82 \pm 0.04) \times 10^{-10}$. The calculation of the NNLO hadronic vacuum polarisation contribution has been achieved for the first time in~\cite{Kurz:2014wya} (see also the evaluation in~\cite{Jegerlehner:2017lbd}), where contributions from five individual classes of diagrams (each with an independent, respective kernel function) were determined. It was found to be $a_{\mu}^{\rm had, \, NNLO \, VP}  =  ( 1.24 \pm 0.01) \times 10^{-10}.$ This estimate is slightly larger than expected and results in a slight increase to the mean value of $a_{\mu}^{\rm had, VP}$. Summing these, the total contribution to the anomalous magnetic moment from the hadronic vacuum polarisation is
\beq \label{eq:amuHVPfull}
a_{\mu}^{\rm had, \, VP}  =  (684.68 \pm 2.42) \times 10^{-10} \, .
\eeq
It should be noted that the negative NLO contribution results in an anti-correlation between its uncertainty and the uncertainty from the LO contribution, consequently resulting in a slight reduction in the overall uncertainty that has been incorporated into equation~\eqref{eq:amuHVPfull}. 

The hadronic LbL contributions, although small compared to the hadronic vacuum polarisation sector, have, in the past, been determined through model-dependent approaches. These are based on meson exchanges, the large $N_c$ limit, ChPT estimates, short distance constraints from the operator product expansion and pQCD. Over time, several different approaches to evaluating $a_{\mu}^{\rm had, \, LbL}$ have been attempted, resulting in good agreement for the leading $N_c$ ($\pi^0$ exchange) contribution, but differing for sub-leading effects. A commonly quoted determination of the LbL contribution is the `Glasgow consensus' estimate of $a_{\mu}^{\rm had, \, LbL}(\text{`Glasgow consensus'})  =  (10.5  \pm 2.6) \times 10^{-10}$~\cite{Prades:2009tw} (alternatively, see~\cite{Melnikov:2003xd,Nyffeler:2009tw,Jegerlehner:2009ry,Colangelo:2014qya}). However, recent works~\cite{Jegerlehner:2015stw,Pauk:2014rta,Nyffeler:2016gnb} have re-evaluated the contribution to $a_{\mu}^{\rm had, LbL}$ due to axial exchanges, where it has been found that this contribution has, in the past, been overestimated due to an incorrect assumption that the form factors for the axial meson contribution are symmetric under the exchange of two photon momenta~\cite{Jegerlehner:2015stw}. Under this assumption, the determination in~\cite{Melnikov:2003xd} previously found the axial vector contribution to be $a_{\mu}^{\rm had, \, LbL; \, axial} =  (2.2  \pm 0.5) \times 10^{-10}$. Correcting this reduces this contribution to $a_{\mu}^{\rm had, \, LbL; \, axial} =  (0.8  \pm 0.3) \times 10^{-10}$~\cite{Jegerlehner:2015stw,Pauk:2014rta}. Applying this adjustment to the `Glasgow consensus' result  (which used a value of the axial vector contribution in~\cite{Prades:2009tw} of $a_{\mu}^{\rm had, \, LbL; \, axial} =  (1.5  \pm 1.0) \times 10^{-10}$), the estimate in~\cite{Nyffeler:2016gnb} finds
\beq\label{PdRV_glasgow-correct}
a_{\mu}^{\rm had, \, LbL} =  (9.8  \pm 2.6) \times 10^{-10} \ \, ,
\eeq
which is the chosen estimate for $a_{\mu}^{\rm had, \, LbL} $ in this work. This result is notably lower than the previously accepted LbL estimates and will incur an overall downward shift on $a_{\mu}^{\rm SM}$. It is, however, still within the original uncertainties when comparing with the original `Glasgow consensus' estimate. Alternatively, it should be noted that the estimate of $a_{\mu}^{\rm had, \, LbL} = (10.2 \pm 3.9) \times 10^{-10}$~\cite{Nyffeler:2016gnb,Nyffeler:2017ohp}, which is a result that is independent of the `Glasgow consensus' estimate, could be employed here. In addition, the recent work~\cite{Colangelo:2014qya} has provided an estimate for the next-to-leading order hadronic LbL contribution. It has found $a_{\mu}^{\rm had, NLO-LbL}  =  (0.3  \pm 0.2) \times 10^{-10}$, which does not alter the hadronic LbL contribution significantly, but is taken into account in the full SM prediction given below.

\begin{figure}[!t] 
  \centering
    \includegraphics[width=0.82\textwidth]{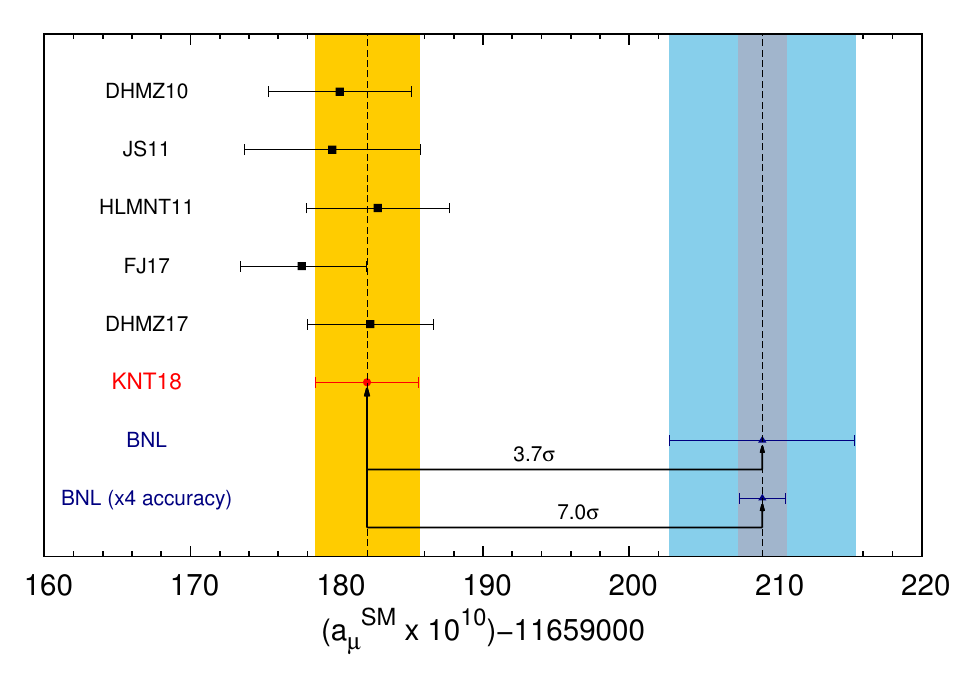}
     \caption{\small A comparison of recent and previous evaluations of $a_{\mu}^{\rm SM}$. The analyses listed in chronological order are: DHMZ10~\cite{Davier:2010nc}, JS11~\cite{Jegerlehner:2011ti}, HLMNT11~\cite{Hagiwara:2011af}, FJ17~\cite{Jegerlehner:2017lbd} and DHMZ17~\cite{Davier:2017zfy}. The prediction from this work is listed as KNT18, which defines the uncertainty band that other analyses are compared to. The current uncertainty on the experimental measurement~\cite{Bennett:2002jb,Bennett:2004pv,Bennett:2006fi,PDG2016} is given by the light blue band. The light grey band represents the hypothetical situation of the new experimental measurement at Fermilab yielding the same mean value for $a_{\mu}^{\rm exp}$ as the BNL measurement, but achieving the projected four-fold improvement in its uncertainty~\cite{Grange:2015fou}.}     \label{amuCompare}
\end{figure} 
Much work has also been directed at the possibility of a model independent calculation of $a_{\mu}^{\rm had, LbL}$ to further consolidate the SM prediction of $a_{\mu}$. One approach involves the measurement of transition form factors by KLOE-2 and BESIII, which can be expected to constrain the leading pseudoscalar-pole ($\pi^0,\eta,\eta'$) contribution to a precision of approximately 15\%~\cite{Nyffeler:2016gnb}. Alternatively, the pion transition form factor ($\pi^0\rightarrow\gamma^*\gamma^*$) can be calculated on the lattice for the same purpose~\cite{Gerardin:2016cqj}. New efforts into the prospects of determining $a_{\mu}^{\rm had, LbL}$ using dispersive approaches are also very promising~\cite{Colangelo:2014dfa,Colangelo:2014pva,Pauk:2014rfa,Colangelo:2015ama,Colangelo:2017qdm,Colangelo:2017fiz}, where the dispersion relations are formulated to calculate either the general hadronic LbL tensor or to calculate $a_{\mu}^{\rm had, LbL}$ directly. These approaches will allow for the determination of the hadronic LbL contributions from experimental data and, at the very least, will invoke stringent constraints on future estimates. Lastly, there has been huge progress in developing methods for a direct lattice simulation of $a_{\mu}^{\rm had, LbL}$~\cite{Pascalutsa:2010sj,Pascalutsa:2012pr,Blum:2014oka,Blum:2015gfa,Blum:2016lnc,Blum:2017cer,Asmussen:2017bup,Gerardin:2016cqj}. With a proof of principle already well established, an estimate of approximately 10\% accuracy seems possible in the near future. Considering these developments and the efforts of the {\em Muon $g-2$ Theory Initiative}~\cite{TGm2} to promote the collaborative work of many different groups, the determination of $a_{\mu}^{\rm had, LbL}$ on the level of the `Glasgow consensus' will, at the very least, be consolidated and a reduction of the uncertainty seems highly probable on the timescales of the new $g-2$ experiments.

Following equation \eqref{amuSMeq}, the sum of all the sectors of the SM results in a total value of the anomalous magnetic moment of the muon of
\beq \label{amuSMfinal}
a_{\mu}^{\rm SM}  =  (11\ 659 \ 182.04  \pm 3.56) \times 10^{-10} \, ,
\eeq
where the uncertainty is determined from the uncertainties of the individual SM contributions added in quadrature. Comparing this with the current experimental measurement given in equation \eqref{amuExp} results in a deviation of
\begin{align} 
\Delta a_{\mu} = (27.06 \pm 7.26)\times 10^{-10}\, ,
\end{align}
corresponding to a $3.7\sigma$ discrepancy. This result is compared with other determinations of $a_{\mu}^{\rm SM}$ in Figure~\ref{amuCompare}. In particular, a comparison with the HLMNT11 estimate given in equation \eqref{amuSM_HLMNT11} shows an improvement in the total uncertainty of $a_{\mu}^{\rm SM}$ of $\sim 27\%$. It should be noted that although, as stated above, the DHMZ17 estimate for $a_{\mu}^{\rm had, \, LO \, VP}$~\cite{Davier:2017zfy} is lower than the value obtained in this work, the estimate of $a_{\mu}^{\rm SM}$ from DHMZ17 is higher than the estimate from this analysis as DHMZ17 choose to use the estimate for the hadronic light-by-light contribution of $a_{\mu}^{\rm had, LbL} = (10.5 \pm 2.6) \times 10^{-10}$~\cite{Prades:2009tw}.

\subsection{Determination of $\alpha(M_{Z}^2)$}\label{sec:alphaMz}

From equation~\eqref{delalphahad_KNT18}, the five-flavour hadronic contribution to $\Delta\alpha(M_Z^2)$ is found to be \allowbreak $\Delta\alpha_{\rm had}^{(5)}(M_Z^2) \allowbreak = (276.11 \pm 1.11) \times 10^{-4}$. Combining this with the leptonic contribution $\Delta\alpha_{\rm lep}(M_Z^2) = (314.979 \pm 0.002) \times 10^{-4}$~\cite{Steinhauser:1998rq,Sturm:2013uka} and the contribution due to the top quark $\Delta\alpha_{\rm top}(M_Z^2) = (-0.7180 \pm 0.0054) \times 10^{-4}$~\cite{Chetyrkin:1995ii,Kuhn:1998ze}, the total value of the QED coupling at the Z boson mass is found in this work to be
\begin{align}
\alpha^{-1}(M_Z^2) & = \Big(1-\Delta\alpha_{\rm lep}(M_Z^2)-\Delta\alpha_{\rm had}^{(5)}(M_Z^2)-\Delta\alpha_{\rm top}(M_Z^2)\Big)\alpha^{-1} \nonumber
\\
\
 & =  128.946 \pm 0.015\, .
\end{align}
Here, the top contribution is determined using the top quark mass, $m_t = 173.1 \pm 0.6$ GeV and the value for strong coupling constant at the mass of the $Z$ boson, $\alpha_s(M_Z) = 0.1182(12)$~\cite{PDG2016}.
A comparison of these results with other determinations of $\Delta\alpha_{\rm had}^{(5)}(M_Z^2)$ and $\alpha^{-1}(M_Z^2)$ is given in Table~\ref{tab:comapredelalpha}. Note that, as discussed in Section~\ref{sec:amuanddelalpha}, the different weighting of the kernel function of the $\Delta\alpha_{\rm had}^{(5)}(M_Z^2)$ dispersion integral compared to the integral for $a_{\mu}^{\rm had, \, LO \, VP}$ results in contributions from higher energy regions  having a larger influence on $\Delta\alpha_{\rm had}^{(5)}(M_Z^2)$ than on $a_{\mu}^{\rm had, \, LO \, VP}$. Therefore, the choice to use either the available inclusive data or pQCD above $\sim2$ GeV as discussed in Section~\ref{sec:DHMZcompare} can have marked differences on the values and errors obtained for $\Delta\alpha_{\rm had}^{(5)}(M_Z^2)$. 
\begin{table}[!t]
\vspace{-0.cm}
\centering
\scalebox{1.0}{
  {\renewcommand{\arraystretch}{1.0}
  \begin{tabular}{|l|c|c|}
\hline													
{\bf Analysis}	&	$\Delta\alpha_{\rm had}^{(5)}(M_Z^2)\times10^{4}$			&	$\alpha^{-1}(M_Z^2)$	\\
\hline	
DHMZ10~\cite{Davier:2010nc}	&	$	275.59 \pm 1.04	$	&	$	128.952 \pm 0.014 	$	\\			
HLMNT11~\cite{Hagiwara:2011af} &	$	276.26 \pm 1.38	$	&	$	128.944 \pm 0.019	$	\\
FJ17~\cite{Jegerlehner:2017zsb}	&	$	277.38 \pm 1.19	$	&	$	 128.919 \pm 0.022	$	\\
DHMZ17~\cite{Davier:2017zfy}	&	$	276.00 \pm 0.94	$	&	$	128.947 \pm 0.012	$	\\			
KNT18 [This work]	&	$	276.11 \pm 1.11	$	&	$	128.946 \pm 0.015	$	\\
\hline													
  \end{tabular} 
  }
}\caption{Comparison of recent and previous evaluations of $\Delta\alpha_{\rm had}^{(5)}(M_Z^2)$ determined from $e^+e^-\rightarrow {\rm hadrons}$ cross section data and the corresponding results for $\alpha^{-1}(M_Z^2)$.}\label{tab:comapredelalpha} 
\end{table}

\section{Conclusions and future prospects for the muon $g-2$}\label{Conclusions}

This analysis, KNT18, has completed a full re-evaluation of the hadronic vacuum polarisation contributions to the anomalous magnetic moment of the muon, $a_{\mu}^{\rm had, \, VP}$, and the hadronic contributions to the effective QED coupling at the $Z$ boson mass, $\Delta\alpha_{\rm had}(M_Z^2)$. These quantities have been determined using the available $e^+e^- \rightarrow {\rm hadrons}$ cross section data as input into corresponding dispersion relations, with an aim to achieve both accurate and reliable results from a predominantly data driven analysis. Since~\cite{Hagiwara:2011af}, all aspects concerning the radiative corrections of the data and the data combination have been reassessed in this work. Specifically, the data are now combined using an iterative, linear $\chi^2$-minimisation developed from a method that has been advocated to be free of bias and that has been studied in detail. Importantly, this data combination method allows for the full use of any available correlated statistical and systematic uncertainties, incorporating experimental covariance matrices in the combination in a bias-free approach. These changes, plus the large quantity of new hadronic cross section data, have resulted in improved estimates for nearly all hadronic channels. This is particularly true for the $\pi^+\pi^-$ channel, where the precision of this final state has improved by approximately one third compared to~\cite{Hagiwara:2011af}, with $a_{\mu}^{\pi^+\pi^-}$ from both analyses in very good agreement. Importantly, the inclusion of recently released neutral data in the $KK\pi$ and $KK2\pi$ channels has removed the need to rely on isospin relations to estimate these final states. In addition, new inclusive hadronic $R$-ratio data from the KEDR collaboration have improved the inclusive data combination. In particular, they have provided the opportunity to reconsider the transition region between the sum of exclusive states and the inclusive data, which has resulted in the transition point being chosen to be at 1.937 GeV in this work, where the different choices for the data input in this point are in agreement within errors. The complete data compilation has yielded the full hadronic $R$-ratio and its covariance matrix in the energy range $m_{\pi}\leq\sqrt{s}\leq 11.2$ GeV. Using these combined data, for the muon $g-2$, this analysis found $a_{\mu}^{\rm had, \, LO \, VP} = (693.26 \pm 2.46)\times 10^{-10}$ and $a_{\mu}^{\rm had, \, NLO \, VP} = (-9.82 \pm 0.04)\times 10^{-10}$. This has resulted in a new estimate for the Standard Model prediction of $a_{\mu}^{\rm SM}  =  (11\ 659 \ 182.04  \pm 3.56) \times 10^{-10}$, which deviates from the current experimental measurement by $3.7\sigma$. For the effective QED coupling, the new estimate in this work of $\Delta\alpha_{\rm had}^{(5)}(M_Z^2)= (276.11 \pm 1.11)\times 10^{-4}$ yields an updated value for the QED coupling at the $Z$ boson mass of $\alpha^{-1}(M_Z^2) = 128.946 \pm 0.015$.

In general, the predictions for $a_{\mu}^{\rm SM}$ have been further scrutinised and are well established. In particular, the improvements in the uncertainty on the level discussed in~\cite{Blum:2013xva} are on track in preparation for the new experimental results from Fermilab and J-PARC. The efforts of the {\em Muon $g-2$ Theory Initiative} and the groups involved within it show great progress and promise in improving the estimate of $a_{\mu}^{\rm SM}$ further. Importantly, it should be noted that there is no indication thus far that the SM prediction does not deviate with the current experimental measurement by more than $3\sigma$. For the hadronic vacuum polarisation contributions, there is scope to further improve the estimates, as new measurements of the $\pi^+\pi^-$ cross section are planned to be released in the near future from BaBar, CMD-3, SND and, possibly, BELLE-2. Also, efforts are currently being made to measure new inclusive $R$-ratio data by BESIII and the experiments at Novosibirsk (SND, CMD-3, KEDR). This, in particular, will benefit the evaluation of $\Delta\alpha_{\rm had}^{(5)}(M_Z^2)$ which, as discussed in Section~\ref{sec:alphaMz}, is more sensitive than $a_{\mu}^{\rm had, \, LO \, VP}$ to the precision of the data from higher energy regions. In addition, lattice determinations of $a_{\mu}^{\rm had, \, VP}$ are rapidly improving. Recent work that combines data from lattice QCD with those from experimental $R$-ratio data have already provided extremely accurate results that are in good agreement with the current estimates from the dispersive method~\cite{Blum:2018mom}. Furthermore, efforts to experimentally measure $a_{\mu}^{\rm had, \, LO \, VP}$ in the space-like region are progressing~\cite{Calame:2015fva,Abbiendi:2016xup} and would provide an alternative check of the results from the dispersive approach. Should any or all these advances reduce the uncertainty of $a_{\mu}^{\rm had, \, VP}$ even further, the improvement of the hadronic light-by-light estimates will become particularly crucial. This is further driven by the fact that the result for $a_{\mu}^{\rm had, \, VP}$ determined in this analysis is the first estimate of the hadronic vacuum polarisation contribution that is more precise than the currently quoted uncertainties for $a_{\mu}^{\rm had, \, LbL}$. However, given these developments in improving the Standard Model prediction of $a_{\mu}$ and the formidable progress made by the new muon $g-2$ experiments at Fermilab and J-PARC, the prospects to either establish the existence of new physics contributing to $a_{\mu}$, or to rule out many new physics scenarios, are highly compelling.

\section*{Acknowledgements}

We would first like to recognise and whole-heartedly thank Alan Martin and Kaoru Hagiwara for their many-year-long collaboration in~\cite{Hagiwara:2003da,Hagiwara:2006jt,Hagiwara:2011af}, and also thank Ruofan Liao for his collaboration in~\cite{Hagiwara:2011af}. We would like to thank Simon Eidelman, Fedor Ignatov, Andreas Nyffeler and the DHMZ17 group (Michel Davier, Andreas Hoecker, Bogdan Malaescu and Zhiqing Zhang) for numerous useful discussions. We would also like to acknowledge the discussions within the {\em Working Group on Radiative Corrections and MC Generators for Low Energies (Radio MonteCarLow)} and {\em The Muon $g-2$ Theory Initiative} concerning this work. The work of Alex Keshavarzi is supported by STFC under the consolidated grant ST/N504130/1. The work of Daisuke Nomura is supported by JSPS KAKENHI grant numbers JP16K05323 and JP17H01133. The work of Thomas Teubner is supported by STFC under the consolidated grants ST/L000431/1 and ST/P000290/1.

\end{document}